\documentclass[12pt]{article}

\usepackage{amssymb,amsmath,amsfonts,eurosym,geometry,ulem,graphicx,caption,color,setspace,sectsty,comment,footmisc,caption,pdflscape,array}
\usepackage[authoryear]{natbib}

\usepackage{hyperref}
\hypersetup{colorlinks=true, allcolors=blue}

\usepackage{booktabs}
\usepackage{threeparttable}    
\usepackage{dcolumn}
    \newcolumntype{d}[1]{D{.}{.}{#1}}

\usepackage[font=small,labelfont=bf]{caption}
\usepackage{subcaption}

\newcommand{\Autoref}[1]{%
  \begingroup%
  \def\chapterautorefname{Chapter}%
  \def\sectionautorefname{Section}%
  \def\subsectionautorefname{Subsection}%
  \def\subsubsectionautorefname{Subsubsection}%
  \def\paragraphautorefname{Paragraph}%
  \def\tableautorefname{Table}%
  \def\equationautorefname{Equation}%
  \def\appendixautorefname{Appendix}%
  \autoref{#1}%
  \endgroup%
}

\usepackage{multibib}
\newcites{supp}{Supplementary References}

\usepackage{rotating}

\newcolumntype{L}[1]{>{\raggedright\let\newline\\arraybackslash\hspace{0pt}}m{#1}}
\newcolumntype{C}[1]{>{\centering\let\newline\\arraybackslash\hspace{0pt}}m{#1}}
\newcolumntype{R}[1]{>{\raggedleft\let\newline\\arraybackslash\hspace{0pt}}m{#1}}

\begin{document}

\begin{titlepage}
\title{Identification of fiscal SVAR-IVs in \\small open economies\footnote{The authors would like to thank Jetro Anttonen, Ilkka Kiema, Samu Kärkkäinen, Niku Määttänen, Antti Ripatti and participants at the 2021 and 2023 IAAE Annual Conferences, the 2022 EEA-ESEM Congress, the 42nd and 43rd Annual Meetings of the Finnish Economic Association and Helsinki GSE Econometrics Workshop for their useful comments and suggestions. Funding provided by the Palkansaajasäätiö is gratefully acknowledged.}}
\author{Henri Keränen\thanks{University of Helsinki, Helsinki GSE, Finnish Economic Policy Council, VATT Institute for Economic Research; Email: \href{mailto:henri.keranen@helsinki.fi}{henri.keranen@helsinki.fi}} \and Sakari Lähdemäki\thanks{Etla Economic Research; Email: \href{mailto:sakari.lahdemaki@etla.fi}{sakari.lahdemaki@etla.fi}}}
\date{\today}
\maketitle
\vspace{0mm}

\begin{abstract}
\noindent We identify fiscal SVAR-IVs by utilizing unexpected variation in the output of trading partner economies, measured by professional forecast errors, to account for the systematic component of fiscal policy. Our identification builds on the small open economy assumption that these forecast errors correlate with output but are exogenous to domestic fiscal policy. In applying our approach to Canada and euro area small open economies we show that the instrument is relevant and find suggestive evidence for its exogeneity. Our baseline estimates for the two-year cumulative spending multiplier are around 1 for Canada and 0.5 for euro area small open economies. \\

\vspace{0mm}
\vspace{0mm}

\bigskip
\end{abstract}
\setcounter{page}{0}
\thispagestyle{empty}
\end{titlepage}
\pagebreak \newpage

\doublespacing

\section{Introduction} \label{sec:introduction}

Estimating the dynamic effects of policy shocks is a central but notoriously hard task in empirical macroeconometrics given the difficulty of credibly identifying exogenous variation in the policy variable \citep{nakamura2018}. To overcome this challenge, modern macro-econometrics increasingly uses external instruments to identify dynamic causal effects \citep{stock2018}. One area where the use of these instruments has become prominent is macro-fiscal literature \citep[e.g., ][]{mertens13, mertens2014reconciliation,nakamura2014fiscal,caldara2017,ramey2018} that provides estimates on the effects of fiscal policy.\footnote{Following the SVAR model of \citet{blanchard2002}, much of the prior macro-fiscal research has relied on using assumptions about timing restrictions in the real-world setting of fiscal policy to identify fiscal shocks. Another prominent identification strategy has been to use sign restrictions, as in \citet{uhlig2009}. Furthermore, another branch of the literature has utilized narratively identified plausibly exogenous policy changes or other shocks in the identification of the effects of fiscal policy. See, for example, \citet{romer2010} and \citet{mertens13,mertens2014reconciliation} in the context of tax policy changes and \citet{ramey2011} and \citet{ramey2018} in which a military-news variable is constructed to identify plausibly exogenous variation in government spending. See \citet{ramey2019} for a comprehensive summary of the macro-fiscal literature. More recently, econometric methods using heteroskedasticity and higher moments to achieve identification have gained prominence \citep[see e.g., ][]{lewis2021identifying}.} Given a valid instrument, researchers can estimate structural relationships in the presence of endogenous variables -- a situation which is emblematic of macroeconomics.

In this paper, we identify structural vector autoregressions with external instruments (SVAR-IV) to study the dynamic effects of fiscal policy in small open economies. While much of the empirical literature focuses on the US economy, the effects of fiscal policy in smaller and more open economies might differ from this benchmark. Moreover, predictions of theoretical models over the effectiveness of fiscal policy in small open economies hinge on specific assumptions over monetary policy and exchange rate regimes. These policy regimes vary across countries and size of the fiscal multiplier across these different settings is, after all, an empirical question.

Our primary contribution to this literature is to introduce an instrumental variable for aggregate output shocks that can be used to identify fiscal shocks in a small open economy setting. As shown by \citet{caldara2017}, non-fiscal proxy variables that are not correlated with the policy shocks of interest can still be used to uncover these shocks within a SVAR model. Proxies that are exogenous to policy shocks (but correlated with other shocks) solve for the identification problem by allowing the econometrician to estimate the systematic component of fiscal policy.\footnote{The systematic component reflects the reaction function of fiscal authorities or, in other words, the implicit fiscal rules that govern how fiscal policy endogenously reacts to other (non-policy) shocks in the model. At a quarterly frequency these reflect, for example, automatic stabilizers that mainly work through output.} As fiscal variables may endogenously react to changes in e.g. aggregate output, one needs to account for this reaction in a SVAR to correctly identify the policy shocks from the reduced form residuals. In practice, the elasticities of fiscal variables to other endogenous variables are estimated by instrumenting the relevant non-fiscal variables with proxies that are unrelated to fiscal shocks.

In this paper we utilize the small open economy setting and propose to use professional forecast errors of trading partner economies as a proxy for output shocks.\footnote{To the best of our knowledge, the idea of using unexpected variation in trading partner economies has not been used in the fiscal SVAR-IV context previously. However, following \citet{rigobon}, some papers studying the cyclicality of fiscal policy \citep[e.g.][]{jaimovich2007procyclicality, ilzetzki2008procyclical, vegh2015tax} have used a similar identification strategy in an effort to estimate the endogenous reaction of fiscal policy to changes in output. Unlike our study, which uses professional forecast errors, these papers have used weighted output growth in trading partners outright but have found in some cases this instrument to be weak. Arguably, forecast errors are also less prone to exogeneity concerns given that they should, from the onset, capture \textit{unexpected} variation in foreign economies. For example, a policymaker aiming to balance business cycles in a small open economy would naturally account for the expected development in foreign economies, thus likely creating a link with fiscal policy.} Given that these errors are unrelated to fiscal policy shocks of a small open economy (exogeneity) while at the same time able to explain output variations in the small domestic economy (relevance), the proposed instrument is plausibly valid. We show empirically that trading partner forecast errors are strongly correlated with the unexpected part of domestic output and that the robust first stage F-statistics when applying the instrument exceed the typical threshold value of 10. These results imply that the proposed instrument is not weak.

Earlier studies have used the utilization adjusted total factor productivity (TFP) series of \citet{fernald2014} as an instrument for output shocks. As already noted by \citet{sims}, the substantial revisions made to this series have also seen its properties change over time. We document similar findings as we find that more recent vintages of the \citet{fernald2014} series seem to have considerably lower correlation with the unexpected part of output compared to what can be found in the \citet{caldara2017} data. Furthermore, by regressing the more recent TFP vintages on proxies of fiscal shocks we find suggestive evidence against the exogeneity assumption of that instrument. This is in contrast to the instrument proposed in this paper for which we find no significant correlation with domestic fiscal proxies. Also, the instrument used in this paper is comparatively simple to construct from observable data and does not require any strong assumptions on structural forms beyond the small open economy assumption.\footnote{Note also that the proposed instrument is not a narrative instrument as, for example, in \citet{romer2010}, \citet{mertens13,mertens2014reconciliation}, or \citet{ramey2018}. That is, we do not choose which shocks are exogenous and which are not. We only assume that the forecasts we use are reasonable and that the resulting forecast errors can proxy unexpected shocks at the quarterly frequency.} In comparison, the construction of utilization-adjusted TFP involves numerous detailed assumptions.

As a second main contribution of this paper, we apply the proposed instrument to study the effectiveness of fiscal policy in two different types of small open economies: Canada and small countries of the euro area. Within our sample period, Canada sets its monetary policy stance independently and operates under a flexible exchange rate. In contrast, the euro area small open economies operate under fixed exchange rates and an exogenous monetary policy since, as individual countries, they are too small to influence the overall policy stance of the monetary union. Therefore, these two distinct types of small open economies provide an interesting comparison on how these differences might affect the effectiveness of fiscal policy.

In constructing the models, we follow the open economy VAR literature \citep[e.g., ][]{kim2008twin,ravn2012consumption,forni2016,klein2019tax} and include variables such as current account balance, real effective exchange rate and interest rate to the reduced form model in order to better account for the dynamics of open economies and to control for the monetary policy stance. In addition, we include forecast variables related to government spending and output to tackle challenges posed by the potential presence of fiscal foresight \citep[see e.g., ][]{leeper2013, forni2014}. To model the dynamics of a representative euro area small open economy we follow \citet{smallbig2013} and estimate a pooled VAR for the euro area sample.

Much of the earlier macro-fiscal research employs the \citet{blanchard2002} type identification which assumes that the contemporaneous elasticity of government spending with respect to output is zero. With the help of the instrument it is possible to estimate this key elasticity instead and thus we are able to compare results with or without this assumption. We find that even small deviations from the zero restriction can yield potentially large differences in spending multiplier estimates. Our baseline estimates for the two year cumulative spending multiplier are slightly above 1 for Canada and around 0.5 for euro area small open economies. Making the \citet{blanchard2002} zero restriction would lead to multiplier estimates that are lower than this baseline: estimate for Canada would roughly half in size while the multiplier for euro area small open economies would be close to zero. Our estimates point to a slightly negative output elasticity in the systematic component of government spending which means that more of the positive co-movement between output and government spending is attributed to policy shocks. This leads to higher spending multiplier estimates when the zero restriction is not imposed.

The finding that a country with a flexible exchange rate in Canada has a larger multiplier goes against the predictions of the traditional Mundell-Fleming framework. This finding could be rationalized, for example, by a monetary policy regime that does not work to offset the effects of fiscal shocks or by smaller import leakages when compared to small euro area economies. On the revenue side, we find that the output effects of (net) revenue shocks are smaller than those of government spending shocks. However, as these revenue shocks also generate a less persistent effect on the government budget balance, revenue side stimulus might appear as fiscally more efficient due to its smaller fiscal costs over time. The estimated impulse responses also suggest that expansionary fiscal shocks not only increase the deficit but also negatively impact the current account balance, lending support to the twin-deficits hypothesis which makes this prediction. Our main results are robust to variations in the reduced form VAR specification as well as to using forecast errors of only the largest G7 economies in the instrument. However, some of the estimated impulse responses concerning small euro area economies appear sensitive to the inclusion of Portugal to the sample.

The rest of the paper is structured as follows. \Autoref{sec:soefiscal} reviews some of the literature on the effects of fiscal policy in small open economies. In \autoref{sec:var} we lay out our empirical strategy, discuss the properties of the proposed instrument and describe the data we use. \Autoref{svar} presents the main results of our analysis along with different robustness checks while \autoref{sec:conclusion} concludes.

\section{Fiscal policy in (small) open economies}\label{sec:soefiscal}

Our research design builds on the small open economy setup and therefore it is important that the model used adequately captures the dynamic features of an open economy. While the canonical 3-variable \citet{blanchard2002} model might still be regularly applied in the literature, it does not necessarily achieve this. In fact, in many of the more recent papers studying the effects of fiscal policy, including \citet{kim2008twin}, \citet{ravn2012consumption}, \citet{forni2016} and \citet{klein2019tax}, even the USA is studied in an open economy setting. These studies find important evidence concerning the effects of fiscal policy that the typical \citet{blanchard2002} model cannot reveal. For example, they provide evidence over the twin deficit hypothesis which suggests that both fiscal deficit and current account deficit increase following fiscal stimulus. Given that all the countries studied in this paper can be considered as small open economies, it is natural to account also for open economy factors when studying the effects of fiscal policy. The open economy features are arguably even more salient for these countries than they are for the USA.

The empirical literature on the effects of fiscal policy in a \textit{small} open economy context is rather sparse. In addition to the aforementioned studies that focus on the USA, some empirical papers study different aspects of fiscal policy using panel data. For instance, \citet{beetsma2011effects} study EU countries, \citet{corsetti2012determines} focus on OECD countries while \citet{smallbig2013} have a panel of 44 countries that are categorized to different types.\footnote{There are also studies that focus on individual countries. See, for example, \citet{ravn2014effects} who study fiscal policy in Denmark and \citet{vcapek2019fiscal} who study Austria.} These papers examine how various factors influence the effectiveness of fiscal policy. For example, both \citet{smallbig2013} and \citet{beetsma2011effects} study whether the degree of openness of a country is linked to the effectiveness of fiscal policy and find evidence towards multipliers being smaller in more open economies. Additionally, these papers as well as \citet{corsetti2012determines} examine whether the exchange rate being either fixed or flexible influences the effectiveness of fiscal policy.\footnote{These papers also study other aspects such as public indebtedness, health of the financial system, and the level of development of a country.} Whereas \citet{beetsma2011effects} are unable to find meaningful differences in output effects, \citet{smallbig2013} and \citet{corsetti2012determines} find that multipliers under fixed exchange rate regimes might be larger.

The empirical literature that studies fiscal policy of open economies mainly refers to two different theoretical frameworks: the Mundell-Fleming framework and the New Keynesian framework. In the former, the exchange rate regime is key to determining the potency of fiscal policy in a small open economy. Fiscal policy under a flexible exchange rate regime is deemed inefficient. An increase in government expenditures puts upward pressure on interest rates which in turn encourages capital inflows. This increases demand for domestic currency which thus appreciates and as a result net exports reduce to offset the effects of fiscal stimulus. In contrast, under fixed exchange rates the Mundell-Fleming framework renders fiscal policy efficient: in order to maintain the fixed exchange rate, the central bank must increase its money supply, ultimately resulting in a rise in aggregate income.

The New Keynesian framework does not provide as clear predictions over the effectiveness of fiscal policy. In this framework, fiscal policy in a small open economy with a flexible exchange rate might be as efficient as under a fixed exchange rate depending on the assumptions over monetary policy. This is because the interactions between fiscal and monetary policy are more comprehensively modeled. However, adding, for example, financial frictions to the New Keynesian framework (see \citet{corsetti2013floats} or \citet{born2013exchange}) tends to make fiscal policy under fixed exchange rates more effective than under flexible exchange rates. Within this framework the real exchange rate typically appreciates after a government spending shock. This goes against the empirical evidence shown, for example, by \citet{kim2008twin}, \citet{ravn2012consumption}, \citet{bouakez2014fiscal} and \citet{klein2019tax}. Moreover, these studies relate to the so-called twin deficit hypothesis. The empirical evidence over this hypothesis is mixed \citep{klein2019tax}. Overall, the effectiveness of fiscal stimulus in small open economies still remains largely an empirical question.

Studying the effects of fiscal policy in both Canada as well as in small open economies of the euro area enables us to compare the relative effectiveness of fiscal policy in two distinct settings. Canada for one can be considered as a more traditional small open economy with a flexible exchange rate that has autonomy over its monetary policy. The euro area's small open economies, on the other hand, are open to trade and capital movements, but due to their size and membership in a large currency union, they cannot control the monetary policy stance and the exchange rates are practically fixed. According to, e.g., \citet{nakamura2014fiscal} it's reasonable to think that these two different types of small open economies might react differently to fiscal policy shocks. Studying the effects of fiscal policy in these two distinct settings can help shed light over how the contrasting economic institutions might influence the effectiveness of policy. Moreover, the fact that fiscal policy has a heightened role as a stabilization tool within a currency union \citep[see e.g., ][]{gali2008optimal} highlights the policy-relevance of this question.

\section{Empirical strategy} \label{sec:var}

\subsection{SVAR specification}

\emph{Reduced form VAR.}--- Our empirical starting point is reduced form VAR model for a vector $y'=[g,r,gdp,cab,rer,srate,defl,f_{\Delta g},f_{\Delta gdp}]$ of nine variables: general government consumption and investment ($g$), government net revenue ($r$), GDP ($gdp$), current account balance as a share of GDP ($cab$), real exchange rate ($rer$), short-term nominal interest rate ($srate$), GDP deflator ($defl$) and the one step-ahead forecasts of growth in $g$ and $gdp$ ($f_{\Delta g}$ and $f_{\Delta gdp}$). Variables $g$, $r$ and $gdp$ enter the model in real per capita terms and in natural logarithms. These three core variables feature across most of the fiscal-SVAR literature since at least \citet{blanchard2002}. The inclusion of $cab$, $rer$, $srate$ and $defl$ to the model is motivated by their role in the dynamics of open economies as in, for example, \citet{kim2008twin}, \citet{forni2016} and \citet{klein2019tax}. For Canada $rer$ is the real exchange rate between Canada and USA while for euro area economies $rer$ is between Germany and USA. Finally, the one-step ahead forecast variables $f_{\Delta g}$ and $f_{\Delta gdp}$ act as controls for possible foresight.\footnote{See, for example, \citet{leeper2013} and \citet{forni2014}. We find that the inclusion of these variables somewhat alters impulse responses for Canada whereas they have a smaller effect in the case of small euro area countries. It has been suggested that inclusion of forward-looking variables like interest rates can also control for the foresight problem \citep[see e.g., ][]{beetsma2011effects}.} In our baseline model we do not linearly detrend the main variables of interest.\footnote{Applied VAR papers vary in how they approach trends in the VAR. In \citet{blanchard2002}, a linear and a quadratic trend are included in the main specification. \citet{caldara2017} detrend some of their endogenous variables. \cite{mertens13} do not add deterministic trends in their main specification, while they test their results' robustness with such a specification. According to \citet{kilian2017}, while a VAR in levels is asymptotically valid even under true cointegration relations, its finite sample bias can be considerable. This bias is even more severe when a deterministic trend is included in the model.}

Accordingly, we specify our reduced form VAR models as follows: \begin{equation}\label{eq:varmodel}  y_{it} = c_{i} + \sum^{p}_{j=1} A_{j} y_{i,t-j} + u_{it}, \end{equation} where $p$ is lag length, $c_i$ is a vector of constants, $A_j$ are the autoregressive coefficient matrices  and $u_{it}$ the reduced form residuals. For Canada, this model is estimated by equation-by-equation OLS. For the euro area small open economies (Finland, Austria, the Netherlands, Belgium and Portugal) we estimate the VAR model in \eqref{eq:varmodel} equation-by-equation in panel form similarly to \citet{smallbig2013} with country fixed effects but with common VAR coefficients on the lag terms.\footnote{The set of euro area small open economies is chosen by first excluding the largest euro area countries, secondly considering the timing of the switch to the euro regime and lastly due to data availability.} Hence we include indexes $i$ in \eqref{eq:varmodel}. These indexes are of course redundant in the model for Canada. In the case of euro area economies we consider $rer$ and $srate$ as exogenous variables given the monetary union context whereas for Canada both of these variables are treated as endogenous.

In lag length selection, we rely on both the standard information criteria and partial autocorrelation functions of the residuals to specify a model that both contains enough information and has no autocorrelation in the residuals. For both the euro area small open economies and Canada, our baseline specification has a lag length of 5.\footnote{In \autoref{robustness} and in \hyperref[sec:online]{Online Appendix}, we also examine robustness to different choices relating to the specification of the reduced form VAR.} For constructing confidence intervals we utilize the residual-based moving block bootstrap proposed by \cite{bruggemann2016} that is shown to be applicable for Proxy-SVARs / SVAR-IVs.\footnote{In this process we also apply \citet{kilian1998} finite sample correction. Bootstrap confidence intervals are constructed using \citet{efron1979} percentiles.}

\emph{Structural form.}--- \citet{caldara2017} develop an analytical framework under which different identification schemes to estimating fiscal multipliers can be considered. In part, their SVAR framework builds on the idea of characterizing the systematic component of fiscal policy and then retrieving shocks to fiscal policy as the unexplained part in the VAR residual of the policy variable. To put their framework in more concrete terms, consider that in the data there is a positive relationship between $policy$ and $output$ but that $policy$ may systematically react to changes in $output$. Any SVAR identification scheme must then decompose this positive co-movement between shocks to $policy$ and other shocks that move $output$ \citep{caldara2017}. This structural decomposition then in turn determines the estimated effect of $policy$ on $output$.

In the \citet{blanchard2002} identification scheme government spending shock is identified under the assumption that the systematic component is zero at the quarterly frequency. It is thought that government spending does not react to other shocks contemporaneously due to implementation lags and thus government spending shocks can be recovered simply as the reduced form residual from the VAR. Similarly, government revenue shocks are identified under the assumption of no contemporaneous policy reaction to other shocks but, in contrast, (net) revenues are allowed to automatically react to output based on institutional knowledge of tax and transfers systems in order to construct the systematic component.

We build our analysis on the observation of \citet{caldara2017} that non-fiscal proxy variables can be used to identify the systematic component of fiscal policy. Suppose we have an instrumental variable $m_{it}$ that satisfies the following conditions:\begin{equation}\label{as1}
E[m_{it} e_{it}^{non{\text -}policy}] = \Gamma \not = 0
\end{equation}
\begin{equation}\label{as2}
E[m_{it} e_{it}^{policy}] = 0,
\end{equation} where $e_{it}^{non{\text -}policy}$ are the non-policy shocks and $e_{it}^{policy}$ are the policy shocks. Given that these familiar conditions of relevance and exogeneity with respect to non-policy and policy hold, one may use $m_{it}$ as an instrument to estimate the contemporaneous elasticities of the policy variables with respect to, for example, output via 2SLS. This alleviates the need for external knowledge over these elasticities.

The strategy of using non-fiscal proxies stands in contrast to some of the other studies in recent fiscal-SVAR literature that directly instrument for policy shocks instead. However, in many cases credible proxies for the policy shocks may be hard to come by. With non-fiscal proxies the identification strategy explicitly hinges on capturing the systematic component of fiscal policy instead. Note that the systematic component here may in principle also contain the policymaker's reaction to non-policy shocks, that is, the \citet{blanchard2002} assumption of no contemporaneous reaction by the policymaker due to implementation lags does not feature in either \eqref{as1} or \eqref{as2}.

Across the paper, we consider two different structural specifications which we label as BP and CK, for \citet{blanchard2002} and \citet{caldara2017} respectively. For both identification schemes we focus on a simple form of a fiscal policy rule in that non-policy shocks can affect fiscal policy variables contemporaneously only through their effect on output. In their application to US data, \citet{caldara2017} argue that this simple form is able to capture well the systematic component of fiscal policy. The only difference between BP and CK identifications in our paper is that in the former we impose a zero restriction on the contemporaneous output elasticity of government spending and investment while in the latter we do not. As discussed, for example, in \citet{caldara2017} and \citet{ravn2014effects}, even small deviations from zero in this parameter can yield considerably different results in the estimated effects of government spending shocks. Here we are able to study these potential differences.

On the revenue side there are no real differences between CK and BP identifications as we estimate the output elasticity of net revenues using the SVAR-IV strategy and our proposed instrument in both cases. We allow for a direct effect from government spending shock to net revenues. That is, we effectively order $g$ before $r$ in the VAR so that we are able to include the spending shock in the regression where the revenue shock is identified. This may lead to different estimates in the revenue equation if the identified government spending shocks from the first step differ between BP and CK identifications. Note that the model is partially identified as we do not uncover other structural shocks from the reduced form residuals.

\subsection{Non-fiscal instrument for output}

\emph{Proposed non-fiscal instrument.}--- We propose to use trading partner forecast errors of output as an instrumental variable for domestic output in small open economies. The essential assumptions needed for the instrument are the following. Firstly, the professional forecasts are sensible in the sense that forecast errors capture meaningful and unexpected variation in trading partner economies. Secondly, these forecast errors have explanatory power on the unexpected changes in aggregate output of the country of interest. Thirdly, fiscal policy shocks of the domestic country do not explain the unexpected variation in its trading partners output as captured by the forecast errors. Formally, the latter two assumptions correspond to equations \eqref{as1} (relevance) and \eqref{as2} (exogeneity), respectively.

While the intuition behind the instruments for Canada and the small euro area countries is exactly the same, the construction of the instrument for these two cases differs somewhat in practice. Roughly speaking, more than 70 percent of Canada's exports are destined to the USA. Accordingly, forecast errors of output for the US economy can be expected to be a good predictor for unexpected movements in Canada's GDP. Contrary to Canada, the exports of the euro area small open economies are more diversified among a number of destination countries. Therefore, to form an instrument for the small euro area countries that would cover roughly a similar share of exports as the instrument for Canada one needs some form of aggregation. In practice we gather forecast errors of output for a number of countries for which these forecasts are available and then weight these errors by their respective export shares. That is, for the small euro area countries our preferred instrument is a weighted average of the trading partners' forecast errors.

\emph{Alternative non-fiscal instrument in the literature.}--- Earlier literature has utilized the quarterly utilization adjusted TFP series of \citet{fernald2014}, which is readily available for the USA. This series is used as an instrument, for example, by \citet{caldara2017} and \citet{angelini} when estimating output elasticities of fiscal variables in a SVAR-IV. Construction of the utilization adjusted TFP series of \citet{fernald2014} involves a number of assumptions. Since TFP growth is that part of aggregate output growth that cannot be explained by changes in inputs, one first needs to, at least implicitly, decide over a specification of the aggregate production function. Furthermore, to form the utilization-adjusted TFP series one needs to adjust the TFP series for capacity utilization. This involves additional decisions over how to account for or how to model/proxy the utilization rate. Clearly any of the choices made in the process of modeling the utilization adjusted TFP series have an effect on the final TFP series. In fact, as documented in \citet{sims} the revisions made to the utilization adjusted TFP series as a result of new releases of data as well as methodological changes made over time seem to have a remarkable effect on the resulting series of \citet{fernald2014}. In contrast, the instrument we propose can be calculated from observable data and should not be subject to such major revisions. See the \hyperref[sec:online]{Online Appendix} for more detailed discussion on the TFP instrument.

\emph{Pretests for relevance and exogeneity.}--- We seek to pretest the validity of the proposed instrument in \autoref{exogrele1}. In Panel A, we regress domestic forecast errors of output on the instrument. This test aims to study whether the instrument has explanatory power on unexpected changes in domestic output. Note that by using a proxy measure for the unexpected changes (professional forecast errors) it is independent of the SVAR model and its exact specification. Columns (1) and (2) show that for both Canada and the euro area small open economies the proposed instrument is significantly related to unexpected movements in output. As a comparison, columns (3)-(5) report results from regressions of \citet{fernald2014} TFP series on GDP forecast errors in the sample used by \citet{caldara2017} and in more recent samples collected in 2023 for the USA.\footnote{Utilization-adjusted TFP growth of \citet{fernald2014} are from the March 7th 2023 vintage and downloaded from the author's website: \url{https://www.johnfernald.net/TFP}.} In column (3) we use the TFP-series from the \citet{caldara2017} dataset while in column (4) we use the more recent version of the same series but restrict the sample period to match the one in column (3). In column (5) we use the more recent version of utilization-adjusted TFP but do not restrict the sample to end in 2006Q4. Unsurprisingly, we find a statistically significant relation between unexpected changes in domestic output and the instrument also in these cases.

As a test for exogeneity, in Panel B of \autoref{exogrele1} we regress the instrument on forecast errors of one-period ahead growth in government spending and investment which, following \citet{auerbach2012}, act as proxy for fiscal shocks. The aim here is to test whether these proxies for fiscal shocks covary with the instrument. Significant correlation between the two is suggestive of the exogeneity assumption not holding. The coefficients for Canada and euro area small open economies are close to zero and statistically insignificant, implying that there is no systematic relation between fiscal proxies and the instrument. In contrast, the results for the USA are mixed. The relationship between the fiscal proxy and the TFP instrument is insignificant in the sample used by \citet{caldara2017} (column (3)) but significant in more recent samples (columns (4) and (5)). Revisions made to the utilization adjusted TFP series over the years seem to have a considerable effect on the estimates, as has also been reported by \citet{sims} in a different setting.

\begin{table}[!htbp]
\caption{Relevance and exogeneity of the instrument.} \label{exogrele1}
\begin{center}
\resizebox{\textwidth}{!}{
	  \begin{tabular}{lccccc}
   \toprule
   \multicolumn{6}{l}{\textbf{Panel A: Relevance}} \\
\midrule
  \multicolumn{6}{r}{\textit{Dependent variable:} Forecast error of $\Delta gdp$}   \\ 
  \cline{2-6} \\[-1.8ex]
                        & CAN                      & EUR          & US(CK)                   & US & US \\  
                    & (1)                       & (2)           & (3)                    & (4)  & (5)\\  
   \midrule 
   Trading partner forecast error instrument         & 0.463$^{***}$             & 0.957$^{***}$ &           &    &   \\   
                    & (0.103)                   & (0.128)       &      &         &   \\   
   $\Delta$ Utilization adjusted TFP  &                &               & 0.085$^{***}$ &  0.052$^{***}$  &  0.056$^{***}$ \\   
                    &                           &               & (0.027)       &  (0.018) &  (0.015) \\   
    \\
   Observations     & 92                        & 382           & 152   & 152         & 204\\  
   Adjusted R$^2$   & 0.188                     & 0.367         & 0.133    & 0.052     & 0.063\\  
   Country FE &                           & $\checkmark$  &                  &      & \\  
   Final observation in sample & 2019Q4 & 2019Q4 & 2006Q4 & 2006Q4 & 2019Q4 \\ 
   \vspace{1mm}\\
\midrule
\multicolumn{6}{l}{\textbf{Panel B: Exogeneity}} \\
\midrule
 &  \multicolumn{5}{c}{\textit{Dependent variable:} Instrument}   \\ 
  \cline{2-6} \\[-1.8ex]
    & CAN & EUR & US(CK)    & US  & US\\
                    & (1)                       & (2)           & (3)           & (4) & (5)\\  
   \midrule 
   Forecast error of $\Delta g$ (OECD)            & -0.048                    & -0.030        &         &  & \\   
                    & (0.107)                   & (0.030)       &    &     &   \\   
   Forecast error of $\Delta g$ (SPF)        &                           &               & -0.414 & 0.721$^{**}$ & 1.09$^{***}$\\   
                    &                           &               & (0.338) & (0.363) & (0.327)\\    
    \\
   Observations     & 92                        & 382           & 101  & 101   & 153\\  
   Adjusted R$^2$   & -0.008                    & -0.010        & 0.007 & 0.025  & 0.064\\ 
   Country FE &                           & $\checkmark$  &       &        & \\  
   Final observation in sample & 2019Q4 & 2019Q4 & 2006Q4 & 2006Q4 & 2019Q4 \\ 
   \bottomrule
   \multicolumn{6}{l}{\scriptsize{$^{***}p<0.01$; $^{**}p<0.05$; $^{*}p<0.1$}}
\end{tabular}
	  }
\end{center}
{\footnotesize{\textit{Notes:} In Panel A each column reports OLS estimates from a regression of professional forecast errors of quarterly growth of GDP on the instruments. For Canada and euro area economies these forecast errors of $\Delta gdp$ are from OECD Economic Outlooks and for the US they are from the Survey of Professional forecasters (SPF). Trading partner forecast error instrument is constructed from US GDP forecast errors (SPF) for Canada and from export-share weighted mean of OECD trading partner forecast errors (OECD) for euro area economies. Utilization-adjusted TFP series of \citet{fernald2014} is either from \citet{caldara2017} dataset (CK in column (3)) or from the March 7th 2023 vintage (columns (4) and (5)) downloaded from the author's website (\url{https://www.johnfernald.net/TFP}). In column (4) we restrict the sample period to match the \citet{caldara2017} sample in column (3). In Panel B, each column reports OLS estimates from a regression of the instruments (now the dependent variable) on professional forecast errors in the growth of the sum of general government consumption and investment. For Canada and euro area economies, these forecast errors of $\Delta g$ are from the OECD Economic Outlooks and for the US they are from the SPF. Heteroskedasticity robust standard errors are used in columns (1), (3), (4) and (5). Column (2) has two-way (Country $\times$ Time) clustered standard errors. \par}}
\end{table}

\emph{Discussion.}--- Above analysis suggests that the proposed instrument is indeed related to unexpected variation in output. Also, when applying the instrument later in \autoref{svar}, we find robust first stage F-statistics that are consistent with the instrument not being weak. Moreover, using a proxy for fiscal shocks we find in \autoref{exogrele1} suggestive evidence in support of the exogeneity assumption holding in our setting.\footnote{Using a proxy shrinkage prior, \citet{keweloh2023estimating} provide evidence of the exogeneity assumption not being fulfilled in fiscal-SVARs of \citet{mertens2014reconciliation} or \citet{caldara2017}. Similar analysis could in principle be applied to the instrument proposed here but is beyond the scope of this paper.} Because the exogeneity condition is crucial but cannot be tested for in the way that relevance can, we provide here a brief discussion of some of the concerns related to the exogeneity of the instrument.

The exogeneity assumption made here bears resemblance to assumptions made in the small open economy (SOE) VAR literature that a small open economy does not affect world demand \citep[see, for example,][]{cushman1997}. However, the exogeneity assumption in Equation \eqref{as2} is arguably \textit{weaker} than the canonical SOE assumption because we only need to assume that unexpected fiscal policy shocks are uncorrelated with unexpected shocks of the foreign trading partners (foreign block). On the contrary, the SOE VARs often assume that the small open economy does not affect the foreign block at all. In the case of Canada, for example, the SOE assumption is often adopted in the US-Canada context. Arguably our instrument is more likely to fulfill the exogeneity assumption in this same context.

One could still be concerned that because of close linkages between one small open economy and another, fiscal shocks in one might cause unexpected changes in the other. This could result in the unexpected shocks of the trading partner being contaminated by domestic fiscal shocks.\footnote{For example, one could be concerned by the possible (contemporaneous) spillover effects of Finnish fiscal policy on the Swedish economy.} To assess this concern, we have compared two different versions of the instrument in the case of euro area small open economies: a baseline version containing data on all available trading partners and one which is constructed as a weighted average of only G7 countries forecast errors. We find that these two versions of the instrument yield similar results. Therefore, it seems possible to exclude the most suspicious countries in terms of exogeneity from the instrument without, necessarily, a large trade-off in instrument relevance. We suggest this concern to be considered on a case-by-case basis.

Another concern to exogeneity could be if there was significant coordination between domestic and foreign fiscal policies. In this case there could possibly be a common component in the instrument and domestic fiscal policy shocks. However, since we are identifying unexpected fiscal policy shocks at the quarterly frequency and the instrument also captures unexpected quarterly variation, this would entail that the unexpected part of fiscal policy was systematically related across countries. This seems unlikely and a \citet{blanchard2002} type assumption about implementation lags of fiscal policy at the quarterly frequency would rule out this concern. Note also that common variation that is not systematically related to fiscal policy of the small domestic economy should be of no concern as it would not threaten exogeneity of the instrument (but plausibly increase instrument relevance).\footnote{In a sense, an econometrician using the empirical approach of this paper would like to observe domestic and foreign economies subject to the same set of non-fiscal shocks but with different fiscal policy shocks.}

\subsection{Data} \label{sec:data}

\emph{VAR variables.}--- We collect data from Statistics Canada, Eurostat and OECD. For $g$, $r$, $gdp$ we rely on national statistical agencies in Statistics Canada and Eurostat while rest of the variables are collected from OECD. We limit ourselves to studying periods with relatively stable macro-institutions, that is, we wish to avoid possible structural breaks. Thus, for Canada, our sample period is 1986Q1-2019Q4 which is motivated by the start of the Great Moderation in the mid 1980s. The sample period for euro area economies is 1999Q1-2019Q4, i.e. the EMU period before COVID-19.\footnote{Due to missing data on some of the variables, the panel is unbalanced.} All data are seasonally adjusted when applicable. For Canada we use real government spending but for European economies we deflate government spending and investment by the GDP deflator since deflators for the individual series are not available for all countries of interest.

\emph{Instrument.}--- For our instrument we collect data on past macroeconomic forecasts of professional forecasters. The forecasts we consider are often reported in levels. Since the level forecasts of real variables rely on base years that change over forecast vintages and are also conditional on the then available and later revised information on past levels of the variables, we transform all level forecasts to log-difference forecasts as follows. Let $F_t[.]$ denote a forecast operator and $v_{t+1}$ is the value of variable $v$ in period $t+1$. Then $F_t[v_{t+1}]$ is the forecast of the value of $v$ in period $t+1$ made in period $t$. Using this notation we can write the forecasted log-difference of GDP in country $i$ made at time $t-1$ as $F_{t-1}[ \Delta gdp'_{it} ]$, where $gdp'$ is the natural logarithm of GDP. Here the apostrophe is marking the fact that, in contrast to the output variable in the VAR model, these forecasts are not in per capita terms. Forecast errors of output are now readily obtained as the difference between the realized log-difference and the forecasted one, i.e. \begin{equation} \Delta gdp'_{it} - F_{t-1}[ \Delta gdp'_{it} ] . \end{equation}

For Canada, we use quarterly forecasts of the US economy from the Survey of Professional forecasters (SPF). In SPF, each quarter after the first release of the previous quarter's GDP figure, a panel of professional forecasters is asked to provide forecasts of several macro-variables of the US economy. The survey is released in the middle of each quarter. In constructing the instrument, we use the mean forecast of one quarter ahead of real GDP. The real GDP forecast is transformed to a log-difference forecast as outlined above.

For euro area economies, our data for the instrument is from the OECD Economic Outlooks (EOs), which contain forecasts of several macroeconomic variables for a number of countries. OECD EOs are published twice a year in June and December. For EOs published before 2003S2, only annual and semi-annual forecasts are available, whereas, from 2003S2 onwards, the EOs contain both annual and quarterly forecasts for a large subset of variables. When quarterly forecasts are not available, we interpolate the semi-annual growth rate (log-difference) forecast over the two quarters. In other words, we assume that the semi-annual growth rate forecast is constant across the period so that we can simply divide the semi-annual log-difference by two into two quarterly log-differences.

Since, unlike in Canada's case, no single country has an overwhelming share in the exports of a typical euro area economy, we combine forecast errors from several countries into a single instrument. In doing so, we weight trading partner forecast errors by their share in domestic exports in order to calculate a weighted average. In doing so we use 4-quarter moving averages of these weights to smooth out possible short-run changes in trade shares. The quarterly data on exports are from OECD International trade statistics.

\section{Results}\label{svar}

This section presents results from the application of the instrument to study fiscal-SVARs for both Canada and Euro area small open economies. Firstly, we report 2SLS estimates of the structural parameters. We then proceed to the output effects of fiscal policy before discussing impulse responses more broadly. We conclude this section with robustness analysis.

\subsection{Structural parameters}\label{sec:parameters}

\autoref{id_table} reports the estimated structural parameters for both the CK (baseline) and BP style identification schemes. In BP, the output elasticity of government spending is set to zero (columns (1) and (5)), while in CK it is estimated (columns (3) and (7)). We find that for both Canada and euro area small open economies the estimates for output elasticity of $g$ are negative but statistically insignificant from $0$. Following subsections will examine the effect of the BP zero restriction on the resulting IRFs and multipliers.

\begin{table}[h]
 \caption{2SLS estimates of the output elasticities of fiscal variables.} \label{id_table}
 \begin{center}
\resizebox{\textwidth}{!}{
	  \begin{tabular}{lcccc|cccc}
   \toprule
     &\multicolumn{4}{c}{Canada}   &\multicolumn{4}{c}{Pooled EUR} \\
     \cmidrule{2-5} \cmidrule{6-9}
    &\multicolumn{2}{c}{BP}   &\multicolumn{2}{c}{CK}  &\multicolumn{2}{c}{BP}  &\multicolumn{2}{c}{CK}  \\
\cmidrule{2-3} \cmidrule{4-5} \cmidrule{6-7} \cmidrule{8-9} 
                                                   & $g$             & $r$             & $g$             & $r$             & $g$                      & $r$                      & $g$                      & $r$\\  
                                                   & (1)           & (2)           & (3)           & (4)           & (5)                    & (6)                    & (7)                    & (8)\\  
   \midrule 
     $gdp$                     &      $\mathbf{0}$         & 3.58$^{**}$   & -0.280        & 3.81$^{**}$   &      $\mathbf{0}$         & 1.42          & -0.529        & 1.44\\   
                         &               & (1.67)        & (0.365)       & (1.74)        &               & (0.986)       & (0.820)       & (1.00)\\   
   $g$                     &               & -0.810        &               &               &               & -0.040        &               &   \\   
                         &               & (0.498)       &               &               &               & (0.121)       &               &   \\  
   $g-a_{g}gdp$                    &               &               &               & -1.03$^{*}$   &               &               &               & -0.061\\   
                         &               &               &               & (0.590)       &               &               &               & (0.130)\\   
    \\
   Standard-Errors & \multicolumn{4}{c}{Newey-West ($3$ lags)} &  \multicolumn{4}{c}{2-way clustered (Country $\times$ Time)} \\ 
   Observations          & 131           & 131           & 131           & 131           & 359           & 349           & 349           & 349\\  
   Adjusted R$^2$        & 0.994         & 0.961         & 0.993         & 0.961         & 0.997         & 0.992         & 0.997         & 0.992\\  
   F-stat. (1st stage), $gdp$ &               & 16.6          & 15.7          & 16.7          &               & 16.0          & 16.6          & 16.1\\  
   Constant                           & $\checkmark$  & $\checkmark$  & $\checkmark$  & $\checkmark$  &            &           &            & \\   
      Country FE                           &  &   &   &   & $\checkmark$           & $\checkmark$           & $\checkmark$           & $\checkmark$\\
   \bottomrule
   \multicolumn{9}{l}{\scriptsize{$^{***}p<0.01$; $^{**}p<0.05$; $^{*}p<0.1$}}
\end{tabular}
	  }
\end{center}
{\footnotesize{\textit{Notes:} This table presents 2SLS estimates on the elasticities of fiscal variables $g$ and $r$ with respect to $gdp$ which is instrumented by trading partner forecast errors of output. All models include 5 lags of VAR variables as controls. BP in columns (1)-(2) and (5)-(6) refers to \citet{blanchard2002} type identification where the output elasticity of government spending ($a_{g}$) is restricted to zero, while CK in columns (3)-(4) and (7)-(8) refers to \citet{caldara2017} type identification where this parameter is estimated using external instruments. Columns (2), (4), (6) and (8) take the elasticity $a_{g}$ (coefficient for $gdp$) from the previous column as given. \par}}
\end{table}

Estimates of the output elasticity of net revenue for Canada are 3.81 (CK, column (4)) or 3.58 (BP, column (2)) and for the euro area small open economies 1.44 (CK, column (8)) or 1.42 (BP, column (6)). For Canada, the estimates of the output elasticity are notably larger than those shown in \citet{perotti2005}, which are derived as in \citet{blanchard2002}. This estimate, 1.86, is the only comparable estimate we find for Canada. According to the results in \citet{caldara2017}, estimating this parameter with a non-fiscal proxy seems to produce notably larger values for this elasticity also for the USA. Also \citet{angelini} estimate large elasticities in a similar framework for the US. For the small euro area countries, the estimates are closer to earlier estimates. For example, \cite{burriel2010fiscal} use an elasticity of 1.54 for the euro area. Importantly, the first stage robust F-statistic reported in \autoref{id_table} are clearly over the rule of thumb value, 10, which indicates that the instrument we propose does not appear to be weak.\footnote{The test statistic we use is the efficient F-test proposed by \citet{olea2013robust}. As argued in \citet{stockandrews2019weak}, this test should be used instead of the standard F-statistic which assumes homoskedasticity when detecting weak instruments. Accordingly, in the case with only one instrument and one endogenous variable it is sufficient to use the following rule of thumb; efficient F statistic $> 10$; or to rely on the critical values provided in \citet{yogo2005}.}

\subsection{Effects of fiscal stimulus}\label{sec:multi}

\emph{Government spending multiplier.}--- We report government spending multipliers in the fashion of \citet{uhlig2009} who calculate present value cumulative fiscal multipliers. The multiplier is calculated as the cumulative sum of the output response divided by the cumulative sum of the response in government spending and it thus accounts for both the dynamics in $gdp$ and $g$. For simplicity, we follow the literature and consider a zero discount rate in the calculation \citep{ramey2019}. Formally we have 
\begin{equation}\label{fmulti}
\mathcal{M}_H = \frac{\sum^{H}_{h=0} IRF(g \rightarrow gdp, h)}{\sum^{H}_{h=0} IRF(g \rightarrow g, h)},
\end{equation}
where $\mathcal{M}_H$ is the cumulative multiplier at horizon $H$ and $IRF(a \rightarrow b,h)$ is the impulse response from variable $a$ to $b$ at horizon $h$. IRFs are in the same units ($\%$ of GDP).

\begin{figure}[h]
\begin{center}
\caption{Cumulative government spending multipliers.}\label{fig:multiplier}
\subfloat[Canada]{\includegraphics[width=0.5\textwidth]{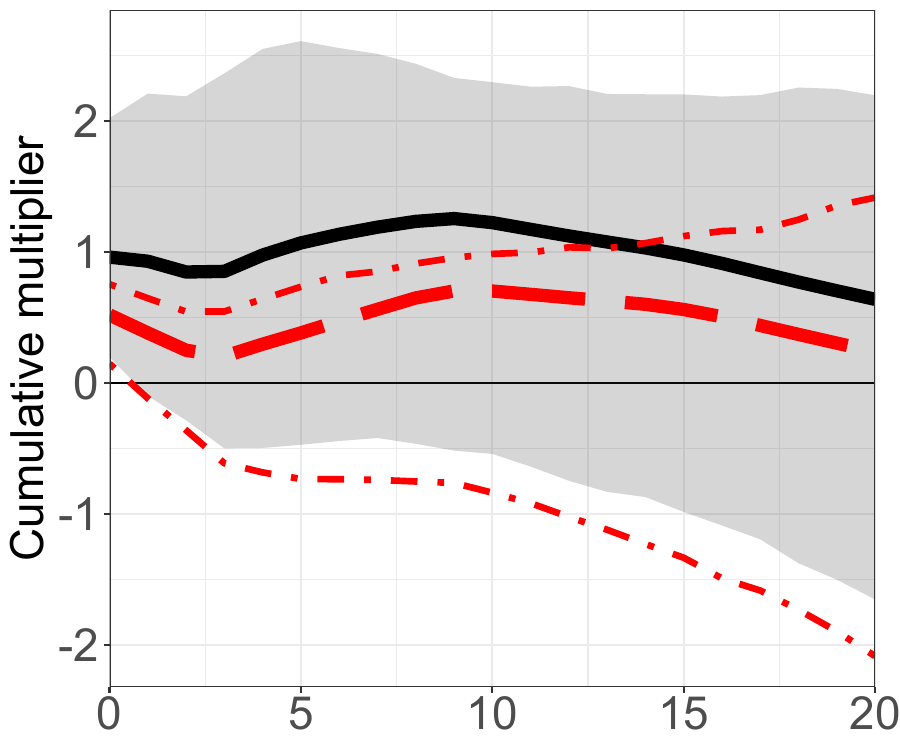}}\hfill
\subfloat[Euro area countries]{\includegraphics[width=0.5\textwidth]{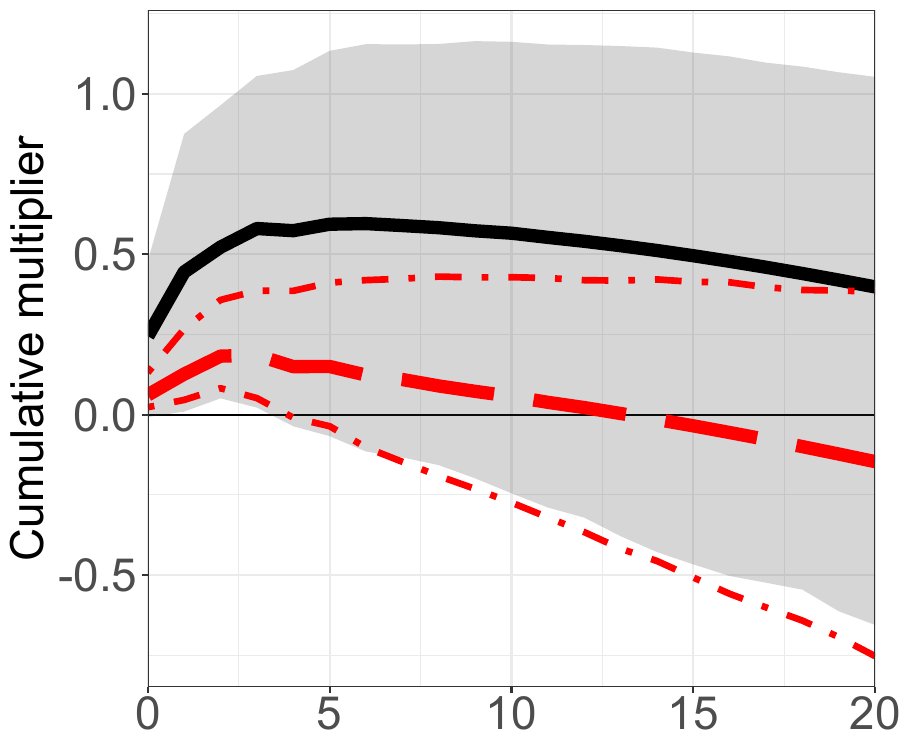}}\hfill
\end{center}
{\footnotesize{\textit{Notes:} This figure plots cumulative government spending multipliers (Equation \eqref{fmulti}) calculated from SVAR impulse responses to government spending shocks for up to 20 quarters from the initial shock. Black line and shaded area represent point estimates and confidence intervals for $CK$ identification while red dashed line and dot-dashed lines represent $BP$ identification. Residual-based moving block bootstrap 0.68 confidence intervals with 1000 draws. Horizontal axis has quarters from 0 to 20. \par}}
\end{figure}

\Autoref{fig:multiplier} plots cumulative multipliers up to horizon of 20 quarters for Canada (panel A) and euro area small open economies (panel B). The black line and shaded area represent point estimates and confidence intervals for CK identification, and the red dashed line and dot-dashed lines represent BP identification. The government spending multiplier for Canada (CK identification) is clearly positive and near 1 for almost the whole period. Using BP identification, the cumulative multiplier is smaller at around 0.5. In both cases the cumulative multiplier starts to decrease slowly after 10 quarters. The euro area small open economies government spending multiplier is small on impact. Using CK identification the cumulative multiplier is around 0.5 for longer horizons. Again, when using BP identification, the cumulative multiplier is smaller and even turns slightly negative after 15 quarters.

Overall, fiscal stimulus from government spending appears more effective in Canada. While point estimates for both are larger than 0, the multiplier for Canada is larger. This contradicts the traditional Mundell-Fleming framework, which predicts that fiscal policy is less efficient in a small open economy with a flexible exchange rate (Canada) than in a small open economy with a fixed exchange rate (small euro area economies).

\emph{Differences between CK and BP identifications.}--- We also find that multiplier estimates differ between BP or CK identifications. This difference is due to the additional zero restriction in BP identification which is not required when the instrument is applied.\footnote{The sensitivity of estimated responses with respect to the output elasticity of government spending has been noted by e.g. \cite{ravn2014effects}.} In CK style identification we estimate output elasticity of government spending using the instrument.

\begin{figure}[h]
\begin{center}
\caption{Output elasticity of $g$ and government spending multiplier.}\label{fig:elas}
\subfloat[Canada]{\includegraphics[width=0.5\textwidth]{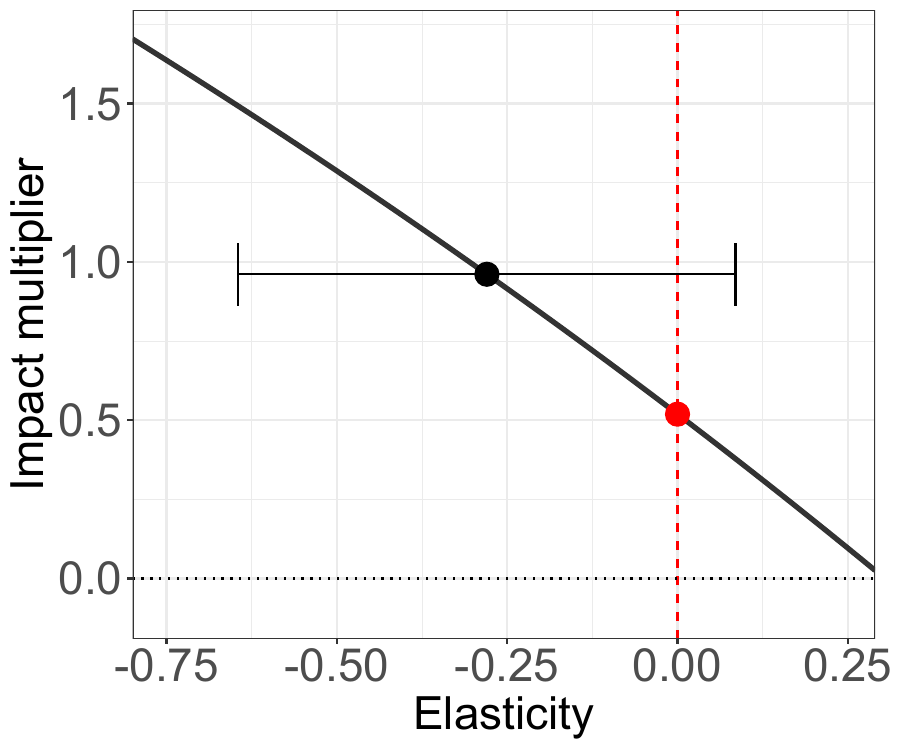}}\hfill
\subfloat[Euro area countries]{\includegraphics[width=0.5\textwidth]{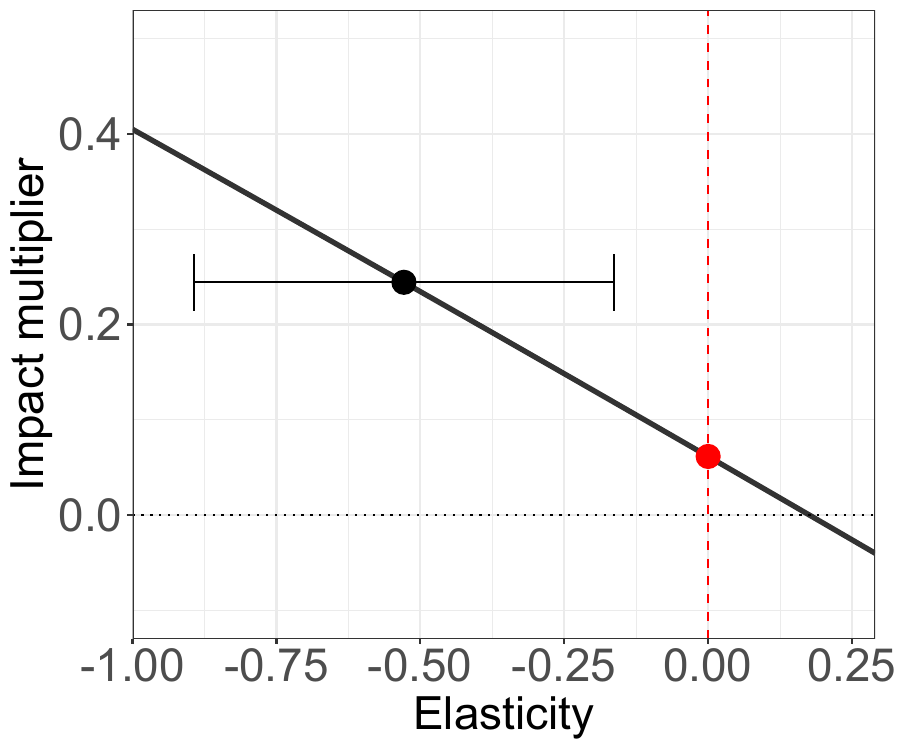}}\hfill
\end{center}
{\footnotesize{\textit{Notes:} This figure plots the impact multiplier of government spending as a function of the elasticity of government spending $g$ with respect to output $gdp$ (solid black line) that is consistent with the reduced form VAR model. Red-dot corresponds to the zero restriction imposed on this elasticity by the $BP$ identification while black-dot is the estimated elasticity along with its one standard error confidence interval from \Autoref{id_table}. \par}}
\end{figure}

To examine the differences between these two identification schemes in detail we depict in \autoref{fig:elas} the impact multipliers of government spending as a function of the output elasticity of government spending. That is, given the reduced form VAR we vary the output elasticity of $g$ in order to trace out the relationship between the two. The impact multipliers across all these structural forms with different output elasticities form a curve in \autoref{fig:elas}. On this curve, the black dot and the associated confidence interval are for the CK identification while the red dot represents the BP zero restriction. Estimates correspond to those in \autoref{id_table}.

Based on \autoref{fig:elas}, the size of this elasticity has a notable effect on the impact multiplier of government spending. Considering Canada, for example, there is a clear difference in the impulse response of GDP to a spending shock between these identification schemes with CK identification producing almost double the initial impact on GDP (close to $1$) than BP identification (roughly $0.5$). However, the output elasticity estimate is not statistically significant from zero. This finding causes a predicament. The BP style zero restriction is well established in the literature but even small deviations from $0$ can lead to markedly different multiplier estimates in a given VAR model. With a valid instrument, one is able to provide an estimate for this elasticity. We get negative point estimates for the elasticity for both Canada and euro area small economies, leading to larger government spending multipliers.

\emph{Stimulus effects of revenue cuts.}--- On the revenue side, fiscal multipliers are not typically reported as in Equation \eqref{fmulti}. Consider for example a tax cut that stimulates the economy. Strong enough positive effect on output combined with a large enough output elasticity of net revenues might mean that the VAR impulse response of net revenues turns positive (as we find for Canada). Thus the negative revenue shock would eventually increase net revenues. This could cause the denominator in \eqref{fmulti} to change signs at some horizon leading to unreasonable cumulative multipliers around this point when the value is close to zero. Similar effect is not typically present on the spending side. For net revenues we, therefore, discuss the effects in terms of impulse responses to a $-1\%$ of GDP net revenue shock instead.

In the model for Canada, a negative net revenue shock stimulates the economy but to a lesser extent than a positive shock to government spending (\autoref{fig:canada_irf}). Whereas a shock to government spending generates an output response of close to unity in the short term (CK identification), a shock that lowers net revenues by $-1\%$ of GDP results in an output response of no larger than $0.4\%$. However, as the net revenue shock has a relatively small effect on budget balance ($r-g$), cumulative fiscal costs of revenue side stimulus appear to be smaller than those of spending side stimulus. In the medium term net revenues are not affected by the revenue shock whereas the effect of a government spending shock on the deficit is more persistent. Using this metric, tax cuts might appear as fiscally more efficient than stimulus spending.

Considering the euro area small open economies, a cut in net revenues seems not to stimulate the economy (\autoref{fig:europe_irf}). Point estimates of the impulse response would rather point to a mildly negative effect on output in the medium term. This somewhat puzzling result, however, does not appear very robust. In \hyperref[sec:online]{Online Appendix}, we show that this result seems to be driven by one country in the sample, namely Portugal. There we also discuss some possible reasons why this country might behave differently compared to rest of the euro area sample. Were Portugal to be excluded from the sample, point estimates of the output response to a cut in net revenues would turn slightly positive. Relatedly, in \autoref{robustness} where we study the robustness of government spending multiplier estimates, leaving Portugal out of the sample alters the results to a degree.

\emph{Differences between Canada and euro area economies.}--- According to our results, both spending and revenue side stimulus have larger output effects for Canada than for the small euro area countries. While not conclusive, this evidence is in contrast to predictions of the Mundell-Fleming framework. There might be a number of reasons why fiscal policy in Canada would be more effective than in small euro area economies. For instance, there might be smaller import leakages due to larger home markets. Also, given that new Keynesian theory highlights the role of monetary policy in determining the effectiveness of fiscal policy, it might be the case that Canadian monetary policy doesn't work to offset fiscal policy. In the next subsection, where we discuss impulse responses more broadly, we find that in Canada the short-term interest ($srate$) does not markedly rise following expansionary fiscal policy.

\subsection{Impulse responses}\label{sec:impulse}

\begin{figure}[!htpb]
\captionsetup[subfigure]{labelformat=empty}
\begin{center}
\caption{SVAR impulse responses to 1\% of GDP fiscal shocks, Canada.}\label{fig:canada_irf}
\subfloat[$g \rightarrow g$]{\includegraphics[width=0.25\textwidth]{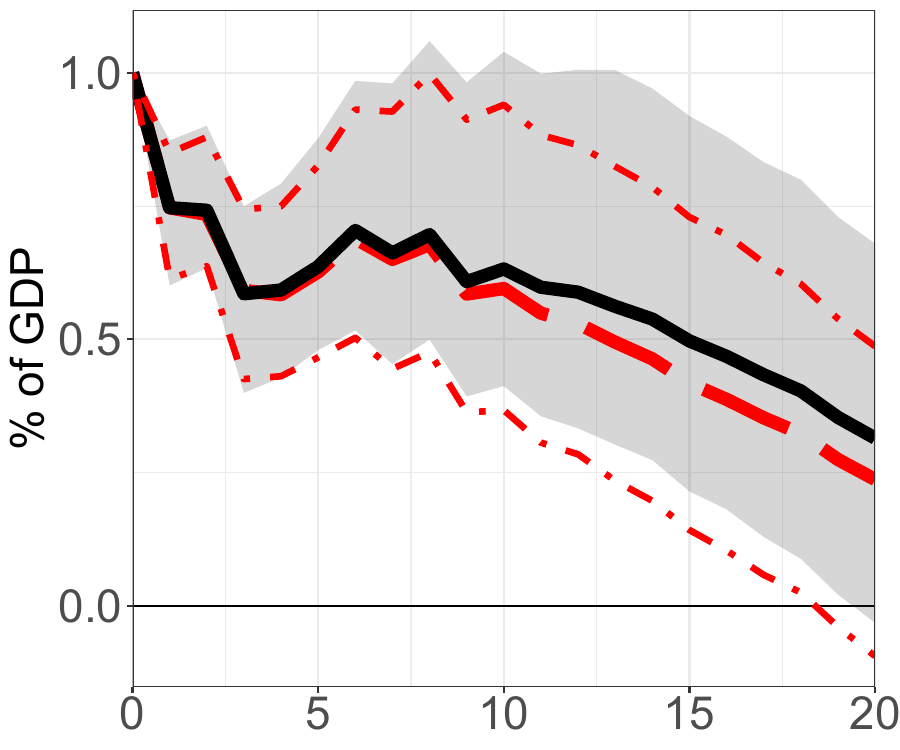}}\hfill
\subfloat[$g \rightarrow r$]{\includegraphics[width=0.25\textwidth]{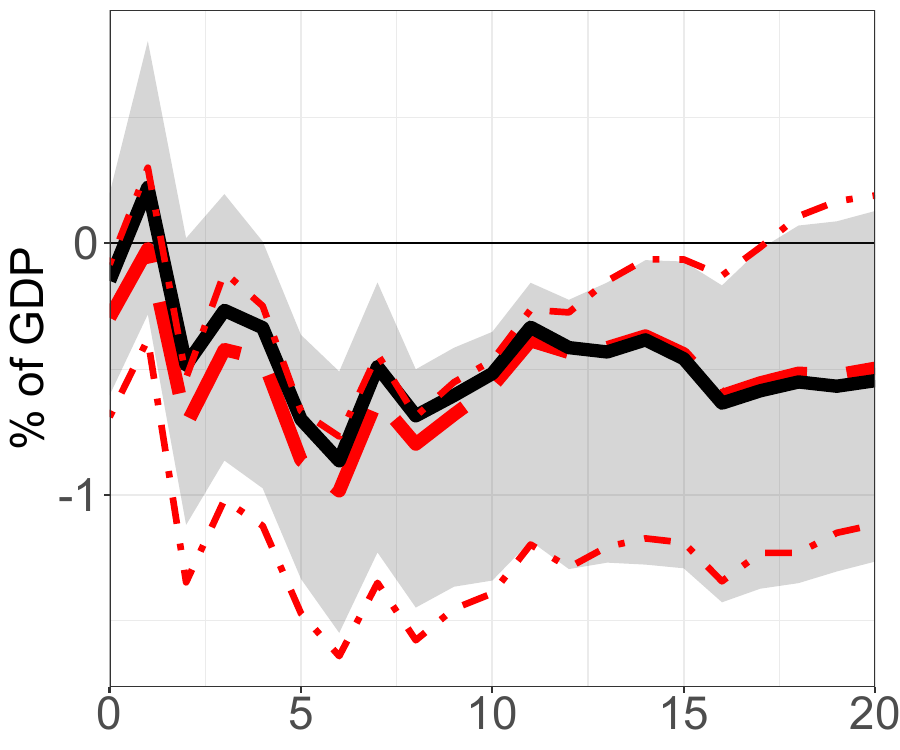}}\hfill
\subfloat[$g \rightarrow (r-g)$]{\includegraphics[width=0.25\textwidth]{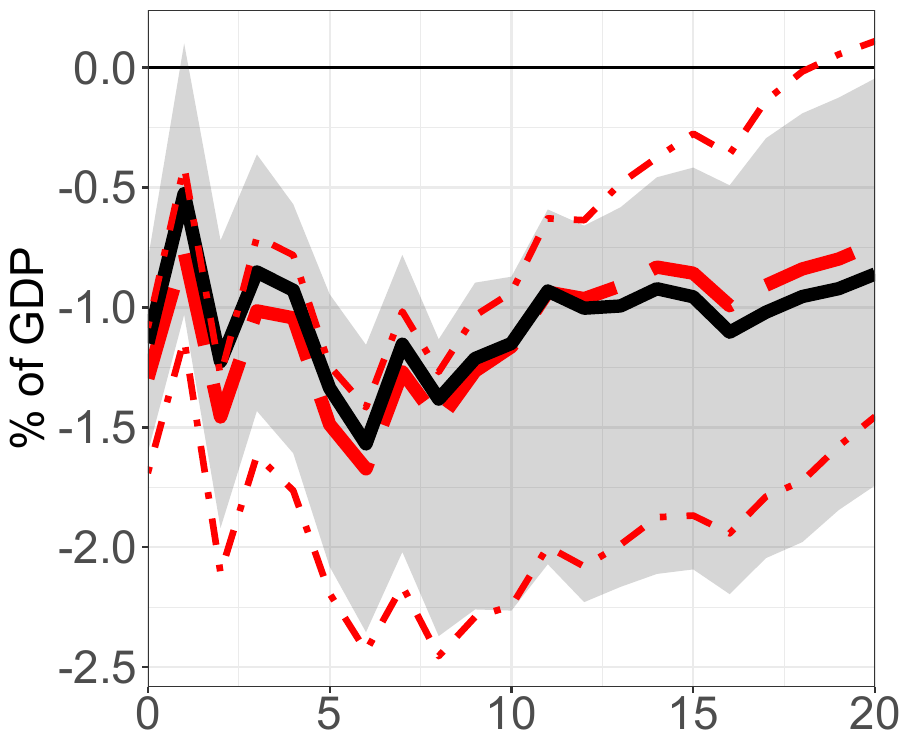}}\hfill
\subfloat[$g \rightarrow gdp$]{\includegraphics[width=0.25\textwidth]{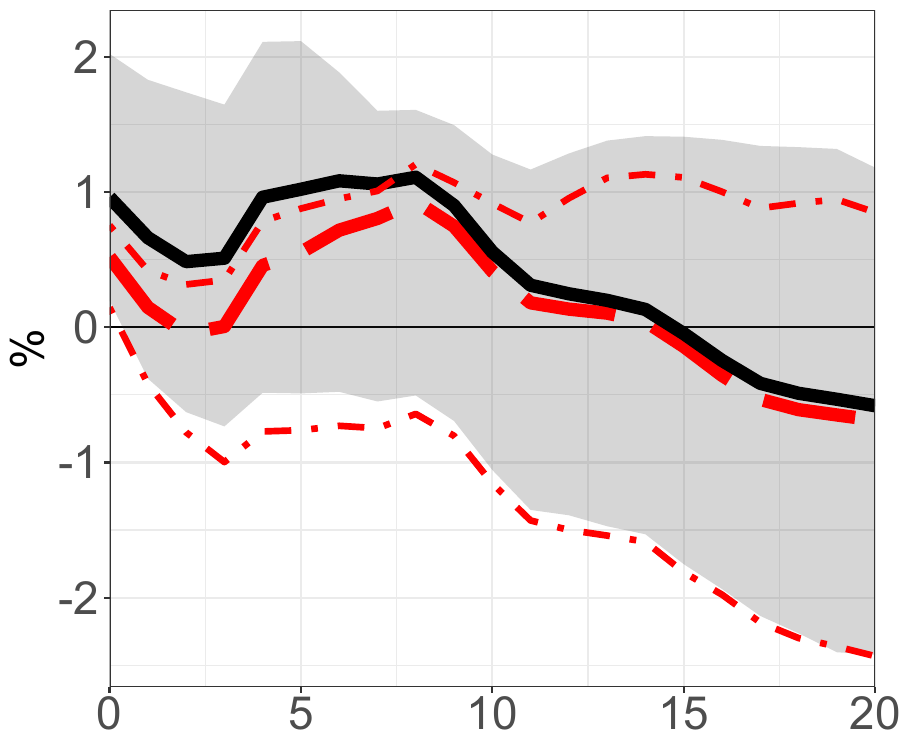}}\hfill
\vspace{2mm}
\subfloat[$g \rightarrow defl$]{\includegraphics[width=0.25\textwidth]{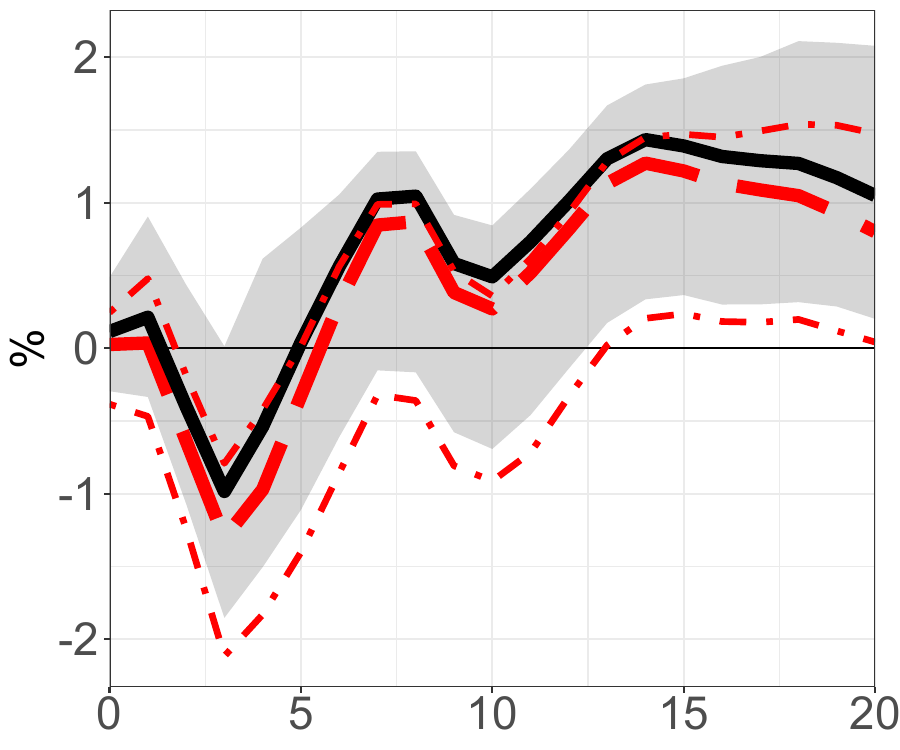}}\hfill
\subfloat[$g \rightarrow rer$]{\includegraphics[width=0.25\textwidth]{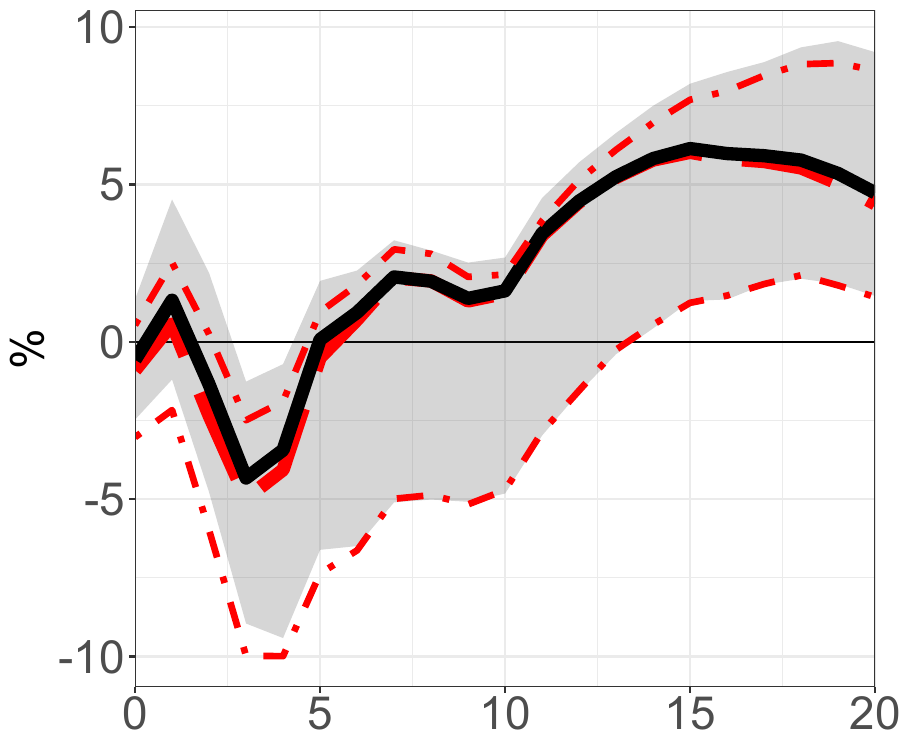}}\hfill
\subfloat[$g \rightarrow cab$]{\includegraphics[width=0.25\textwidth]{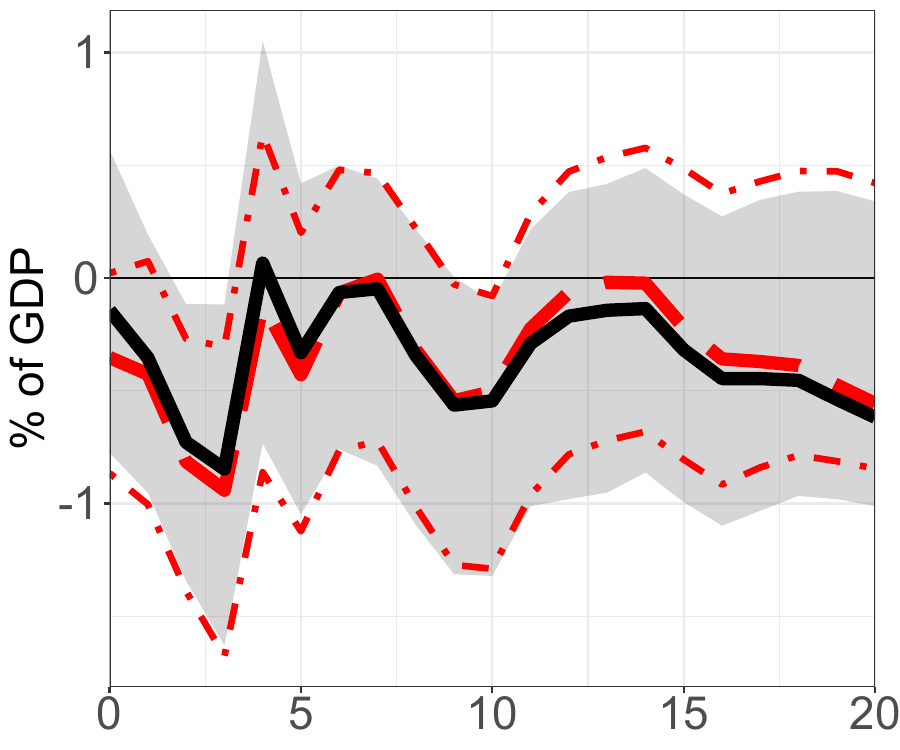}}\hfill
\subfloat[$g \rightarrow srate$]{\includegraphics[width=0.25\textwidth]{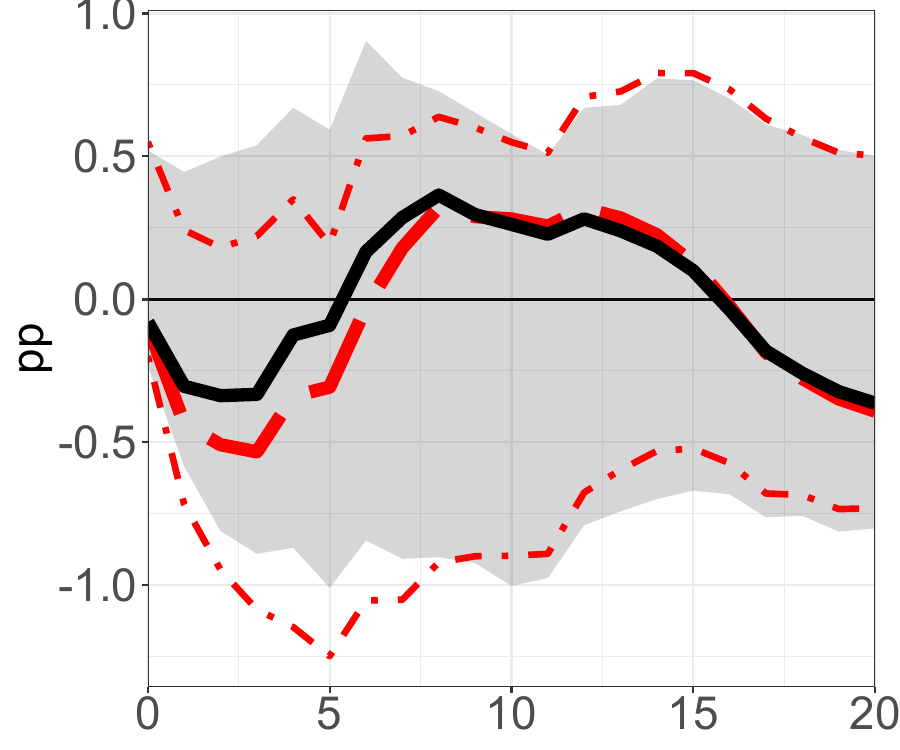}}\hfill

\vspace{8mm}

\subfloat[$r \rightarrow g$]{\includegraphics[width=0.25\textwidth]{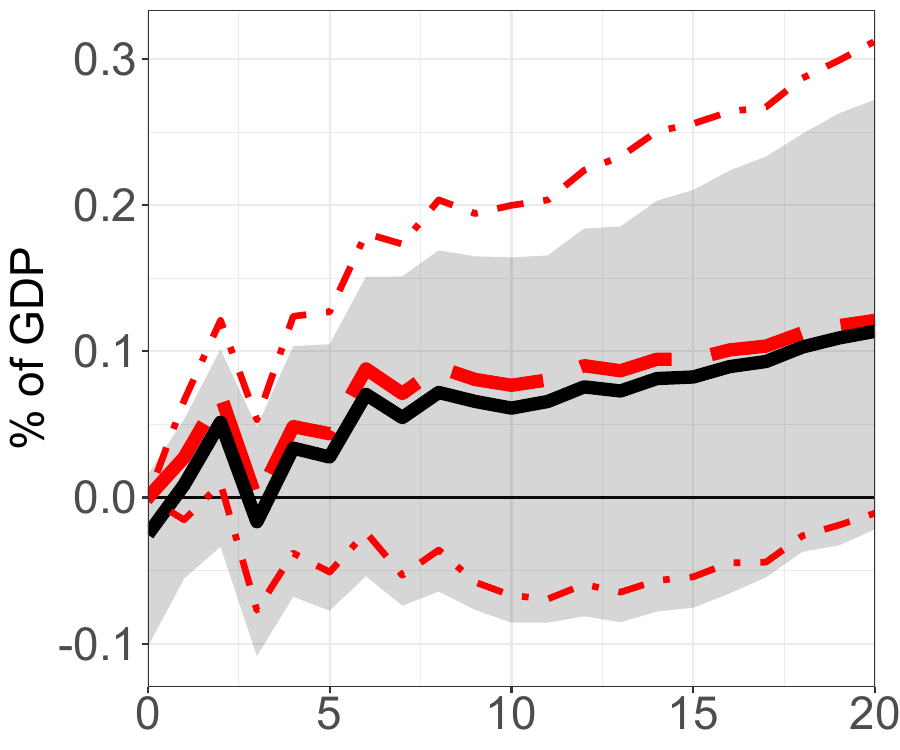}}\hfill
\subfloat[$r \rightarrow r$]{\includegraphics[width=0.25\textwidth]{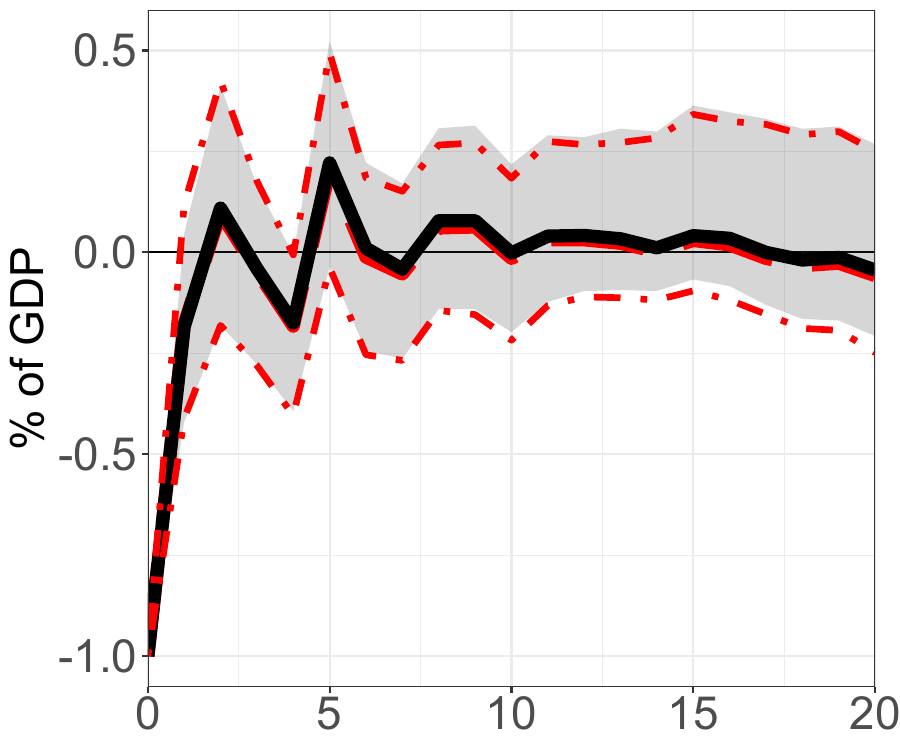}}\hfill
\subfloat[$r \rightarrow (r-g)$]{\includegraphics[width=0.25\textwidth]{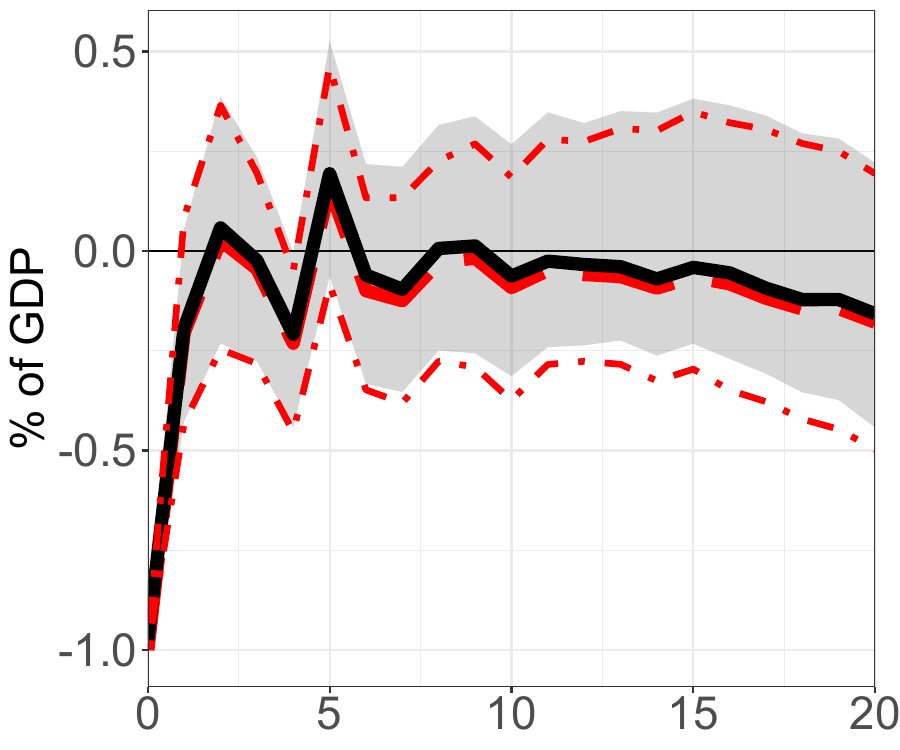}}\hfill
\subfloat[$r \rightarrow gdp$]{\includegraphics[width=0.25\textwidth]{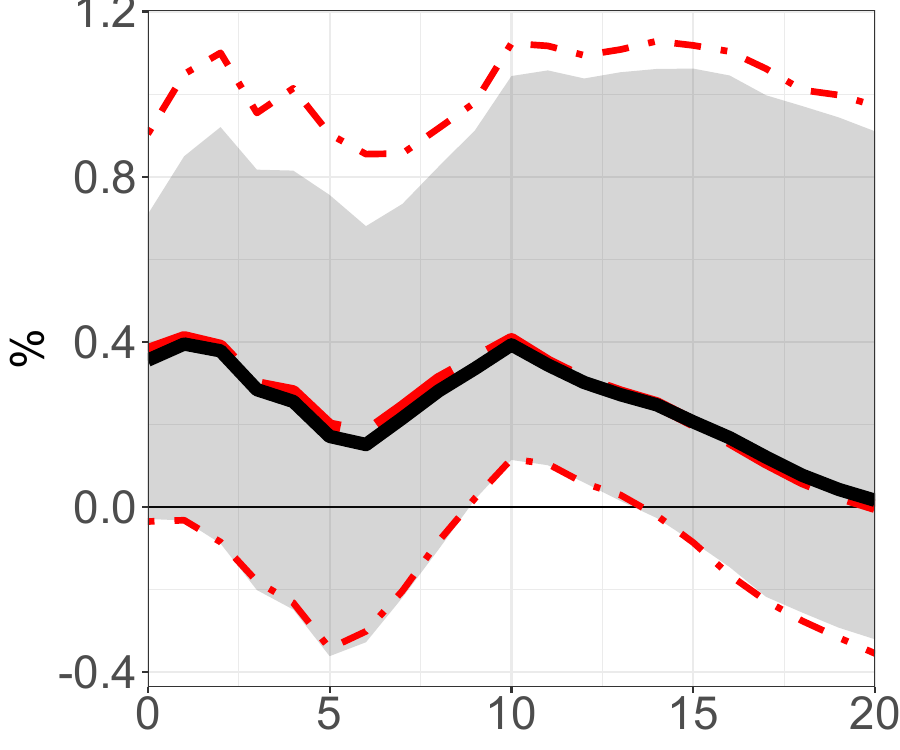}}\hfill
\vspace{2mm}
\subfloat[$r \rightarrow defl$]{\includegraphics[width=0.25\textwidth]{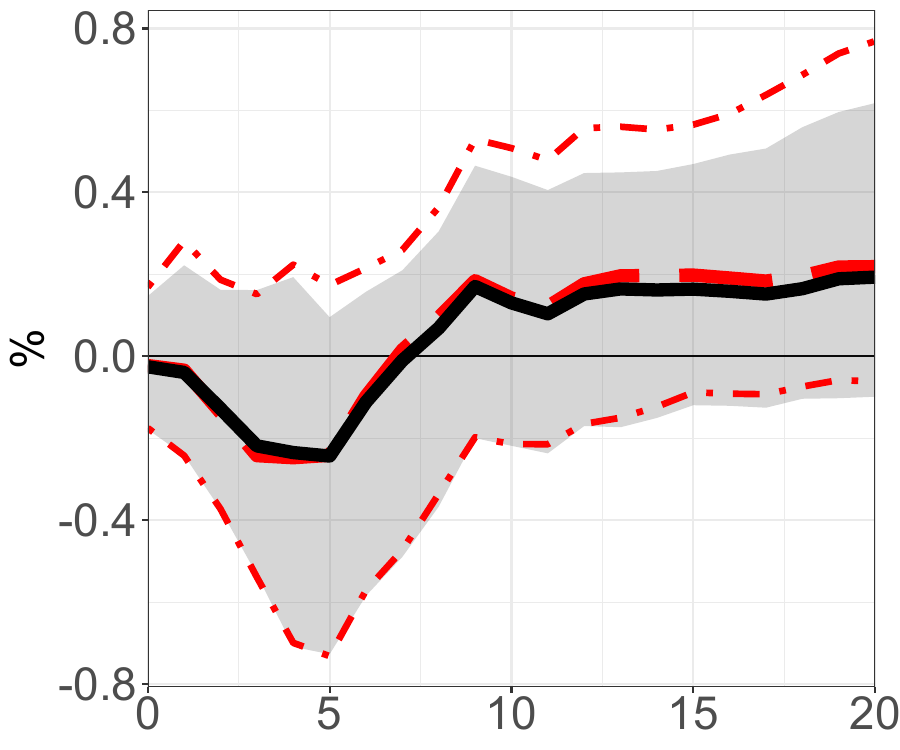}}\hfill
\subfloat[$r \rightarrow rer$]{\includegraphics[width=0.25\textwidth]{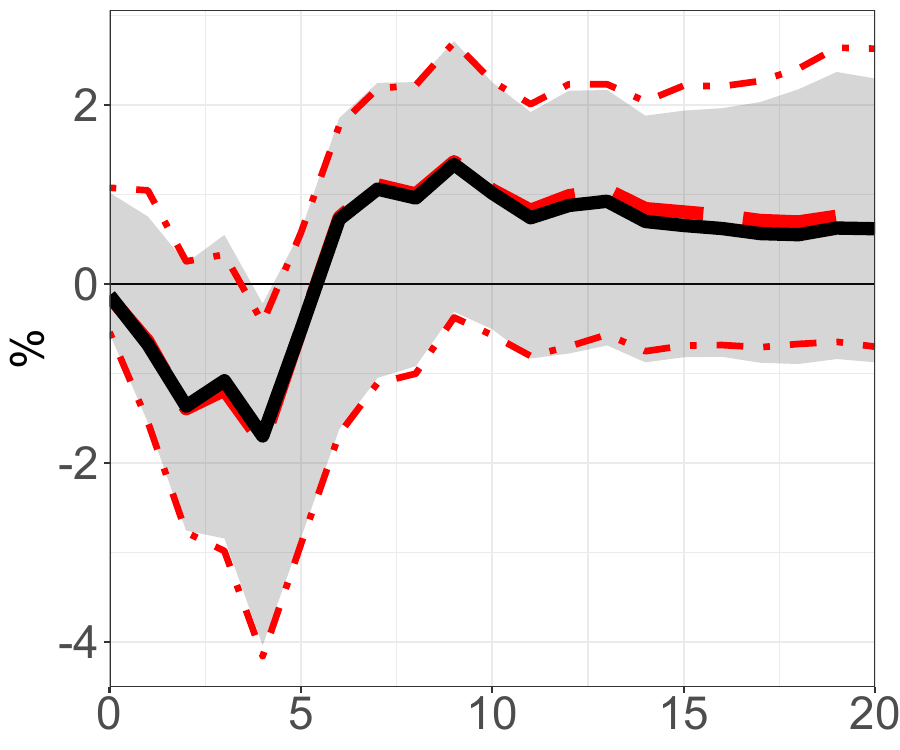}}\hfill
\subfloat[$r \rightarrow cab$]{\includegraphics[width=0.25\textwidth]{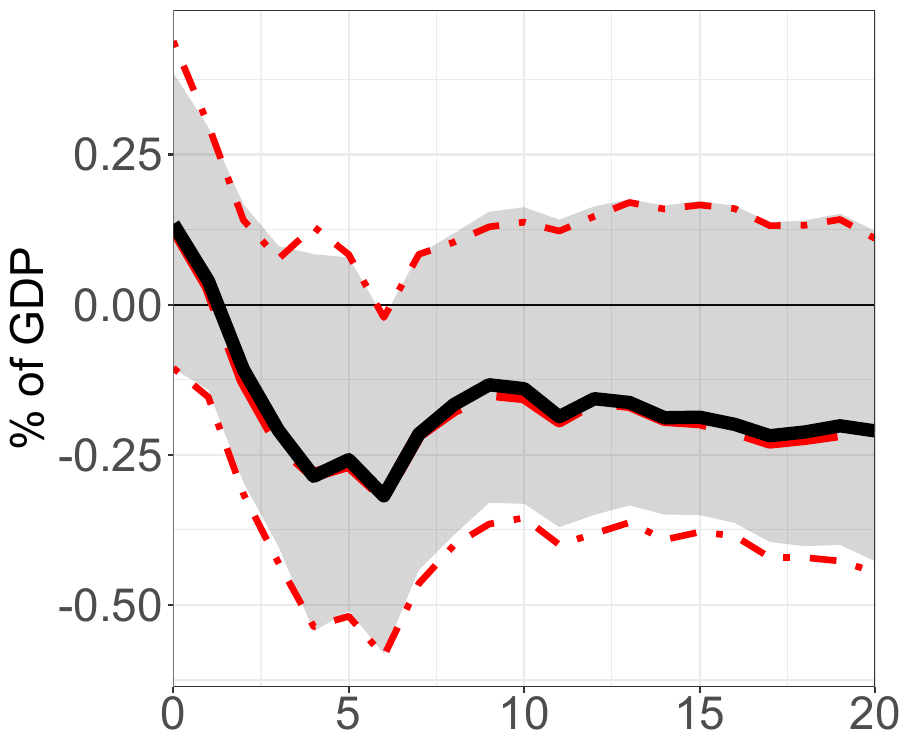}}\hfill
\subfloat[$r \rightarrow srate$]{\includegraphics[width=0.25\textwidth]{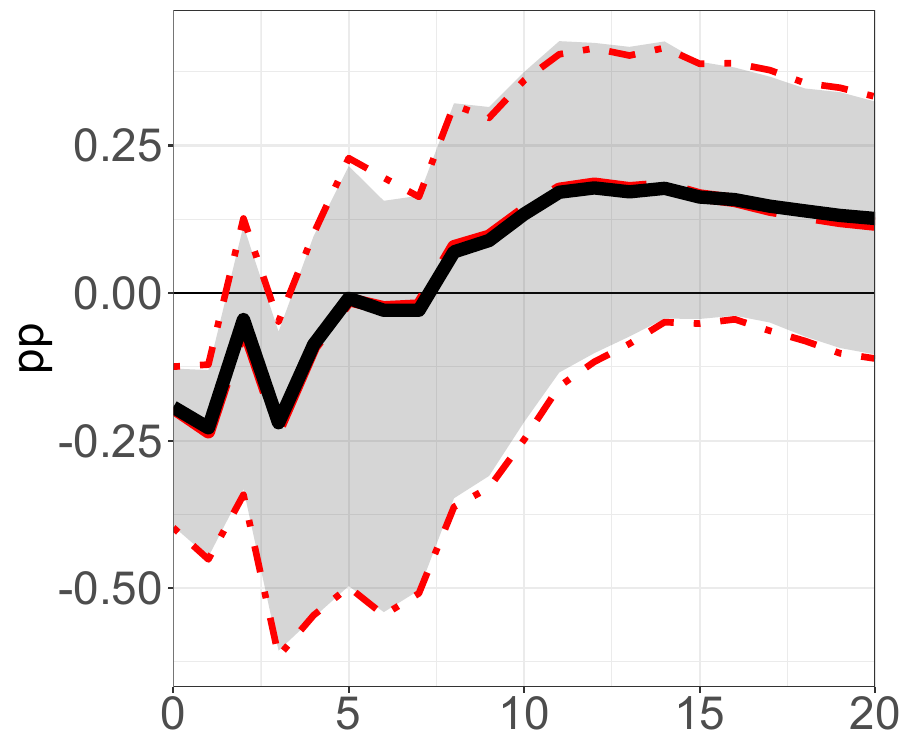}}\hfill
\end{center}
{\footnotesize{\textit{Notes:} This figure plots SVAR impulse responses to fiscal shocks that either increase $g$ by 1\% of GDP or decrease $r$ by 1\% of GDP. VAR specification has 5 lags of $g$, $r$, $gdp$, $defl$, $rer$, $cab$, $srate$, $f_{\Delta g}$ and $f_{\Delta gdp}$ as defined in the text. Sample is 1986Q1-2019Q4. Black line and shaded area represent impulse responses and confidence intervals for $CK$ identification while red dashed line and dot-dashed lines represent $BP$ identification. Residual-based moving block bootstrap 0.68 confidence intervals with 1000 draws. Horizontal axis has quarters from 0 to 20. \par}}
\end{figure}

\begin{figure}[!htpb]
\captionsetup[subfigure]{labelformat=empty}
\begin{center}
\caption{Pooled SVAR impulse responses to 1\% of GDP fiscal shocks, euro area countries.}\label{fig:europe_irf}
\subfloat[$g \rightarrow g$]{\includegraphics[width=0.25\textwidth]{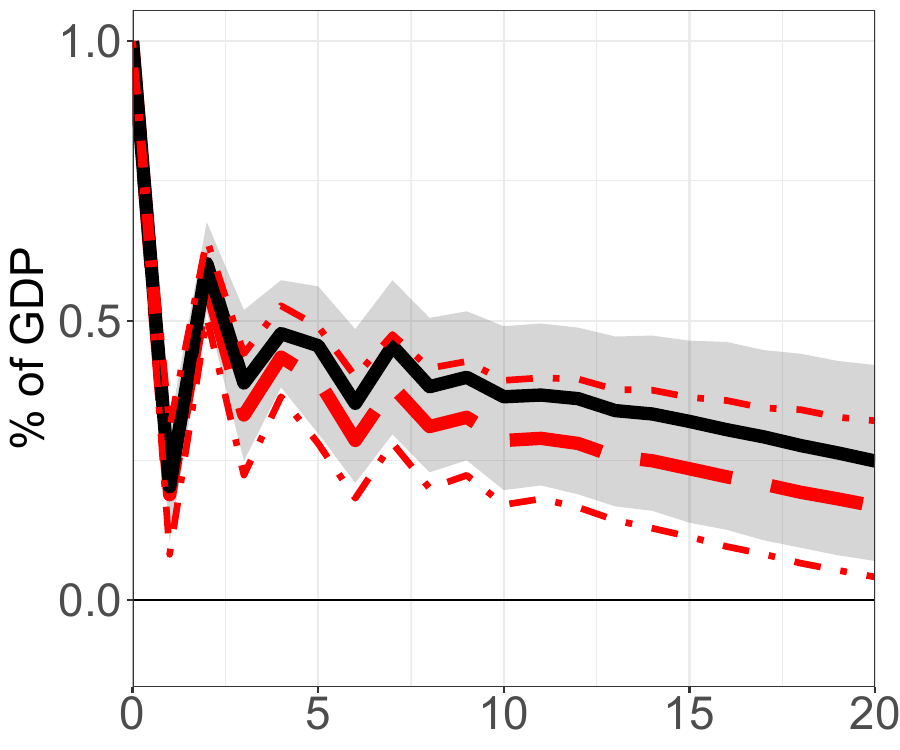}}\hfill
\subfloat[$g \rightarrow r$]{\includegraphics[width=0.25\textwidth]{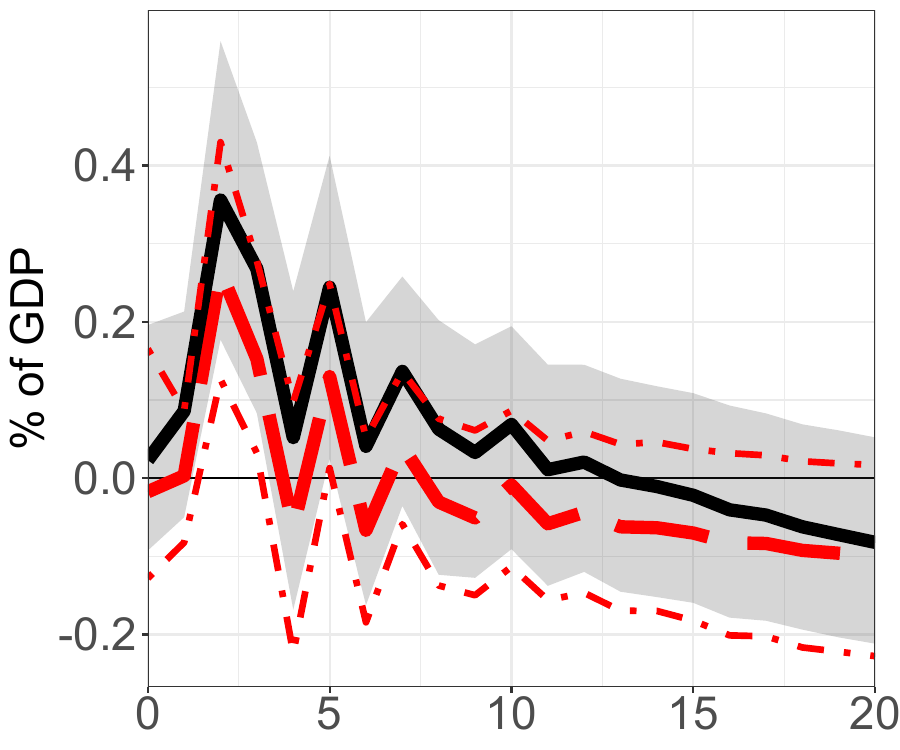}}\hfill
\subfloat[$g \rightarrow gdp$]{\includegraphics[width=0.25\textwidth]{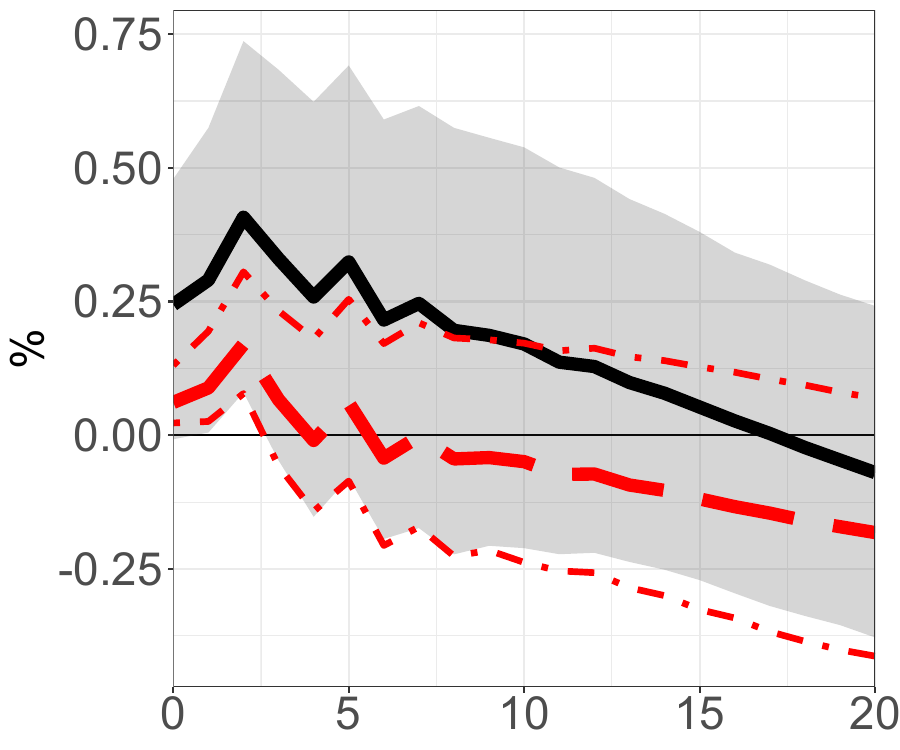}}
\vspace{2mm}
\subfloat[$g \rightarrow defl$]{\includegraphics[width=0.25\textwidth]{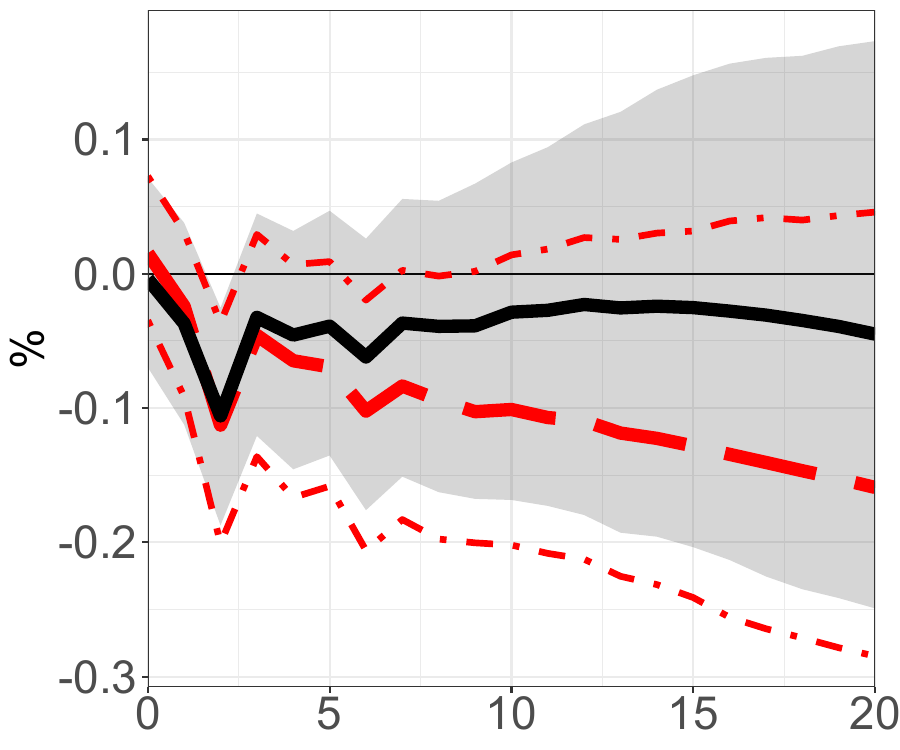}}\hfill
\subfloat[$g \rightarrow (r-g)$]{\includegraphics[width=0.25\textwidth]{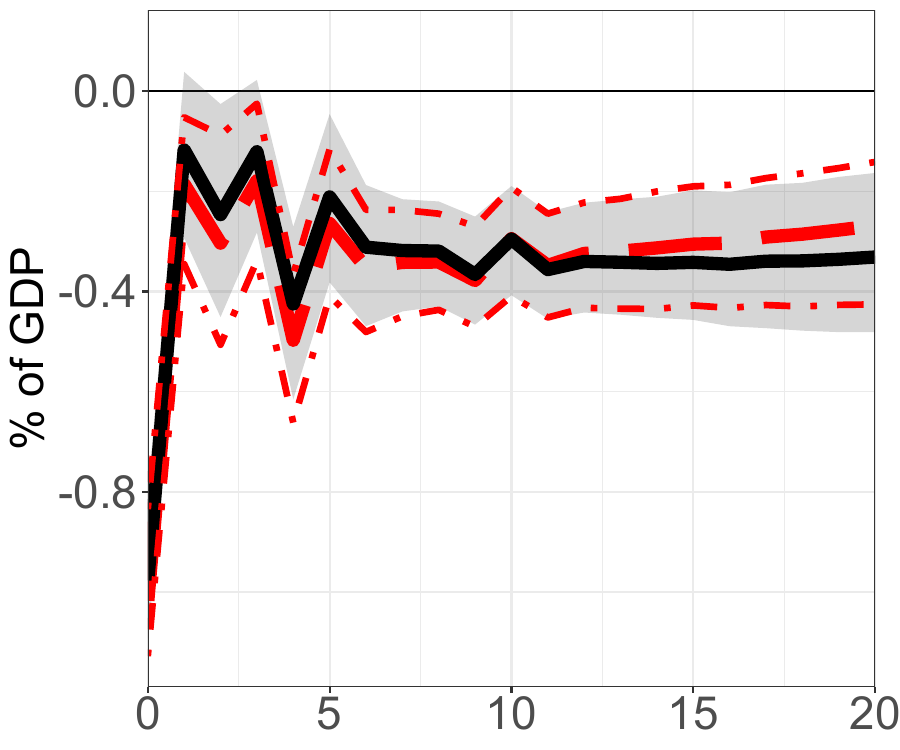}}\hfill
\subfloat[$g \rightarrow cab$]{\includegraphics[width=0.25\textwidth]{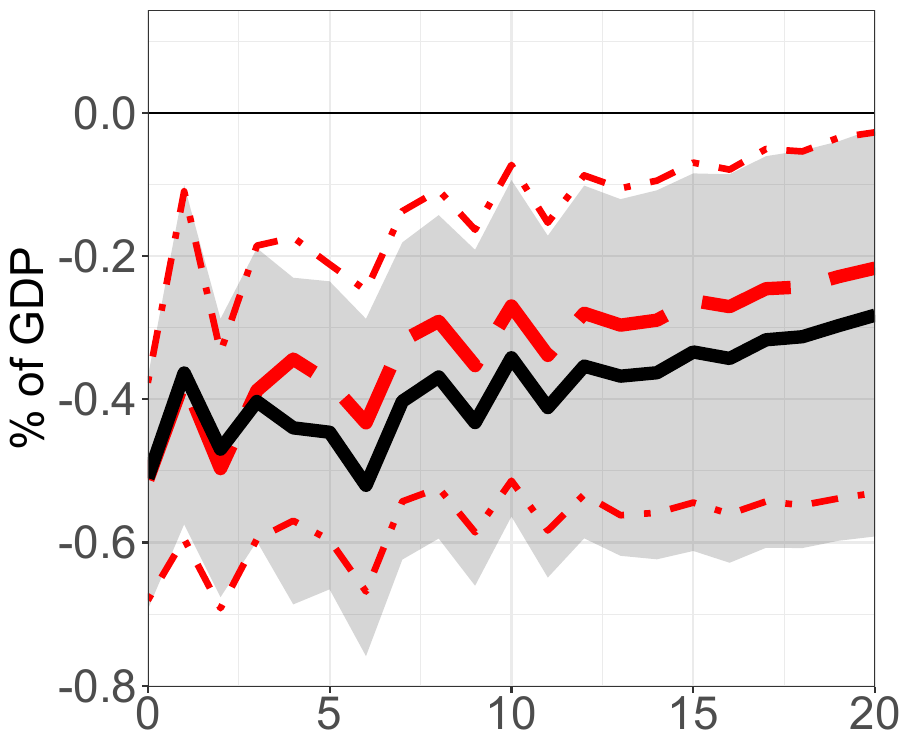}}

\vspace{8mm}

\subfloat[$r \rightarrow g$]{\includegraphics[width=0.25\textwidth]{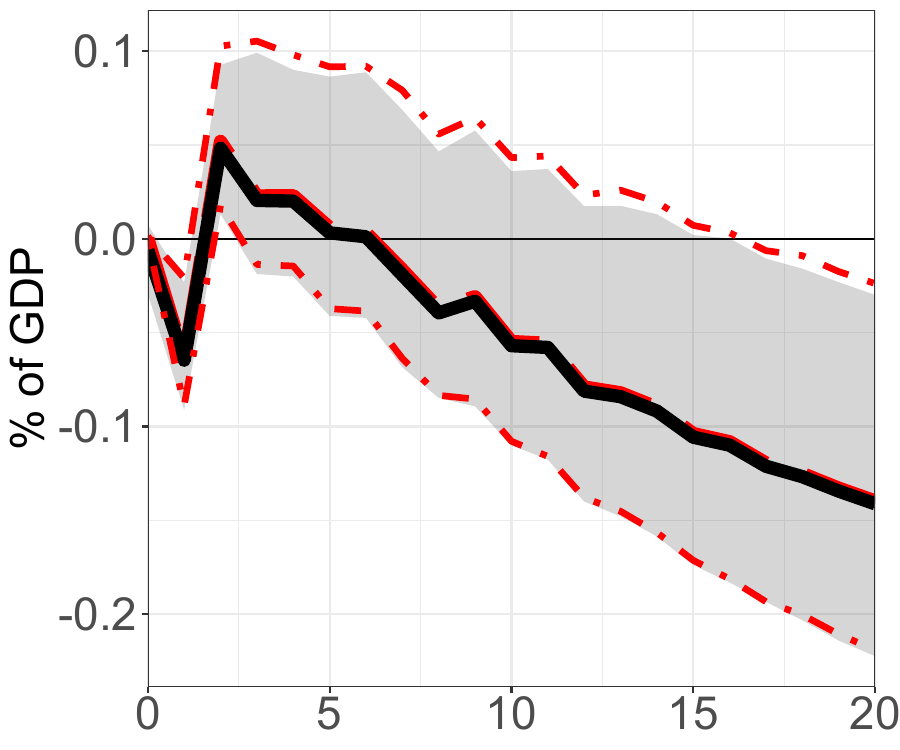}}\hfill
\subfloat[$r \rightarrow r$]{\includegraphics[width=0.25\textwidth]{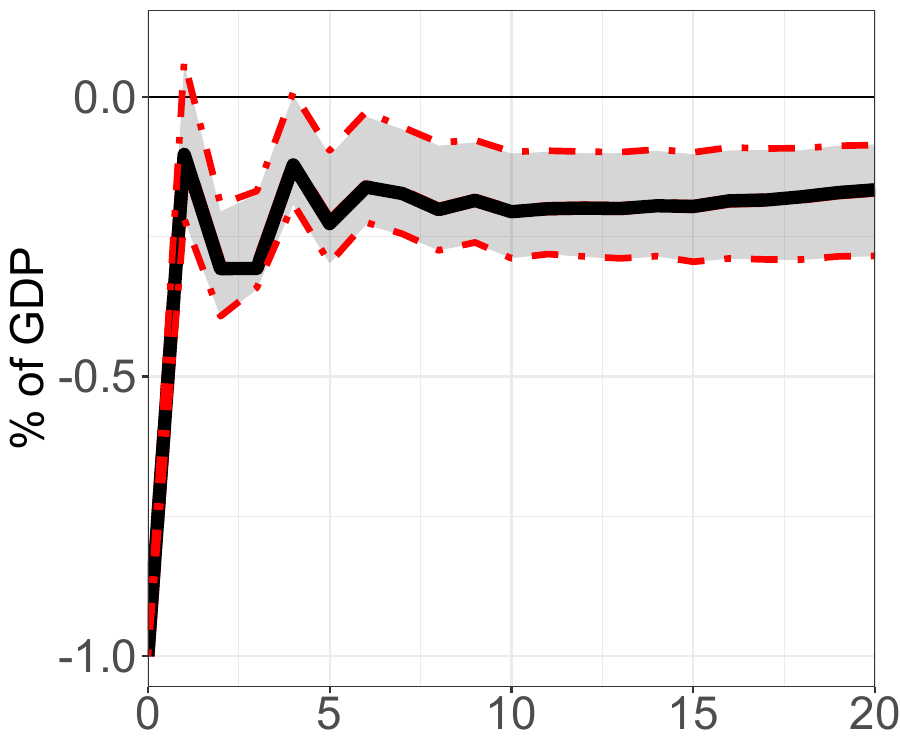}}\hfill
\subfloat[$r \rightarrow gdp$]{\includegraphics[width=0.25\textwidth]{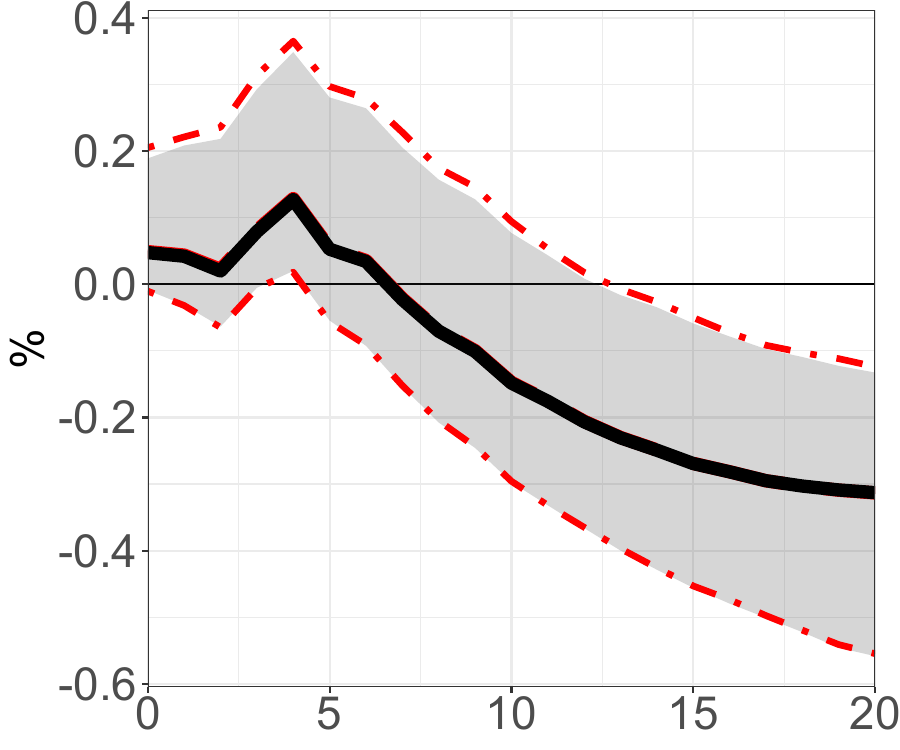}}
\vspace{2mm}
\subfloat[$r \rightarrow defl$]{\includegraphics[width=0.25\textwidth]{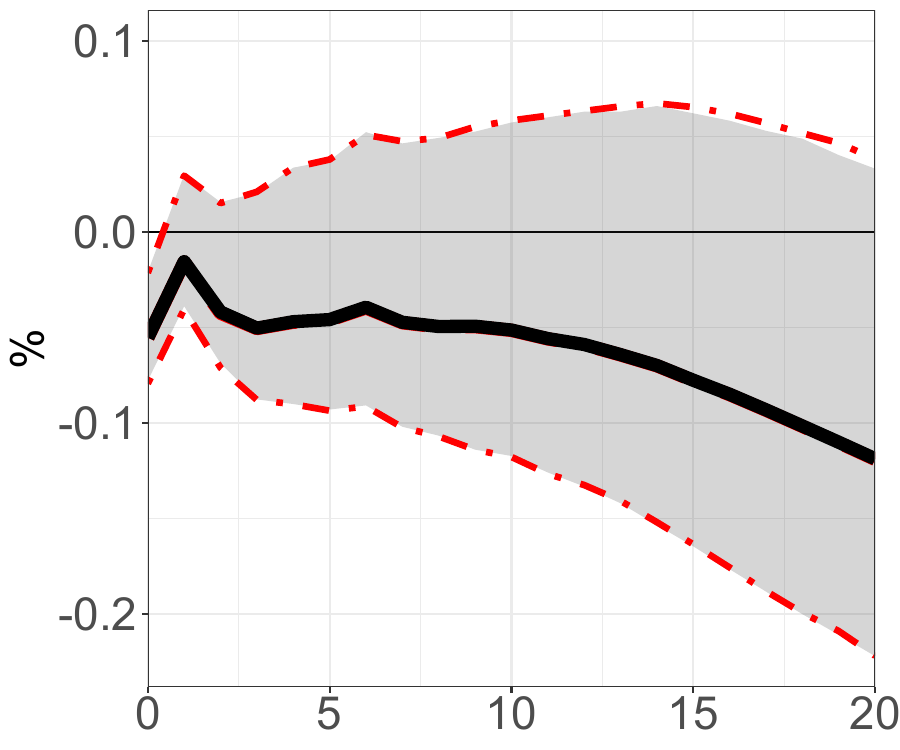}}\hfill
\subfloat[$r \rightarrow (r-g)$]{\includegraphics[width=0.25\textwidth]{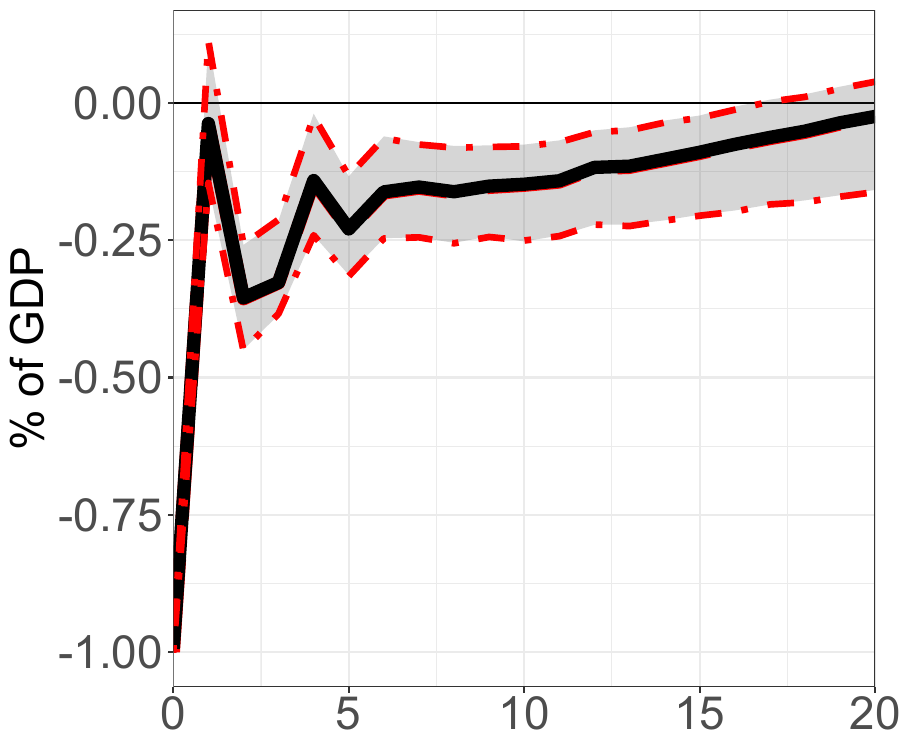}}\hfill
\subfloat[$r \rightarrow cab$]{\includegraphics[width=0.25\textwidth]{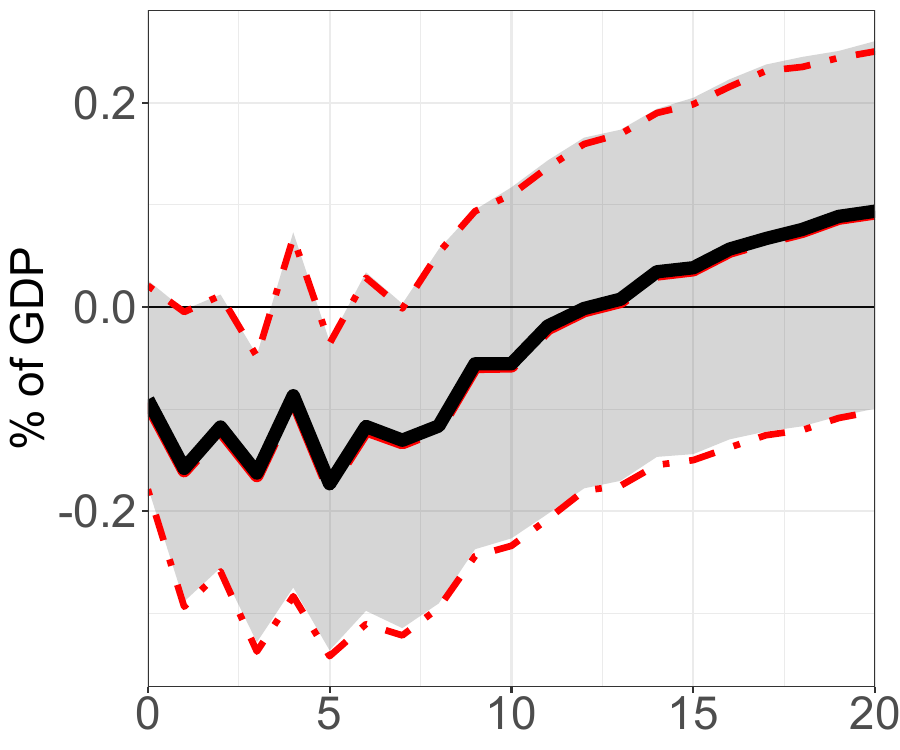}}

\end{center}
{\footnotesize{\textit{Notes:} This figure plots pooled SVAR impulse responses to fiscal shocks that either increase $g$ by 1\% of GDP or decrease $r$ by 1\% of GDP. VAR specification has 5 lags of $g$, $r$, $gdp$, $defl$, $rer$, $cab$, $srate$, $f_{\Delta g}$ and $f_{\Delta gdp}$ as defined in the text. $rer$ and $srate$ are considered as exogenous. Sample is 2001Q1-2019Q4 for Austria and 1999Q1-2019Q4 for Belgium, Finland, Portugal and the Netherlands. Black line and shaded area represent impulse responses and confidence intervals for $CK$ identification while red dashed line and dot-dashed lines represent $BP$ identification. Residual-based moving block bootstrap 0.68 confidence intervals with 1000 draws. Horizontal axis has quarters from 0 to 20. \par}}
\end{figure}

\autoref{fig:canada_irf} and \autoref{fig:europe_irf} depict impulse responses for Canada and the euro area small open economies using both BP and CK style identifications of the structural form. The black lines and shaded areas represent impulse responses and ($68\%$) confidence intervals for CK identification and red dashed lines and dot-dashed lines represent BP identification. Impulse responses are plotted up to 20 quarters ahead. The impulse responses represent the impact of the endogenous variables to a $1\%$ of GDP fiscal shock (either $+g$ or $-r$). We mainly focus here on the CK identification which we consider to be the baseline.

\emph{Canada: $+g$ shock.}--- A shock to government spending initially raises GDP but after 15 quarters the impulse response turns negative. We find a negative effect on GDP deflator during the first year but after 5 quarters the effect on inflation turns positive. The real exchange rate follows a similar pattern. Current account starts to decrease straight from the first quarter after the shock and while varying over time the impulse response stays on balance negative. The response of nominal short-term interest seems not to markedly deviate from zero although the response somewhat wavers around it. Government net revenues decrease after the spending expansion. This in turn results in a budget deficit ($r-g$) after a government spending shock. Thus we find that a government spending shock results in both a budget and a current account deficit. This evidence is in line with the twin-deficits hypothesis.

These results are in some respects consistent with the Mundell-Fleming framework. The current account decreases which implies that capital flows in. At the same time, the real exchange rate appreciates. Also, the short-term interest rate, while it initially slightly drops, rises after 5 quarters before again decreasing. There are some distinct features also. In the standard Mundell-Fleming framework inflation is not affected as monetary response offsets the inflationary pressures from the fiscal expansion. However, here it seems that a government spending shock has positive effect on inflation in the medium term. Moreover, GDP rises on impact and the effect stays positive for several years. However, in the long run, the impulse response turns negative. Overall, it seems that the decrease of the current account is not enough to fully offset the effects of expansionary fiscal policy as the standard Mundell-Fleming framework would suggest.

\emph{Canada: $-r$ shock.}--- A negative shock to net revenue slightly raises GDP and the effect seems to last almost 5 years. The shock also mildly raises inflation roughly after two years. The real exchange rate first decreases but roughly after two years after the shock the response turns positive. Current account response turns negative after a few quarters. The budget balance response on the other hand quickly recovers from the initial negative shock to close to zero in a matter of quarters. This pattern is not as clearly aligned with the twin-deficits hypothesis as the responses to a spending shock while not necessarily clearly against it either as the overall effect on budget balance still seems negative. The response of nominal short-term interest rate is slightly negative at first before turning mildly positive roughly after two years. However, the response does not seem to markedly deviate from zero overall. The effects of this expansionary revenue shock on $defl$, $rer$, $cab$, and $srate$ are qualitatively quite similar to the spending shock but more muted.

\emph{Euro area small open economies: $+g$ shock.}--- A shock to government spending clearly raises GDP at first (CK identification). Quite quickly, however, the impulse response starts to slowly decrease towards zero. The overall effect is nevertheless positive and lasts for several years. A government spending shock seems to have a small negative effect on domestic inflation. The impulse response for current account is clearly negative for the whole period and there is also a negative effect on the budget balance meaning that the results are again in line with the twin-deficits hypothesis. Overall, it seems that fiscal stimulus raises GDP in the short run. However, the decrease in current account indicates that at least part of the stimulus might spill over to other countries. That is, imports rise which dilutes the effect of fiscal stimulus.

\emph{Euro area small open economies: $-r$ shock.}--- A negative net revenue shock has a close to zero positive effect on output at first but later on, the effect turns slightly negative. The shock lowers government revenues quite persistently and, after a few quarters, government spending also starts to decrease. The overall effect on the budget deficit is negative. In addition, a shock that lowers net revenues seems to add deflationary pressures. For the current account, the impulse response decreases at first but in the longer run there is no effect and some of the point estimates turn somewhat positive. Thus, in this case, evidence over the twin-deficits hypothesis is inconclusive.

\subsection{Robustness}\label{robustness}

\begin{figure}[!h]
\begin{center}
\caption{Robustness of cumulative government spending multipliers.}\label{fig:robustness}
\subfloat[Canada, lags]{\includegraphics[width=0.5\textwidth]{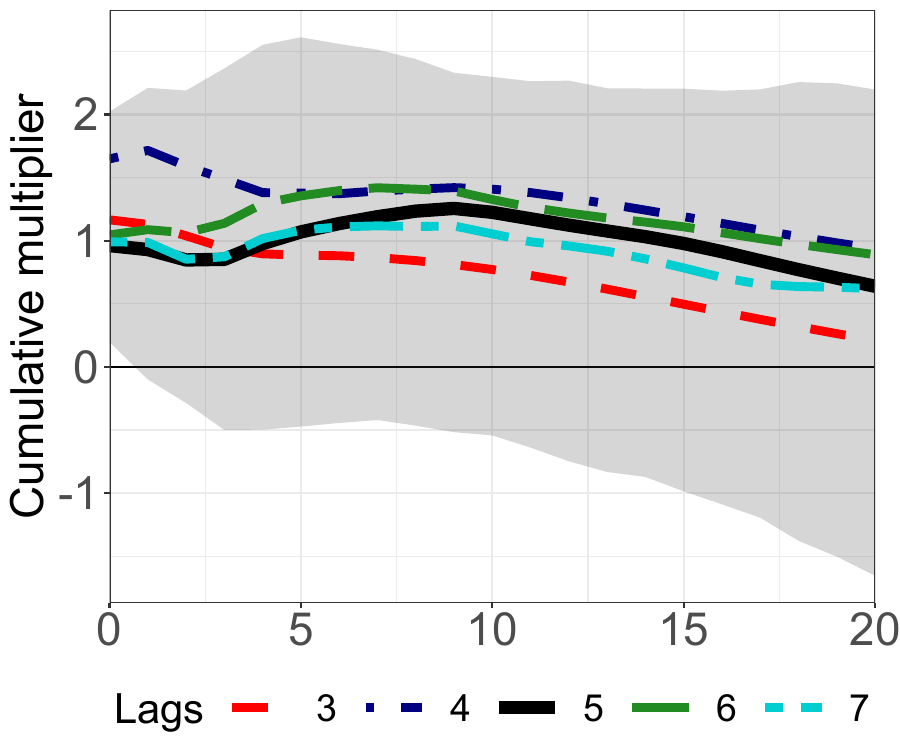}}\hfill
\subfloat[Euro area countries, lags]{\includegraphics[width=0.5\textwidth]{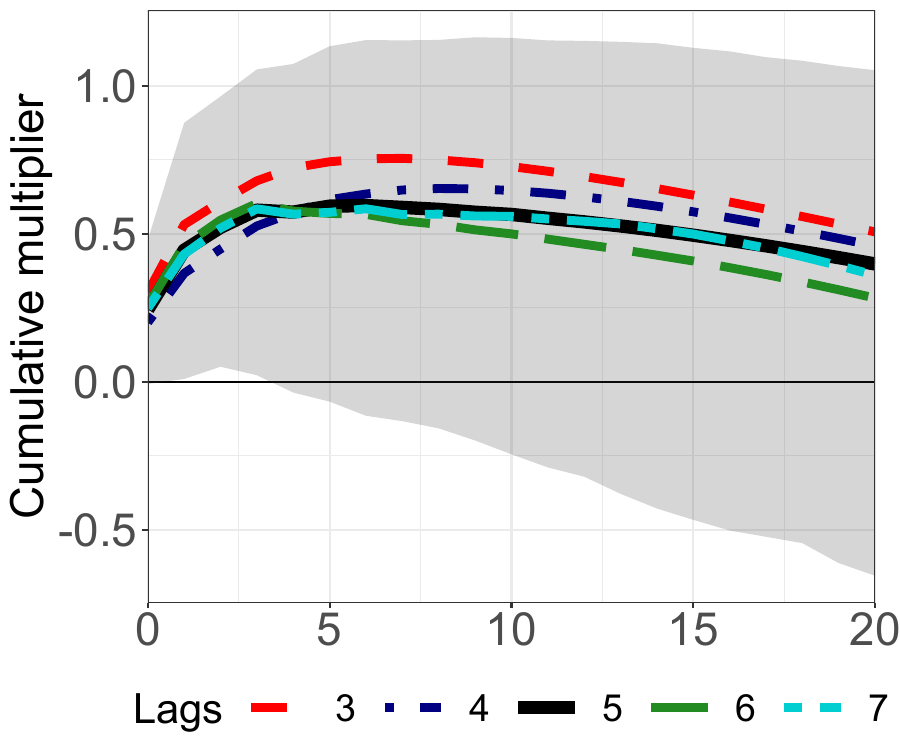}}\hfill \\
\vspace{5mm}
\subfloat[Euro area countries, leave one country out]{\includegraphics[width=0.5\textwidth]{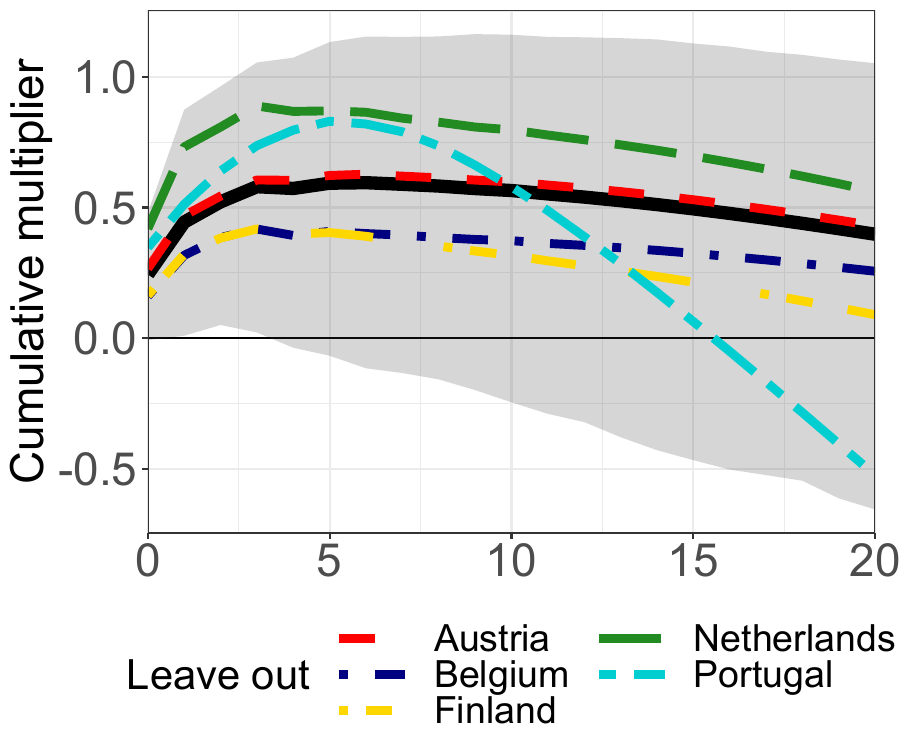}}\hfill
\end{center}
{\footnotesize{\textit{Notes:} This figure plots cumulative government spending multipliers (Equation \eqref{fmulti}) calculated from SVAR impulse responses to government spending shocks for up to 20 quarters from the initial shock. Black line and shaded area represent point estimates and confidence intervals for baseline $CK$ identification (5 lags). Residual-based moving block bootstrap 0.68 confidence intervals with 1000 draws. Horizontal axis has quarters from 0 to 20. \par}}
\end{figure}

\emph{Reduced form VAR specification.}--- Evidently, choices over the reduced form VAR specification also have an effect on the impulse responses and, thereby, estimates of the fiscal multiplier. In \hyperref[sec:online]{Online Appendix}, where we repeat the estimation using different choices of the VAR lag length, we find rather similar impulse responses across different specifications. Overall, the estimated impulse responses using different lag lengths mostly fall on the 0.68 confidence interval of the baseline specification. Differences in impulse responses seem to be driven more by the choice of identification scheme than by changes to the reduced form VAR specification. We also examine how the different lag choices affect government spending multipliers. As shown in \autoref{fig:robustness}, different lag lengths have a quite small effect on the spending multipliers in both cases, Canada (panel a) and euro area countries (panel b).

\emph{Leaving one country out of the euro area panel.}--- The small open euro area economies studied here are similar in the sense that they are all small open developed economies with relatively similar types of institutions and importantly all belong to the same monetary union. They, nevertheless, are in some aspects also different. By adopting the empirical approach of \citet{smallbig2013} on the reduced form, we are implicitly assuming that the countries still behave similarly even if there might be some underlying differences in the magnitudes of the impulse responses and, thereby, also in the size of the fiscal multiplier. The pooled model uses each country's variation to estimate impulse responses for a representative small open economy of the currency union.

To examine how sensitive our results are to the influence of individual countries we do the following: we exclude one country at a time from the sample and examine how this affects the government spending multiplier. The results from this robustness check are depicted in \autoref{fig:robustness} (panel c). Indeed there is some variation in the size of cumulative multipliers. However, all multipliers are within the 0.68 confidence interval of the baseline. Furthermore, the shapes of the cumulative multipliers are rather similar except when Portugal is excluded from the sample. This could reflect different dynamics that were experienced by Portugal, one of the GIIPS countries, especially after the financial crisis. Without Portugal, cumulative multipliers are somewhat larger in the short-term compared to the baseline, but at larger horizons cumulative multipliers converge faster towards zero and even turn negative at around 4 years. \hyperref[sec:online]{Online Appendix} discusses possible explanations that could explain Portugal's distinctive dynamics.

\section{Conclusion} \label{sec:conclusion}

In this paper, we propose an instrument for aggregate output that is based on professional forecast errors in trading partner economies and apply it in fiscal SVAR-IVs to disentangle policy shocks from endogenous variation in output. The instrument is suited to a small open economy setting where policy shocks of the domestic economy do not cause contemporaneous forecast errors in its trading partners while, at the same time, unexpected developments in these trading partners do cause variation in domestic output.

Compared to the utilization-adjusted TFP series of \citet{fernald2014}, which prior literature has used as a proxy for output shocks, the proposed instrument has some desirable properties. It does not rely on strong modeling assumptions that are needed to construct the TFP series and to further adjust it for capacity utilization. Instead, the instrument we propose is constructed from observable data on forecast errors. Revisions to the series of \citet{fernald2014} has also seen its properties change over time \citep{sims} and the revised series does not seem to yield results consistent with those reported in the prior literature.

Using the instrument we find that estimates of the government spending multiplier are sensitive to small differences in the output elasticity of government spending which the typical \citet{blanchard2002} identification restricts to zero. According to our point estimates, the impact multiplier roughly doubles when the output elasticity is not restricted to zero. In our baseline without the zero restriction, we find cumulative spending multipliers of around $1$ for Canada and $0.5$ for small euro area economies. Furthermore, regardless of the specification, we find that fiscal stimulus seems more effective in Canada than in the small open economies of the euro area.

Our results also suggest that while spending shocks have larger outright output effects than revenue shocks, they also seem to have higher medium-run fiscal costs due to their more persistent effect on the deficit. Furthermore, considering Canada, our results lend support to the twin-deficits hypothesis as both the fiscal and the current account deficits increase following expansionary policy. For the small euro area countries, the results are less conclusive, while perhaps also more in favor of the twin-deficits hypothesis than against it.

Finally, we would like to note that the proposed instrument can potentially be useful in other contexts as well. Considering SVAR models, one possible step further would be to use our instrument together with additional identifying information, for example, using non-gaussian features of the data \citep{lanne2021gmm,keweloh2021generalized}. This could result in stronger identification and reduce parameter uncertainty. Furthermore, there is no particular reason why one would be limited to utilizing this instrument in estimating fiscal multipliers in SVAR models. As an example, the instrument and structural identifications we utilize can as well be applied in local projections \citep{jorda2005estimation,plagborg}.

\clearpage
\addcontentsline{toc}{section}{References}    

\singlespacing
\setlength\bibsep{0pt}
\bibliographystyle{econ-aea}
\bibliography{references}

\clearpage

\onehalfspacing

\renewcommand{\thefigure}{A\arabic{figure}}
\setcounter{figure}{0}

\renewcommand{\thetable}{A\arabic{table}}
\setcounter{table}{0}

\renewcommand{\theequation}{A\arabic{equation}}
\setcounter{equation}{0}

\setcounter{page}{1}

\section*{Online Appendix} \label{sec:online}
\addcontentsline{toc}{section}{Online Appendix}

\subsection*{Utilization adjusted TFP and Caldara and Kamps (2017) specification using different vintages of data }\label{sec:tfp_vintages}

As alluded to in the main text, the utilization adjusted TFP series of \citetsupp{fernald2014} has seen many revisions over time. Being a purified Solow residual, revisions to measures in output and its components obviously lead to revisions also in the utilization-adjusted TFP. However, largest revisions to this series arguably emanate from changes to the actual methodology rather than from changes in the underlying data. As shown by \citetsupp{sims},  these revisions have a large effect on the empirical results in their study on the effects of news shocks.

\citetsupp{fernald2014} constructs the utilization adjusted TFP as follows. First, the log change in TFP is constructed as \begin{equation}\label{eq:TFP}
    \Delta \ln TFP_t = \Delta \ln Y_t - \alpha_t \Delta \ln K_t - (1-\alpha_t) \Delta \ln L_t,
\end{equation}
where log change in output $\Delta \ln Y_t$ is measured as the equally weighted average of log changes in gross value added in the business sector (expenditure side) and in gross domestic income less non-business output (income side); $\alpha_t$ is capital's share of income; $\Delta \ln K_t$ is capital input growth calculated from disaggregated quarterly investment data and $\Delta \ln L_t$ is labor input growth measured as the sum of growth in business sector hours worked and change in labor quality/composition.

Secondly, this measure of technological change is adjusted for capacity-utilization to filter out business cycle fluctuations. As capacity-utilization itself is not observable, \citetsupp{fernald2014} uses observable margins to proxy for changes in utilization. Accordingly, employment and industry capital (extensive margin) are quasi-fixed but hours per worker, labor effort and capital use (intensive margin) can be adjusted without cost. Because firms operate on all margins simultaneously changes in hours per worker, which are more readily observable than other intensive margins, arguably works as a proxy also for overall utilization-adjustment. The utilization-adjustment is thus written as follows:
\begin{equation}
    \Delta \ln \widehat{util}_t = \Sigma_i \kappa_i \hat{\beta}_i \Delta \ln h_{it}^c
\end{equation}
where $h_{it}^c$ denotes hours per worker in industry $i$, $\kappa_i$ represents industry specific weights and $\hat{\beta}_i$ is an estimate of industry specific factors of proportionality estimated using demand-side shocks as instruments. With an estimate for changes in utilization, the Solow residual from Equation \eqref{eq:TFP} can be adjusted to arrive at a measure of utilization-adjusted TFP:

\begin{equation}
    \Delta \ln TFP_t^{util} = \Delta \ln TFP_t - \Delta \ln \widehat{util}_t 
\end{equation}

It is worth noting that when using a TFP series as an instrument for output one is effectively using a filtered series of output as an instrument for output. This perhaps helps to ensure the relevance of the instrument but also raises concerns about the series' exogeneity. Based on the estimates in \autoref{exogrele1}, it appears that the more recent vintages of the TFP series of \citetsupp{fernald2014} are predicted by government spending forecast errors of professional forecasters. Given how the adjusted TFP series is constructed, it relies critically on the filtering process to correctly filter out all demand-driven variation in the original output series. According to \citetsupp{sims}, the TFP measure is likely confounded by business cycle fluctuations due to imperfect measurement of factor utilization.\footnote{\citetsupp{tfpcomin} use firm surveys on capacity utilization as a more direct proxy of utilization which arguably could result in a less confounded TFP series. Moreover, \citetsupp{tfpcomin} illustrate how the zero-profit assumption as well as ignoring adjustment costs can bias the calculation of TFP. By relaxing the zero-profit assumption they show that one underestimates long-run TFP growth and overestimates its volatility and cyclicality if one uses the conventional method of calculating TFP growth.} This in turn might also result in correlation between the TFP series and government spending shocks.

To study how the methodological changes and data revisions concerning the TFP series potentially affect the fiscal SVAR estimates we do the following exercise: for each vintage of \citetsupp{fernald2014}'s TFP series we re-estimate the main elasticites in \citetsupp{caldara2017} and examine whether these change from vintage to vintage. For comparison we also replicate the results in \citetsupp{caldara2017} using the endogenous variables and TFP series from their original dataset. In each case we estimate the Bayesian VAR using the MATLAB programs provided in the supplementary files of \citetsupp{caldara2017}.

When re-estimating the elasticities with different vintages of \citetsupp{fernald2014} TFP series, we also retrieve the then available vintages of the endogenous variables used by \citetsupp{caldara2017}. Using data from different vintages that were published far away from each other could be problematic because of data revisions. Real-time data for the endogenous variables are downloaded from ALFRED. For each TFP vintage we take the publication date and download the latest available series for endogenous variables before that date. Therefore, these versions of the endogenous variables should correspond to those that would have been available at the time of each TFP vintage and thus the estimates should correspond to those one would have estimated at that time using the most recent data available. Old vintages of \citetsupp{fernald2014} TFP series are available from the author's website: \url{https://www.johnfernald.net/TFP}. To calculate per-capita values, population data is the same as in \citetsupp{caldara2017}. The sample period is held fixed at $1950Q1-2006Q4$ as in \citetsupp{caldara2017}.

\begin{figure}[!htpb]
\vspace{-1cm}
\begin{center}
\caption{Estimates of output elasticity of federal tax revenues in the \citet{caldara2017} specification using different vintages of data.}\label{fig:ck_r}
\includegraphics[width=\textwidth]{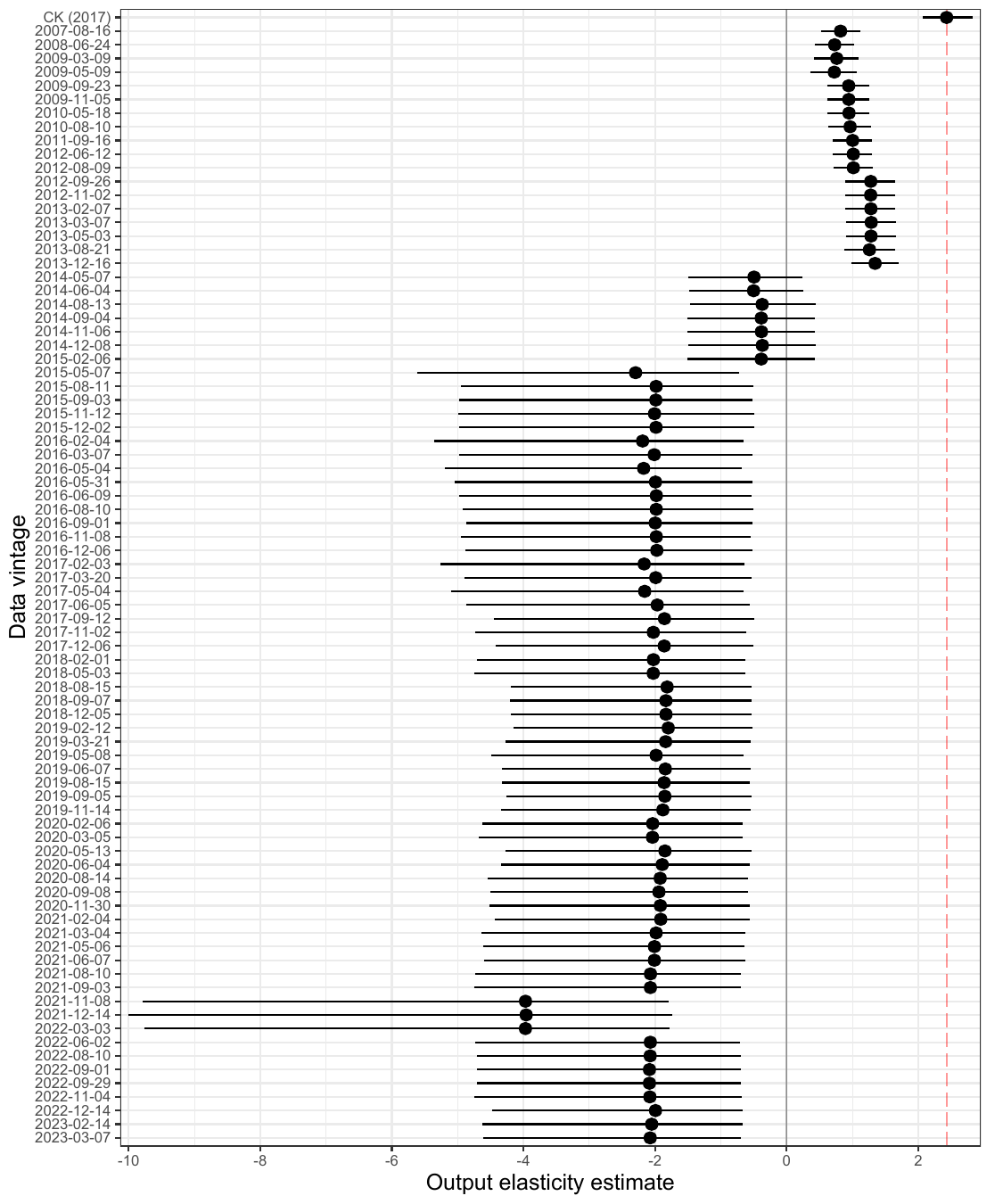}\
\end{center}
{\footnotesize{\textit{Notes:} Each point represents coefficient estimates of the output elasticity of federal tax revenues in the \citet{caldara2017} specification using their replication codes but different vintages of data. Upmost coefficient (and the dashed line) corresponds to the estimate using the original data, whereas other estimates use data available on each date given in the vertical axis. These dates are associated with different vintages of the \citet{fernald2014} TFP-series. Each coefficient is plotted with the 95\% credible set. \par}}
\end{figure}

\begin{figure}[!htpb]
\vspace{-1cm}
\begin{center}
\caption{Estimates of output elasticity of government spending and investment in the \citet{caldara2017} specification using different vintages of data.}\label{fig:ck_g}
\includegraphics[width=\textwidth]{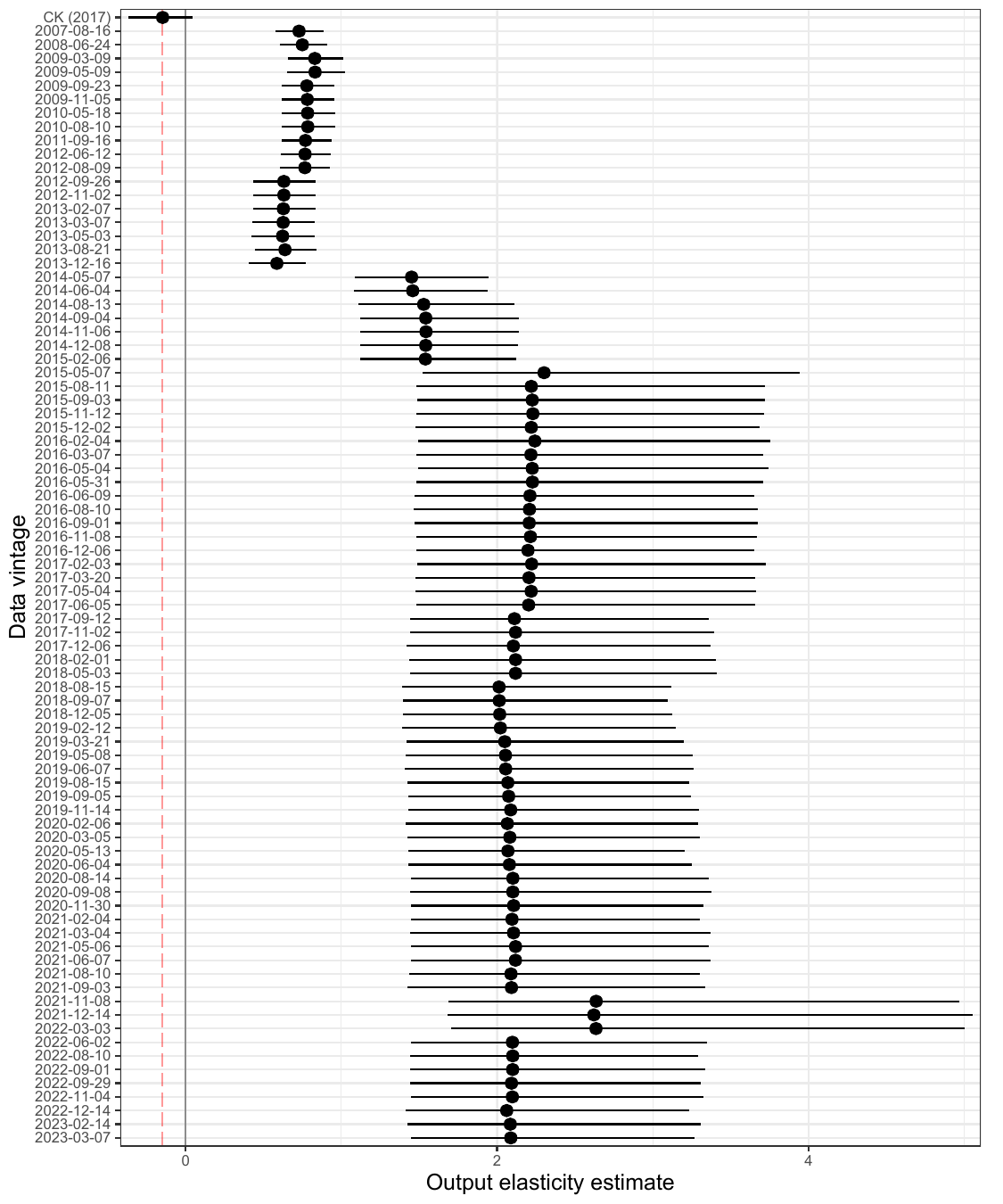}\
\end{center}
{\footnotesize{\textit{Notes:} Each point represents coefficient estimates of the output elasticity of government spending and investment in the \citet{caldara2017} specification using their replication codes but different vintages of data. Upmost coefficient (and the dashed line) corresponds to the estimate using the original data, whereas other estimates use data available on each date given in the vertical axis. These dates are associated with different vintages of the \citet{fernald2014} TFP-series. Each coefficient is plotted with the 95\% credible set. \par}}
\end{figure}

\autoref{fig:ck_r} plots estimates of output elasticity of federal tax revenue and \autoref{fig:ck_g} plots estimates of output elasticity of government spending and investment for different vintages of data. In both cases we find that the estimates for the first vintages are rather different than estimates for the newer vintages. Considering the output elasticity of government spending and investment the estimated coefficient for the newer vintages is around 2 times as large as the estimated coefficient for the older vintages. Considering the output elasticity of federal tax revenue even the sign of the estimates changes. In both cases the estimated coefficients stabilizes somewhere during 2015. After 2015 the estimates stay rather similar expect for a few vintages. Note also that the results in \citetsupp{caldara2017} are somewhat different than the results we get with the corresponding vintage of \citetsupp{fernald2014}'s TFP series. This is because the corresponding TFP series within \citetsupp{fernald2014}'s vintages is not exactly the same TFP series that is included in the dataset of \citetsupp{caldara2017}.

\begin{figure}[!htpb]
\vspace{-1cm}
\begin{center}
\caption{First stage robust F-statistics in the \citet{caldara2017} specification using different vintages of data.}\label{fig:ck_fstat}
\includegraphics[width=\textwidth]{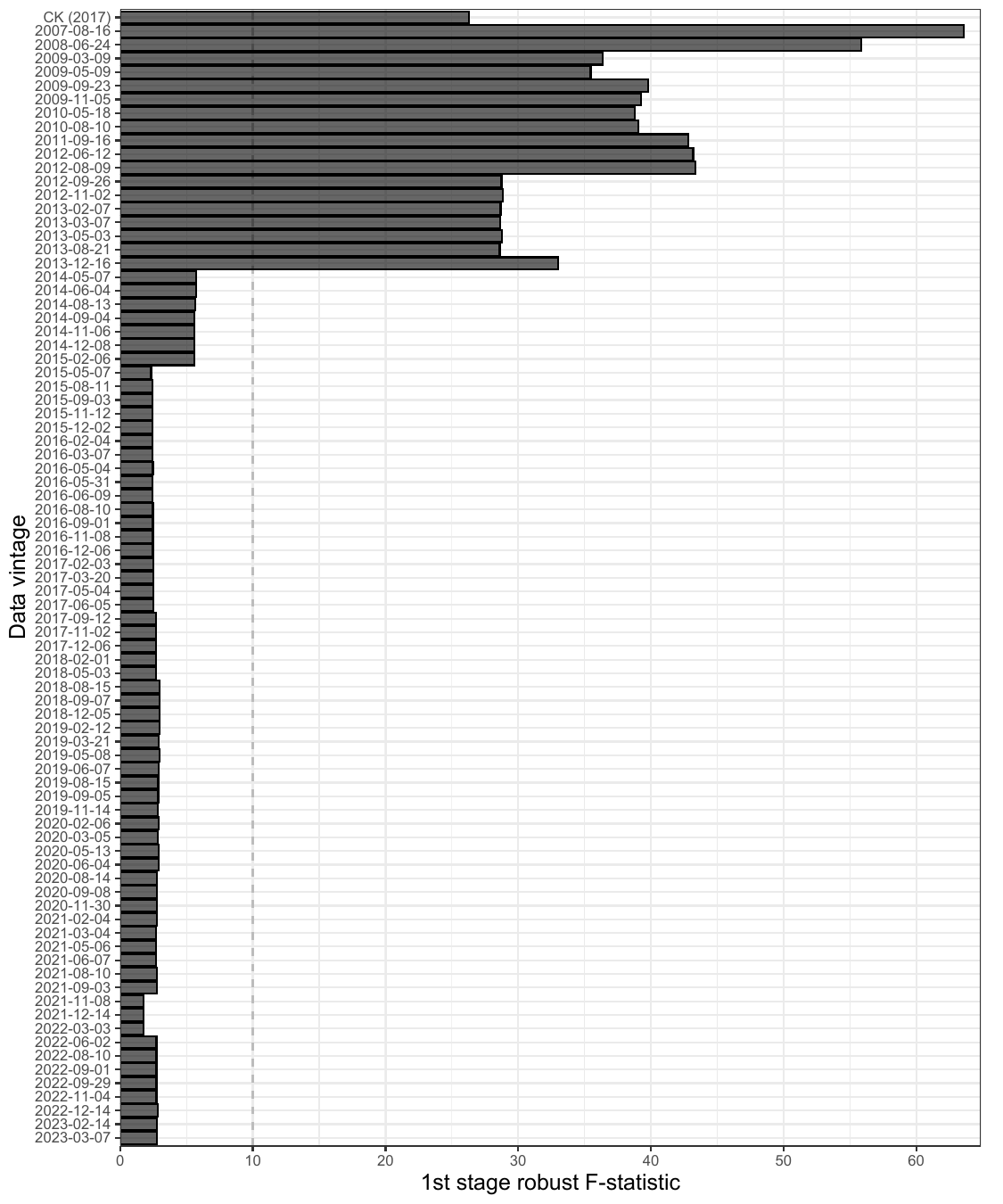}\
\end{center}
{\footnotesize{\textit{Notes:} Each bar plots the first stage F-statistics from replicating Table 2 of \citet{caldara2017} using different vintages of data. Upmost bar corresponds to the estimate using the original data, whereas other estimates use data available on each date given in the vertical axis. These dates are associated with different vintages of the \citet{fernald2014} TFP-series. \par}}
\end{figure}

Furthermore, \autoref{fig:ck_fstat} gives the 1st stage robust F-statistics for each estimate of output elasticity of federal tax revenue reported in \autoref{fig:ck_r}. The F-statistics for the older vintages indicate that the TFP instrument is highly relevant whereas the newer vintages of the series indicate that the relevancy of the instrument is rather modest. Also the robust F-statistics stabilizes after the year 2015. All in all, the different data vintages considered have a large effect on the estimates and the associated F-statistic of the SVAR model \citetsupp{caldara2017} utilize.

\subsection*{Robustness to reduced form VAR specification}\label{sec:appendixf}

\begin{figure}[!htpb]
\vspace{-1cm}
\captionsetup[subfigure]{labelformat=empty}
\begin{center}
\caption{SVAR impulse responses to 1\% of GDP fiscal shocks, Canada.}\label{fig:lagrobust_can}
\subfloat[$g \rightarrow g$]{\includegraphics[width=0.25\textwidth]{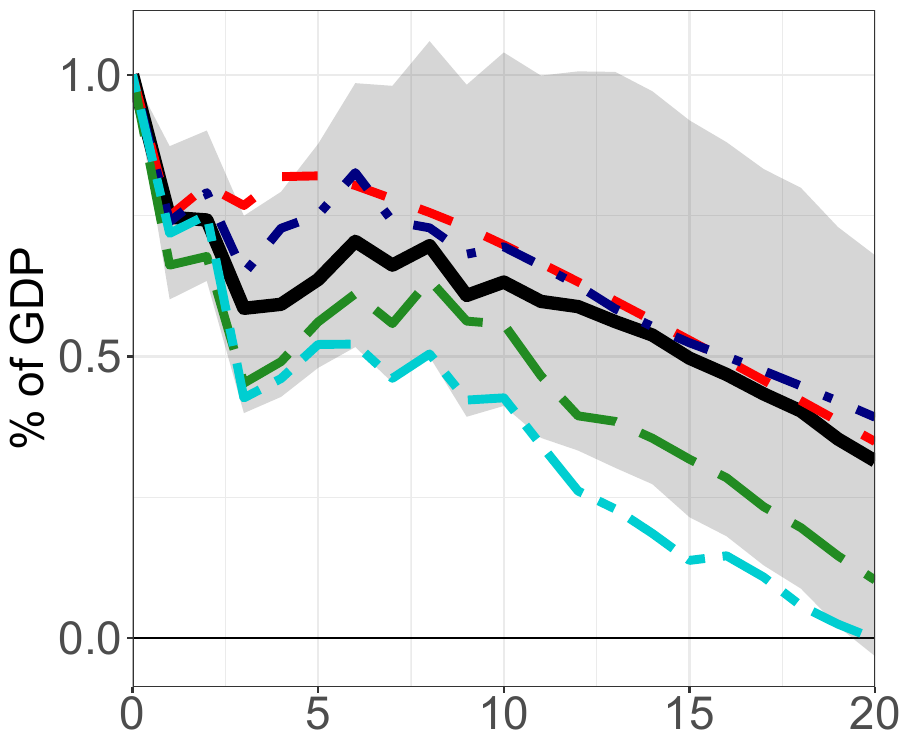}}\hfill
\subfloat[$g \rightarrow r$]{\includegraphics[width=0.25\textwidth]{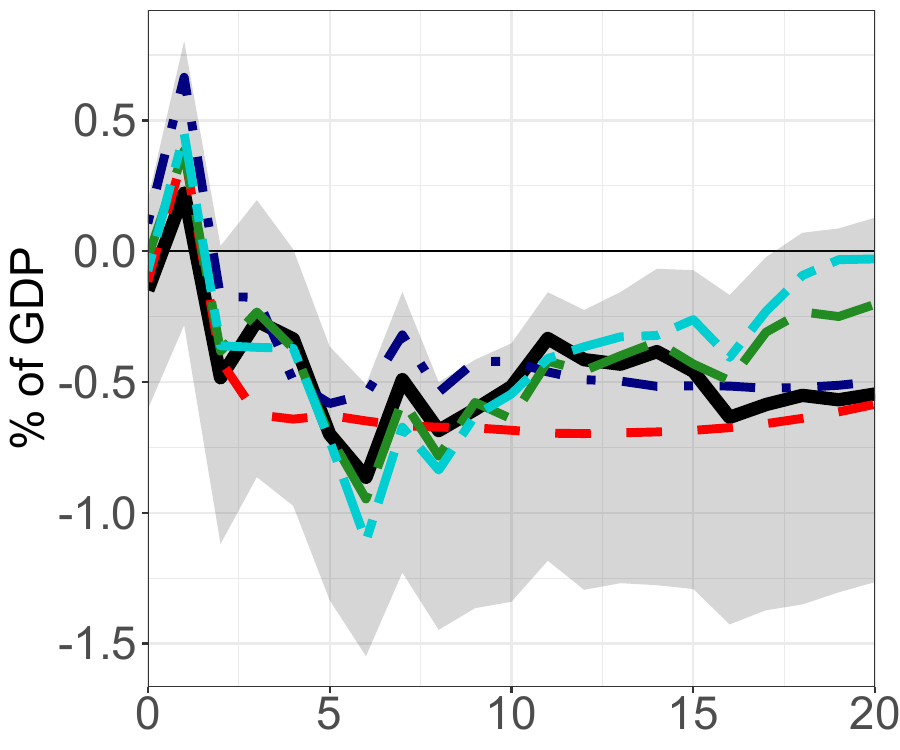}}\hfill
\subfloat[$g \rightarrow (r-g)$]{\includegraphics[width=0.25\textwidth]{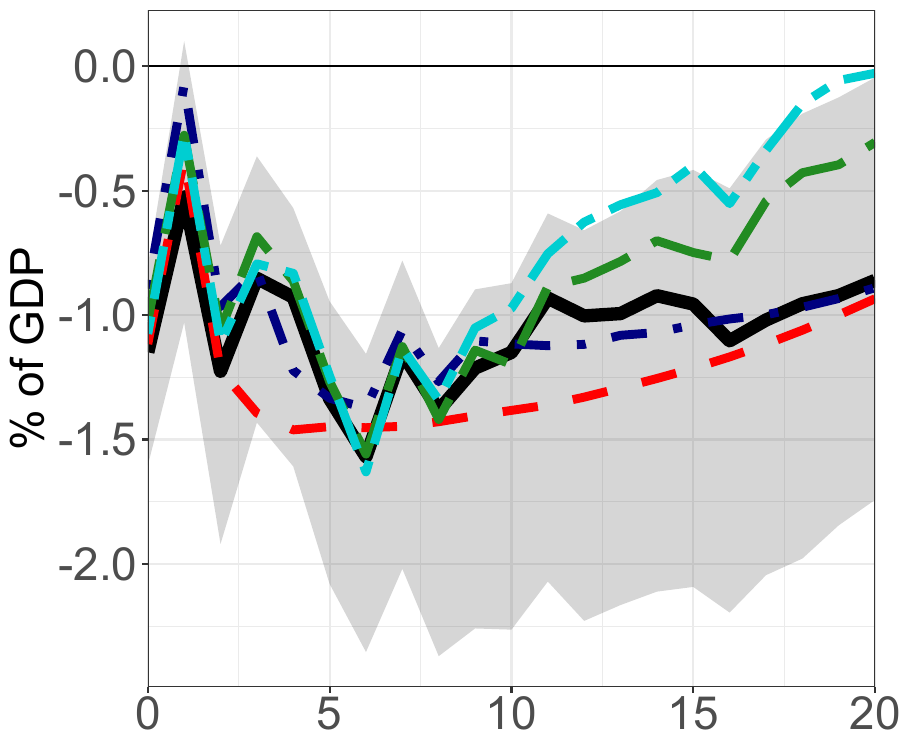}}\hfill
\subfloat[$g \rightarrow gdp$]{\includegraphics[width=0.25\textwidth]{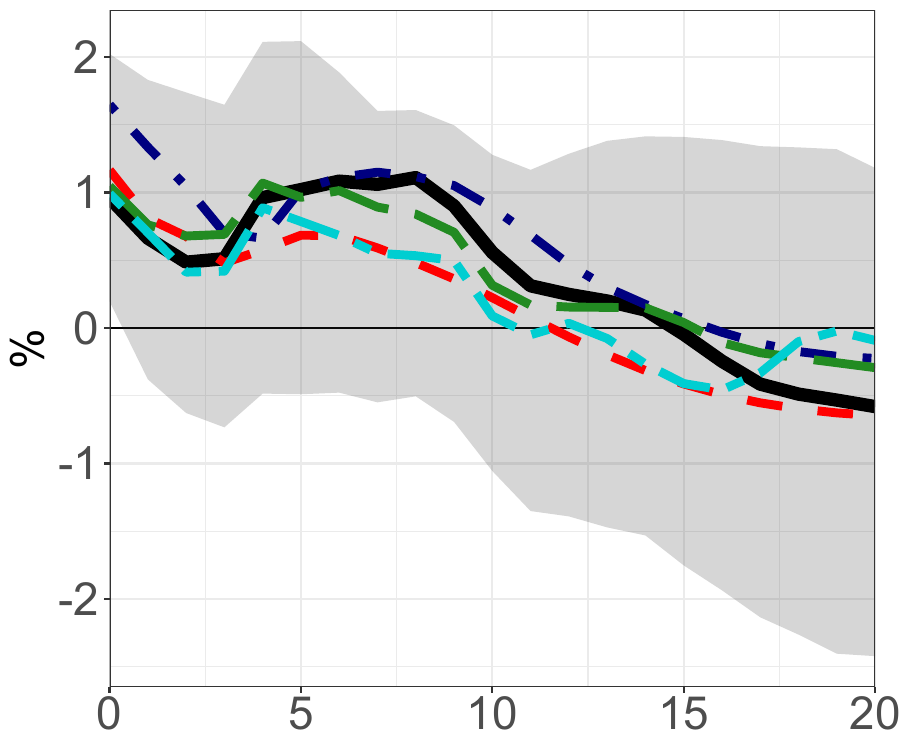}}\hfill
\vspace{2mm}
\subfloat[$g \rightarrow defl$]{\includegraphics[width=0.25\textwidth]{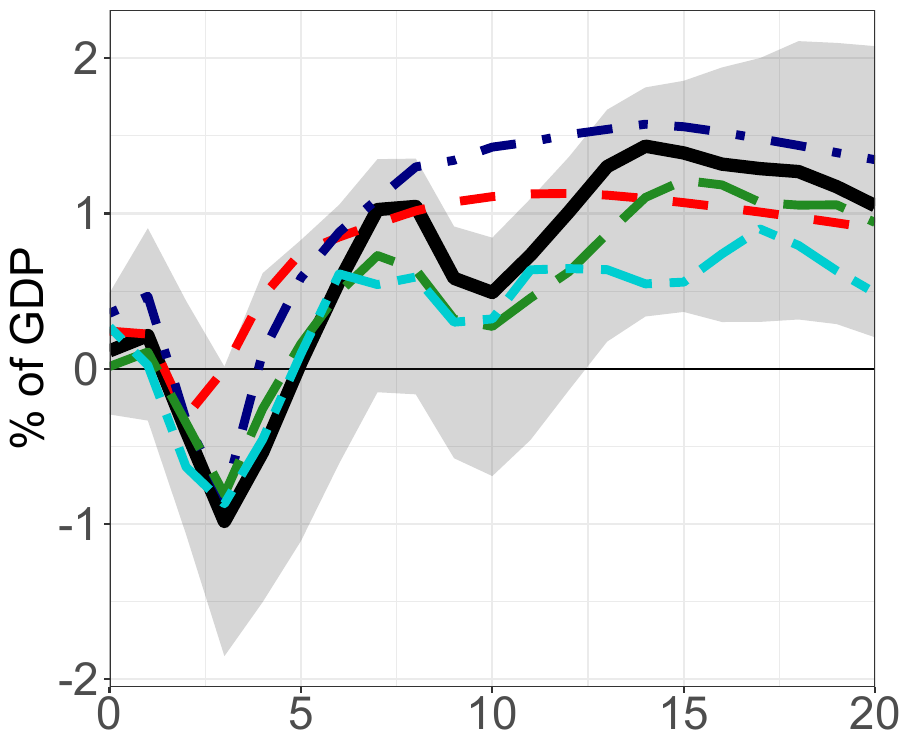}}\hfill
\subfloat[$g \rightarrow rer$]{\includegraphics[width=0.25\textwidth]{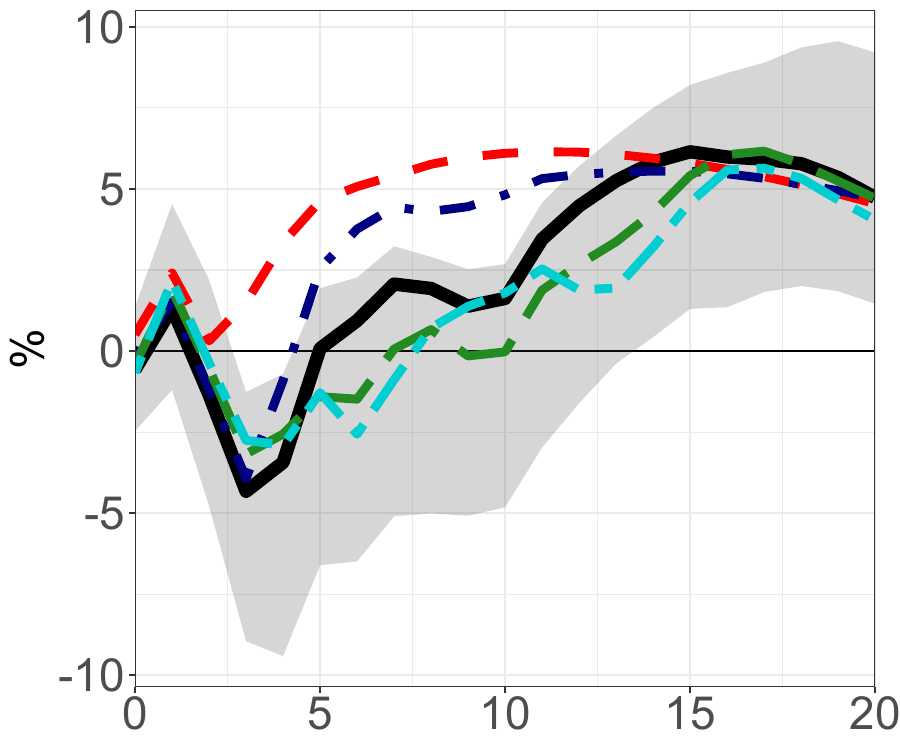}}\hfill
\subfloat[$g \rightarrow cab$]{\includegraphics[width=0.25\textwidth]{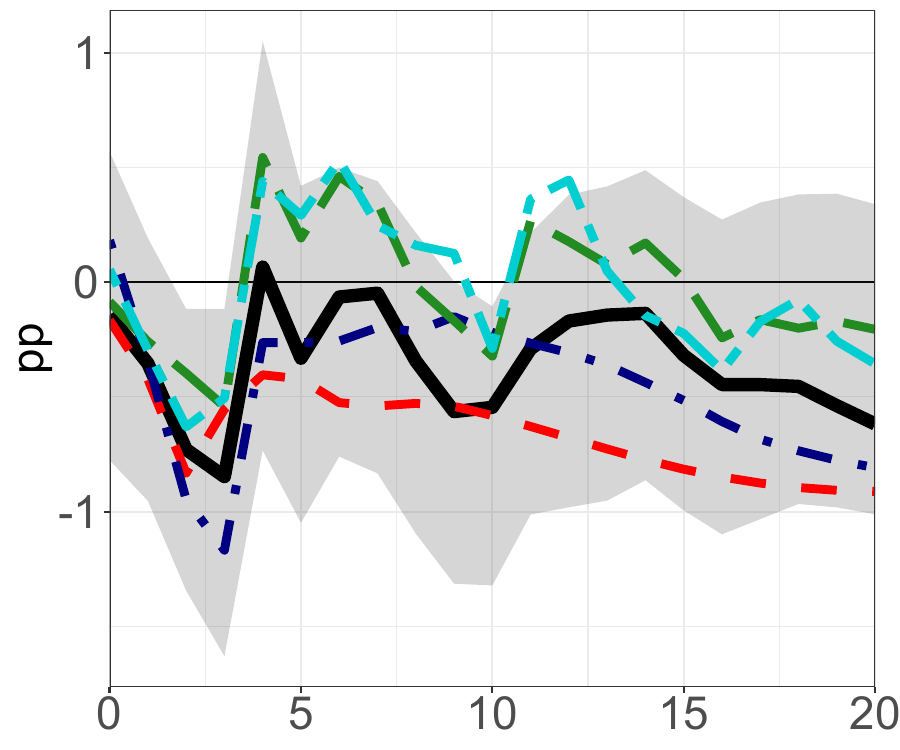}}\hfill
\subfloat[$g \rightarrow srate$]{\includegraphics[width=0.25\textwidth]{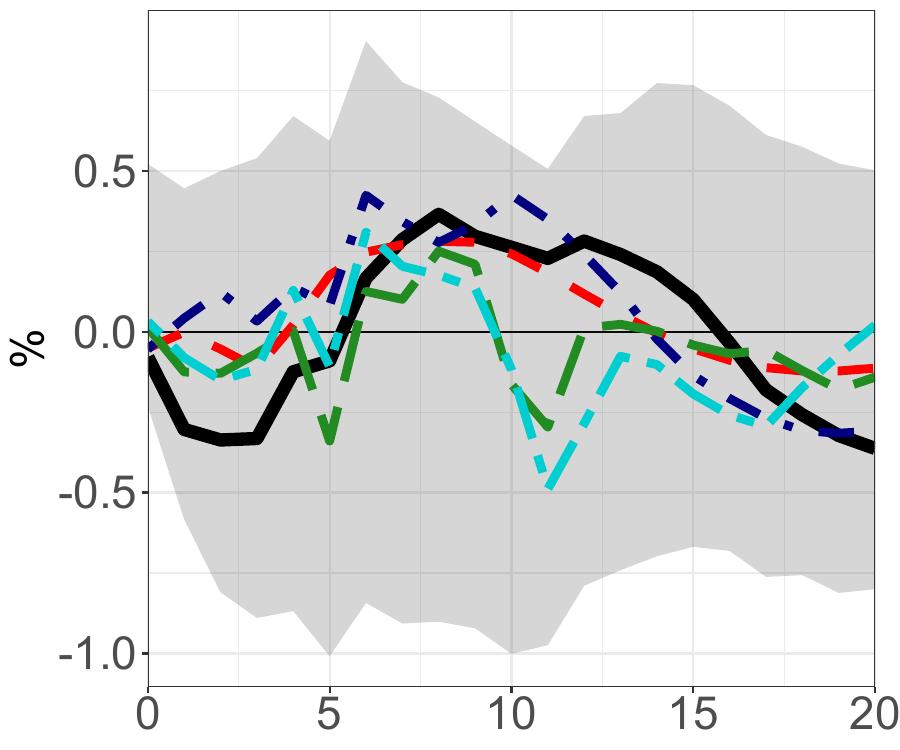}}\hfill

\vspace{8mm}

\subfloat[$r \rightarrow g$]{\includegraphics[width=0.25\textwidth]{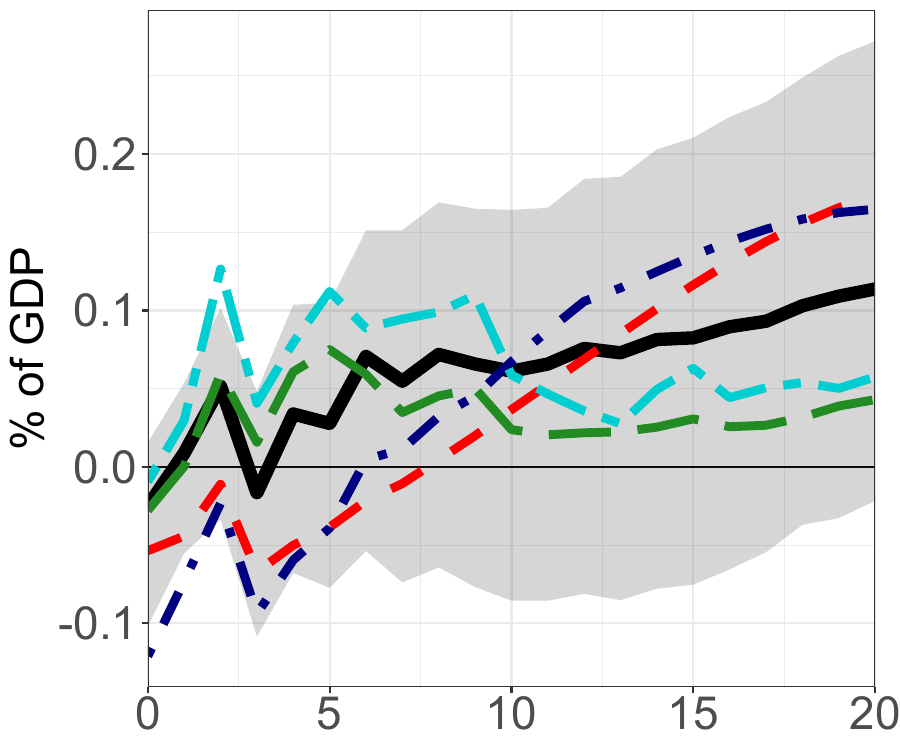}}\hfill
\subfloat[$r \rightarrow r$]{\includegraphics[width=0.25\textwidth]{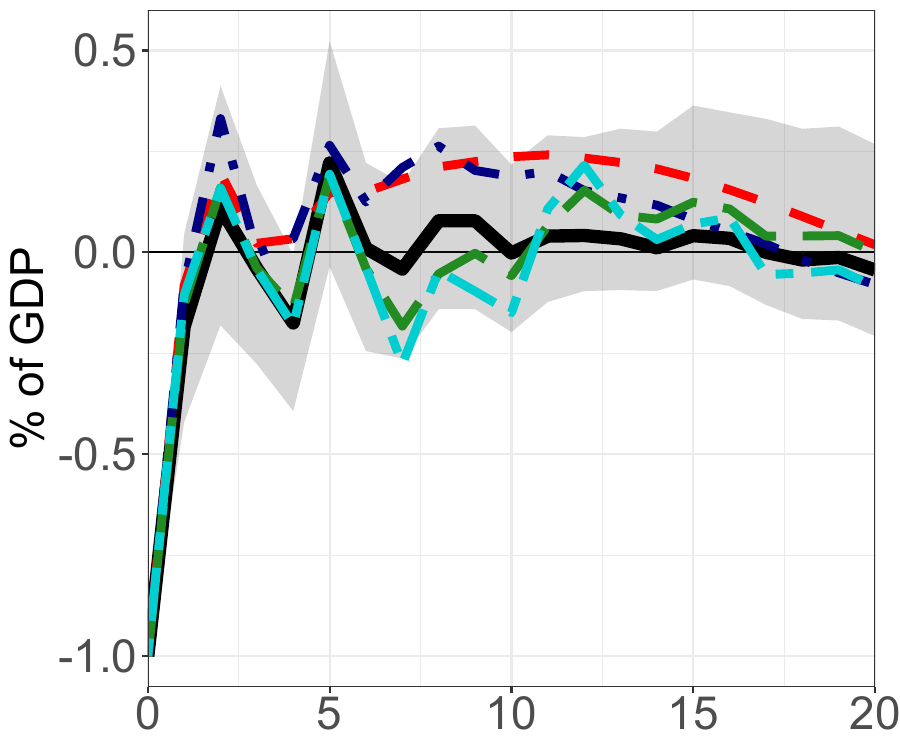}}\hfill
\subfloat[$r \rightarrow (r-g)$]{\includegraphics[width=0.25\textwidth]{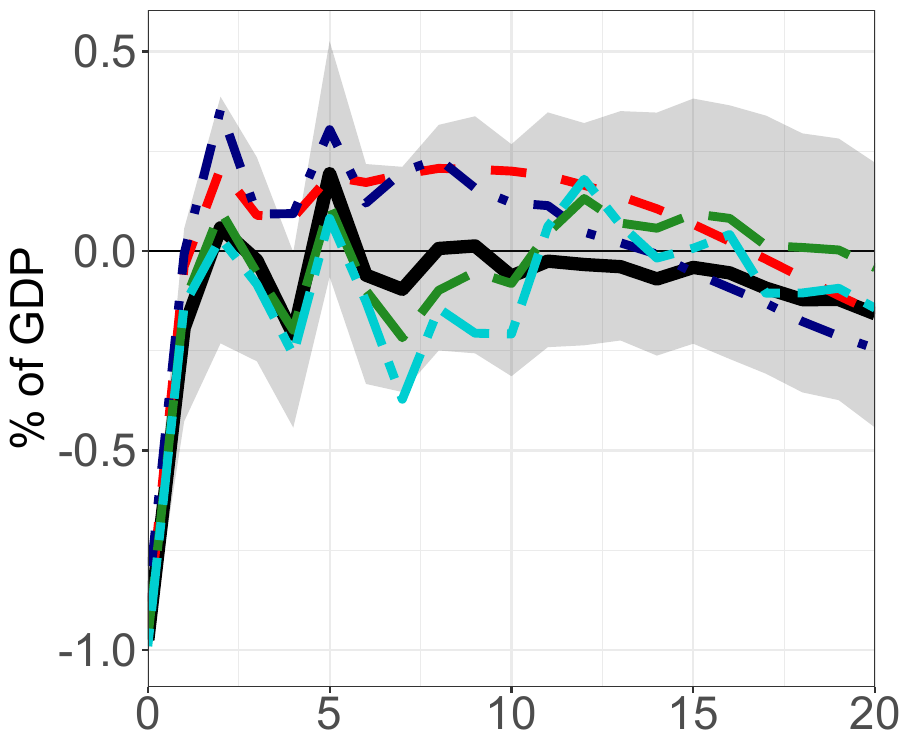}}\hfill
\subfloat[$r \rightarrow gdp$]{\includegraphics[width=0.25\textwidth]{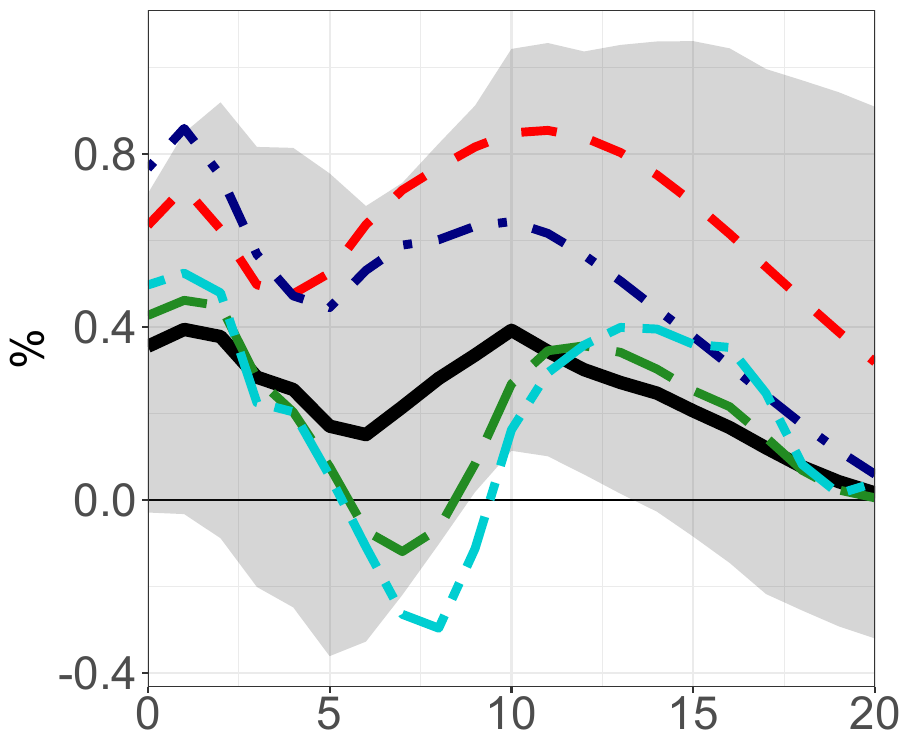}}\hfill
\vspace{2mm}
\subfloat[$r \rightarrow defl$]{\includegraphics[width=0.25\textwidth]{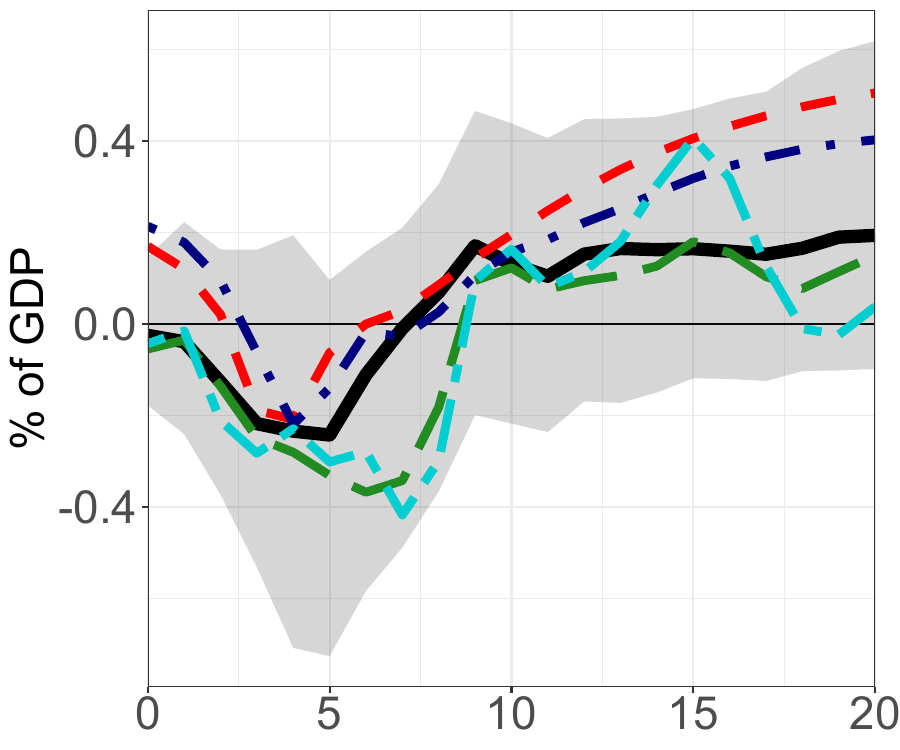}}\hfill
\subfloat[$r \rightarrow rer$]{\includegraphics[width=0.25\textwidth]{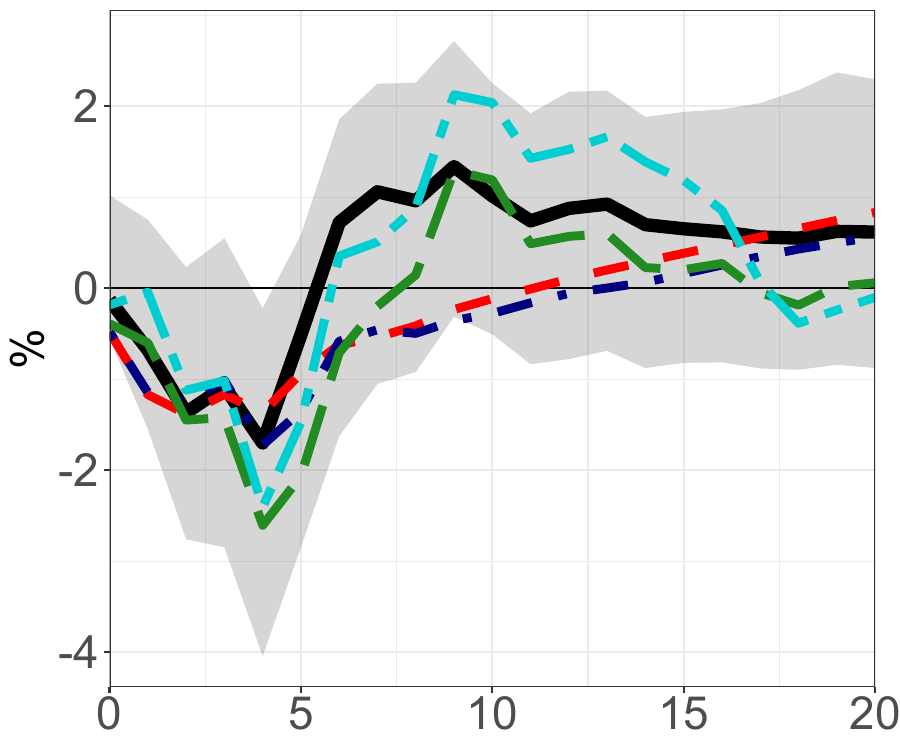}}\hfill
\subfloat[$r \rightarrow cab$]{\includegraphics[width=0.25\textwidth]{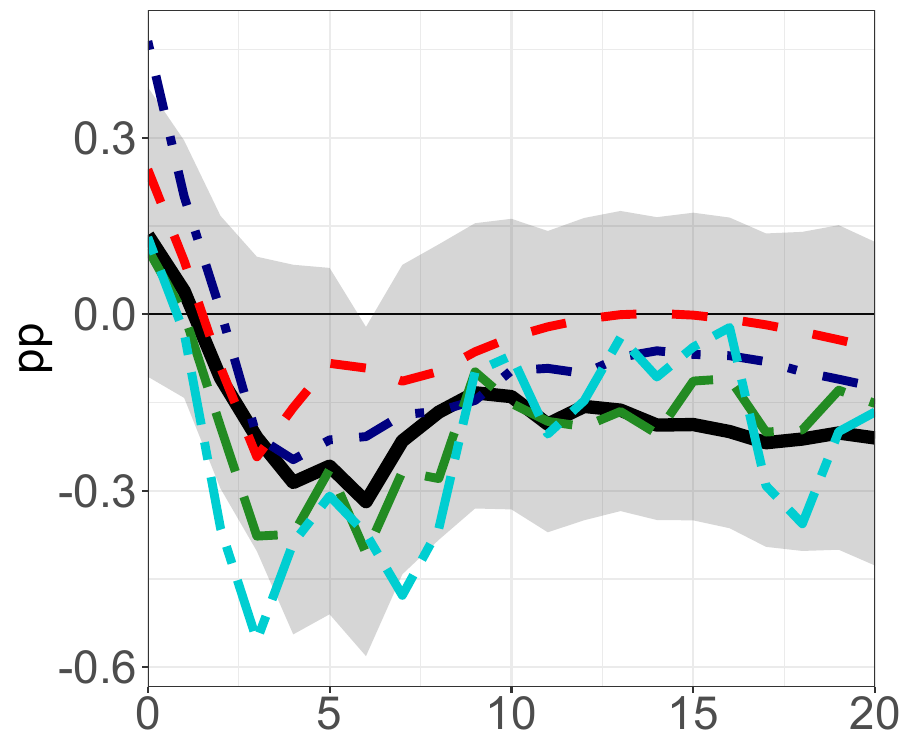}}\hfill
\subfloat[$r \rightarrow srate$]{\includegraphics[width=0.25\textwidth]{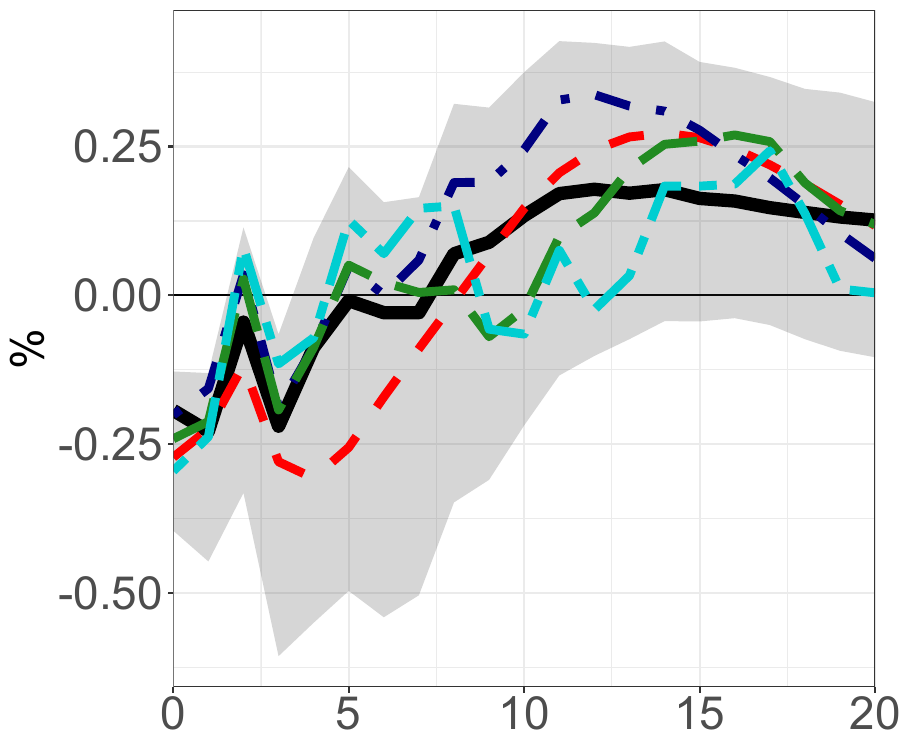}}\hfill

\subfloat[]{\includegraphics[width=0.6\textwidth]{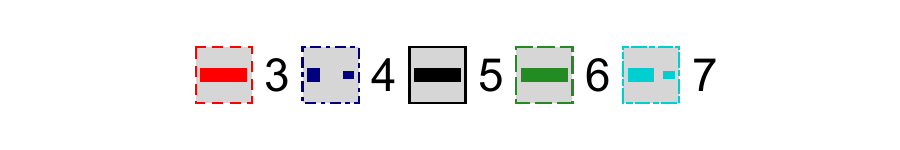}}\hfill

\vspace{-2mm}

\end{center}
{\footnotesize{\textit{Notes:} This figure plots SVAR impulse responses to fiscal shocks that either increase $g$ by 1\% of GDP or decrease $r$ by 1\% of GDP. Baseline VAR specification has 5 lags of $g$, $r$, $gdp$, $defl$, $rer$, $cab$, $srate$, $f_{\Delta g}$ and $f_{\Delta gdp}$ as defined in the text. Sample is 1985Q1-2019Q4. Black line and shaded area represent impulse responses and confidence intervals for baseline $CK$ identification while colored lines are IRFs using different lag lengths in the reduced form VAR. Moving block bootstrap 0.68 confidence intervals are for baseline specification (5 lags).  Horizontal axis has quarters from 0 to 20. \par}}
\end{figure}

\begin{figure}[!htpb]
\vspace{-2cm}
\captionsetup[subfigure]{labelformat=empty}
\begin{center}
\caption{SVAR impulse responses to 1\% of GDP fiscal shocks, Euro area countries.}\label{fig:lagrobust_eur}
\subfloat[$g \rightarrow g$]{\includegraphics[width=0.25\textwidth]{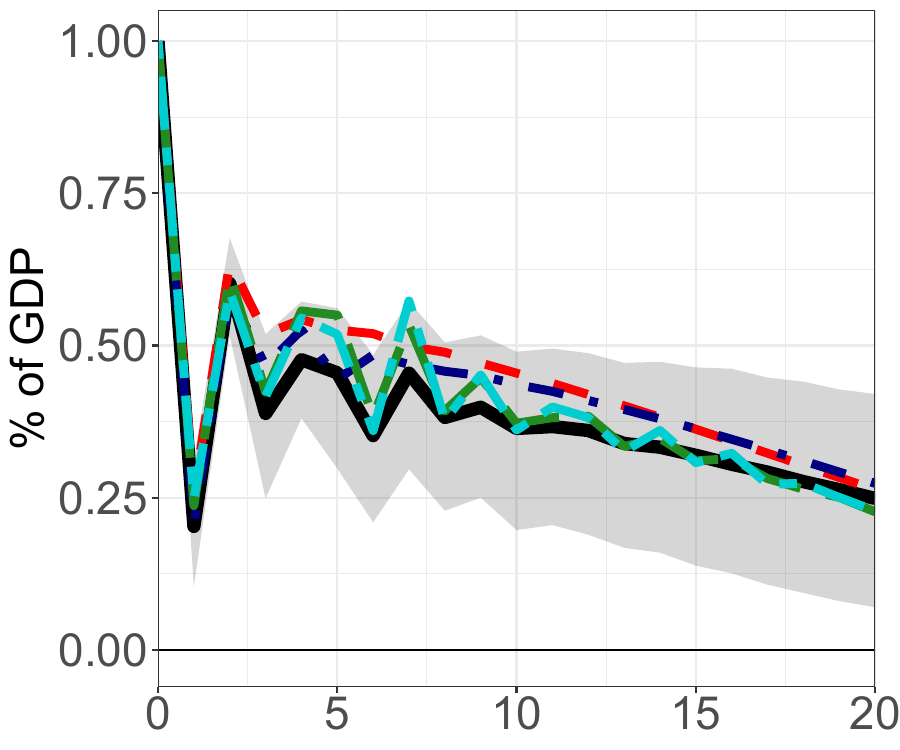}}\hfill
\subfloat[$g \rightarrow r$]{\includegraphics[width=0.25\textwidth]{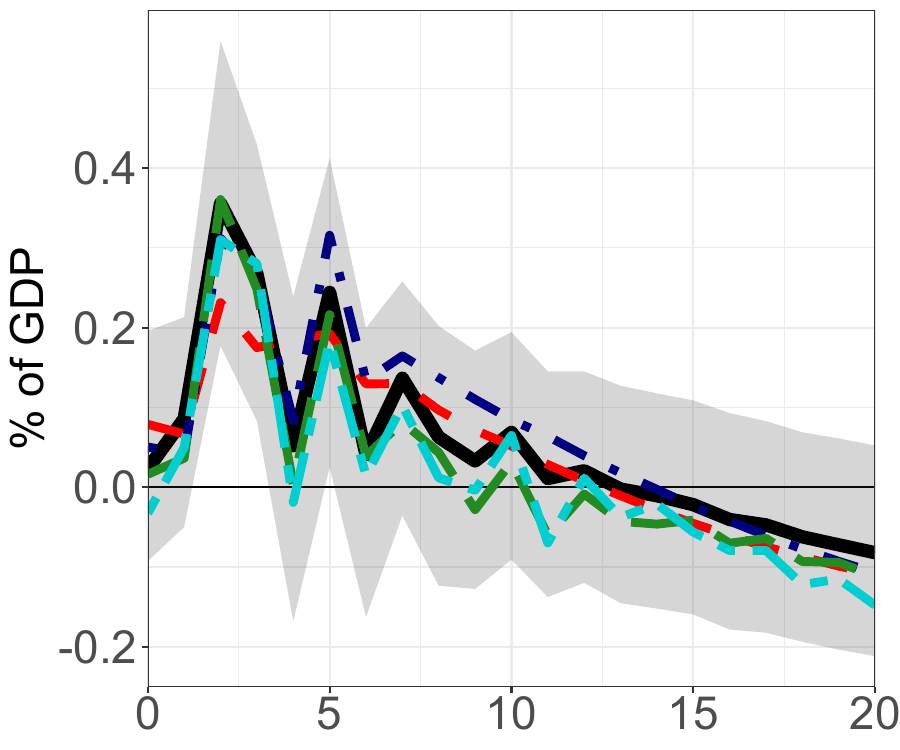}}\hfill
\subfloat[$g \rightarrow gdp$]{\includegraphics[width=0.25\textwidth]{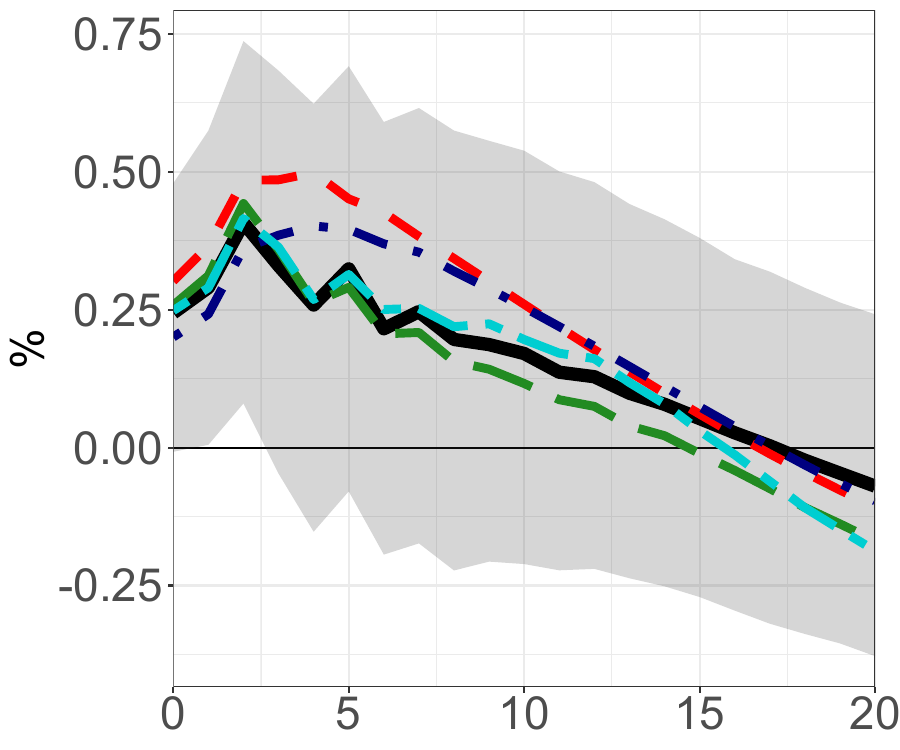}}
\vspace{1mm}
\subfloat[$g \rightarrow defl$]{\includegraphics[width=0.25\textwidth]{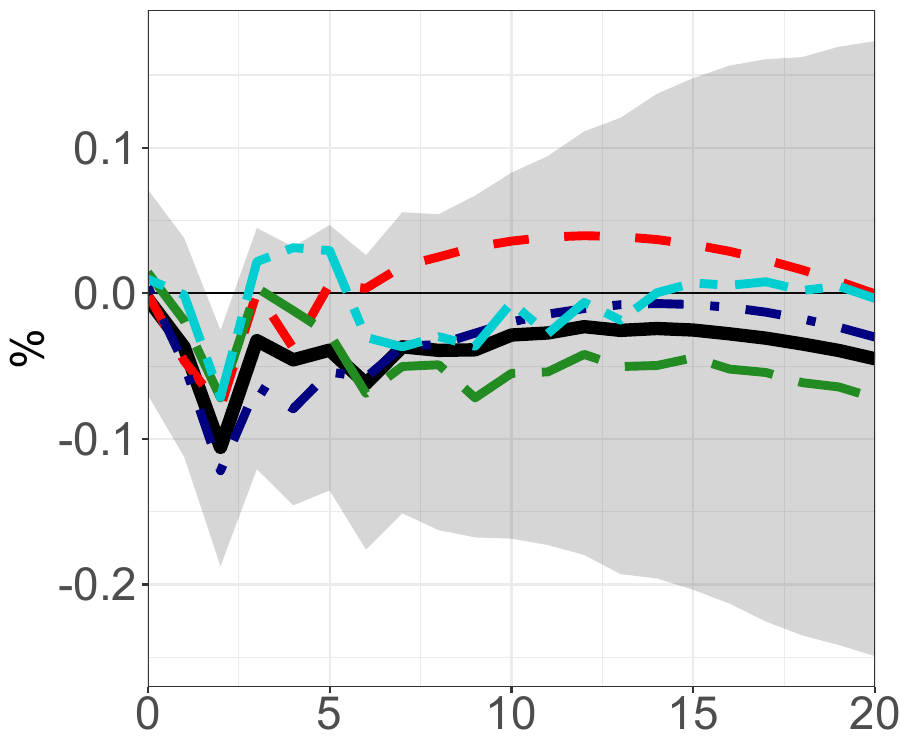}}\hfill
\subfloat[$g \rightarrow (r-g)$]{\includegraphics[width=0.25\textwidth]{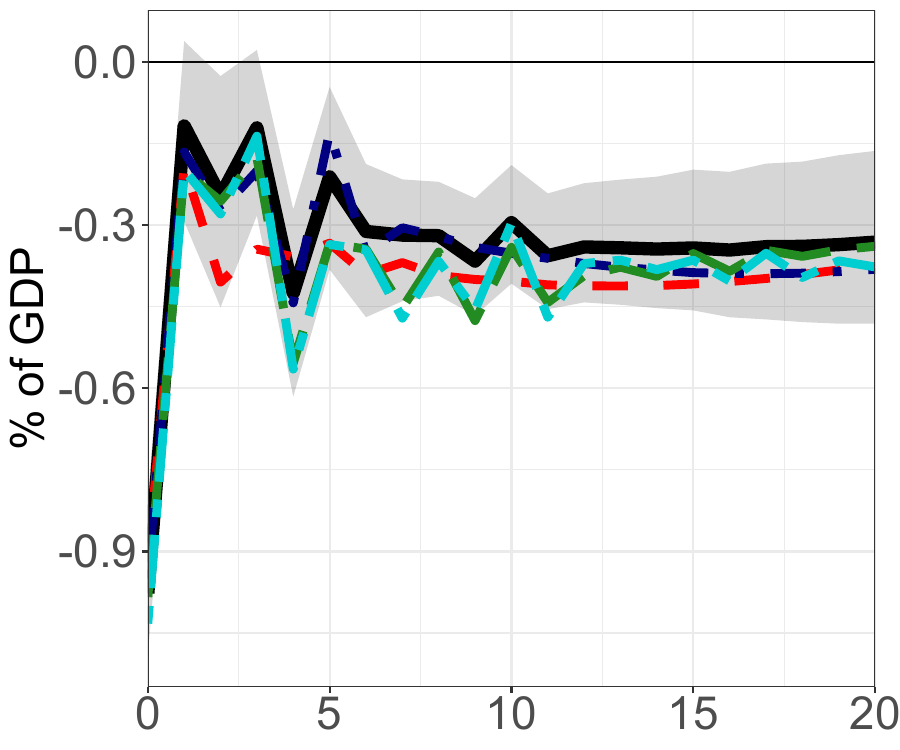}}\hfill
\subfloat[$g \rightarrow cab$]{\includegraphics[width=0.25\textwidth]{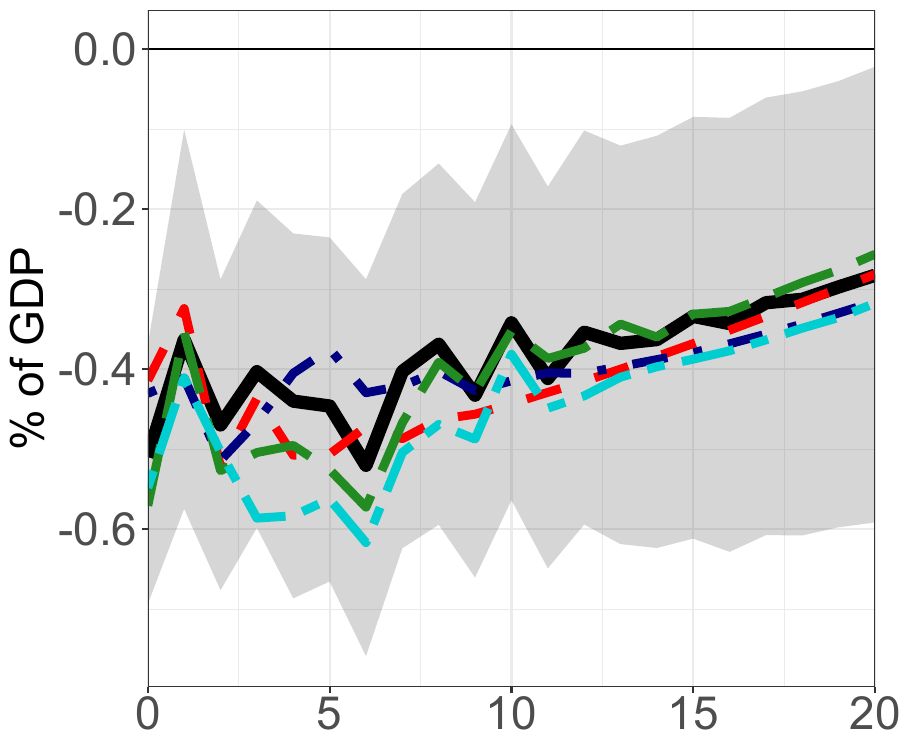}}

\vspace{8mm}

\subfloat[$r \rightarrow g$]{\includegraphics[width=0.25\textwidth]{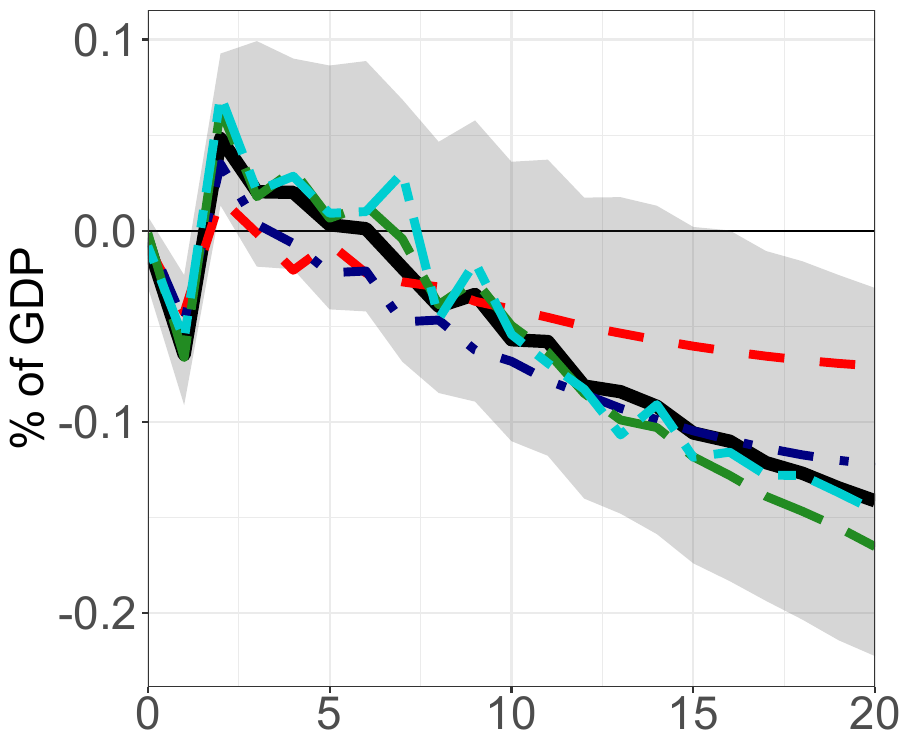}}\hfill
\subfloat[$r \rightarrow r$]{\includegraphics[width=0.25\textwidth]{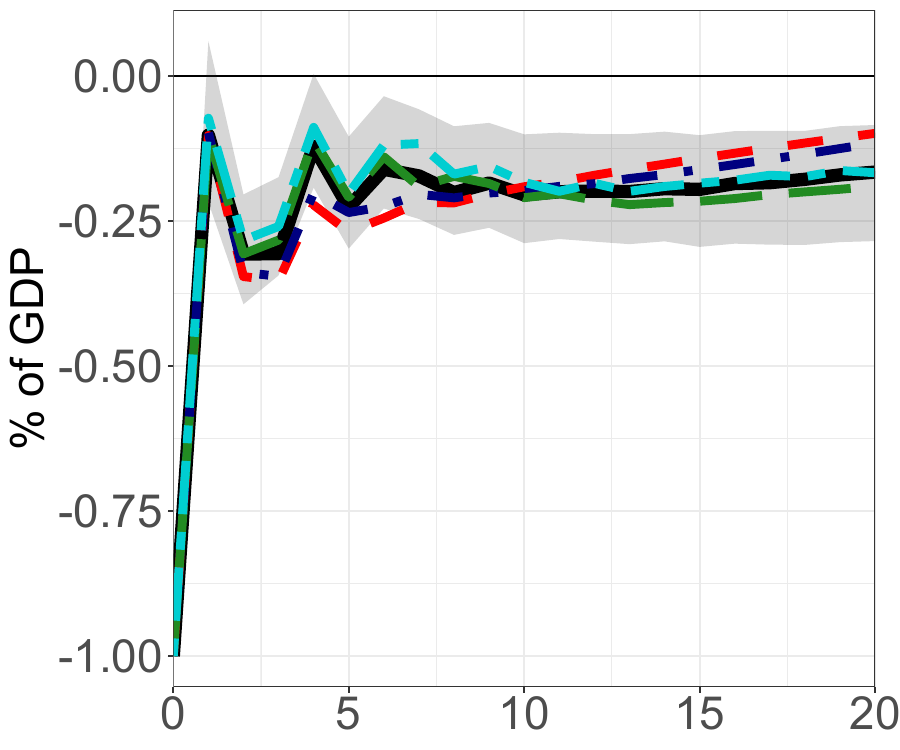}}\hfill
\subfloat[$r \rightarrow gdp$]{\includegraphics[width=0.25\textwidth]{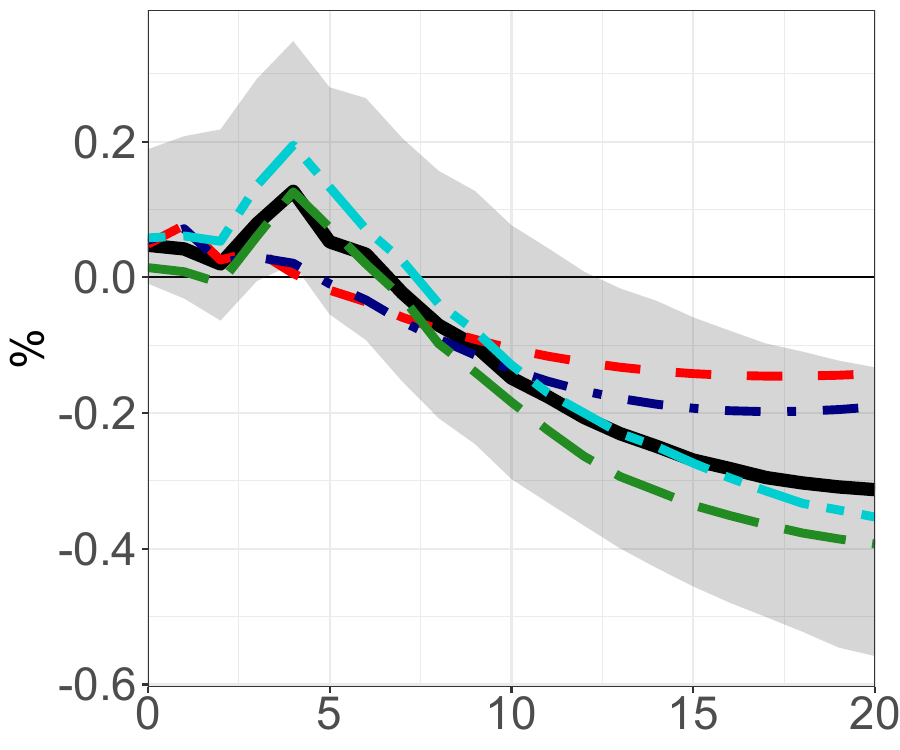}}
\vspace{1mm}
\subfloat[$r \rightarrow defl$]{\includegraphics[width=0.25\textwidth]{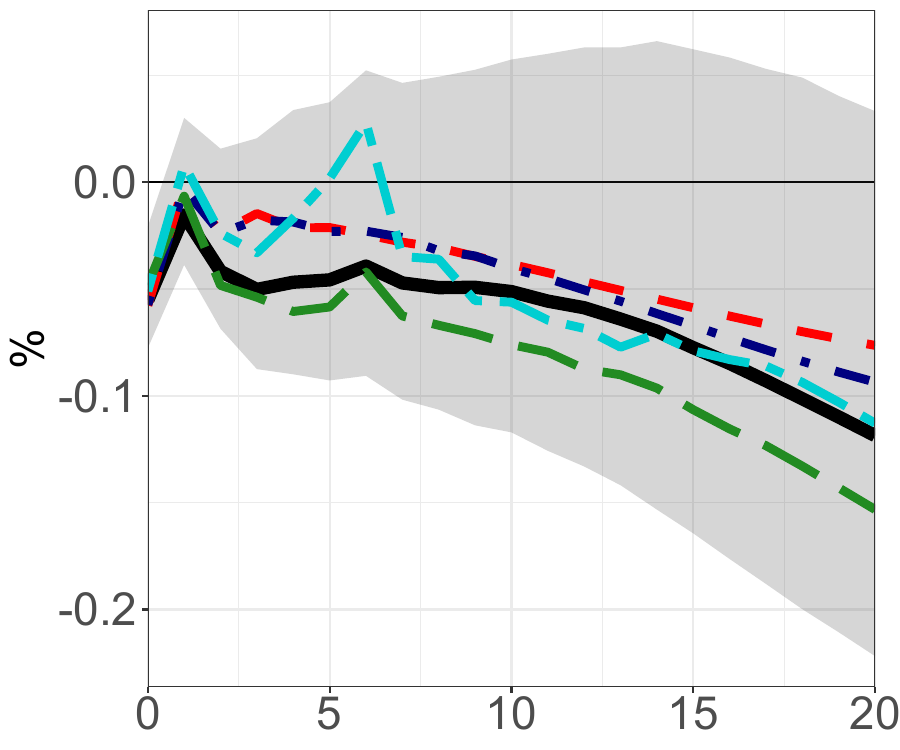}}\hfill
\subfloat[$r \rightarrow (r-g)$]{\includegraphics[width=0.25\textwidth]{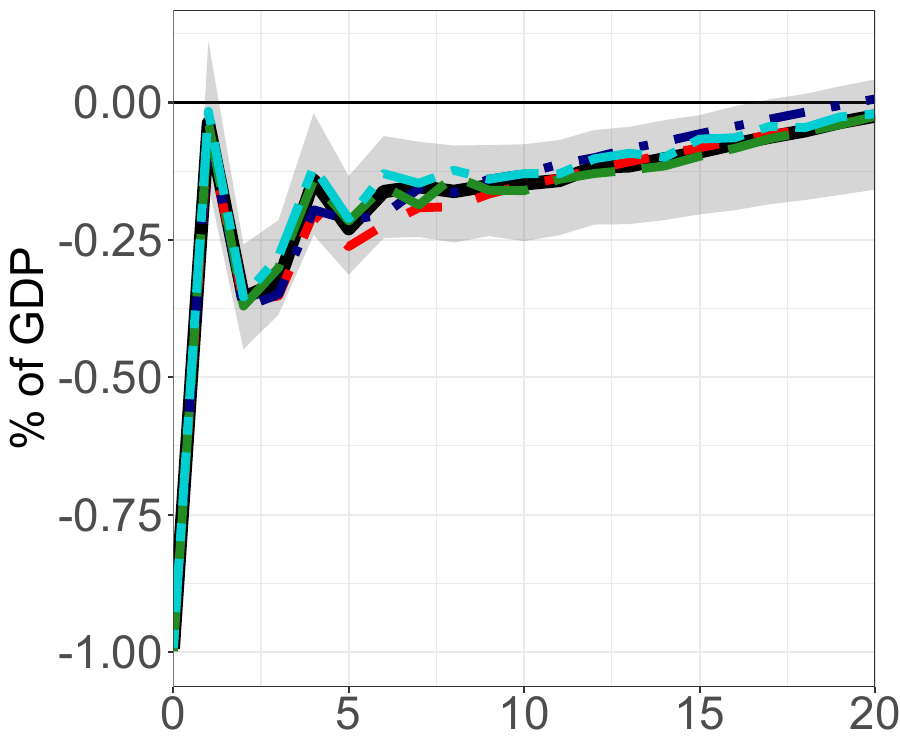}}\hfill
\subfloat[$r \rightarrow cab$]{\includegraphics[width=0.25\textwidth]{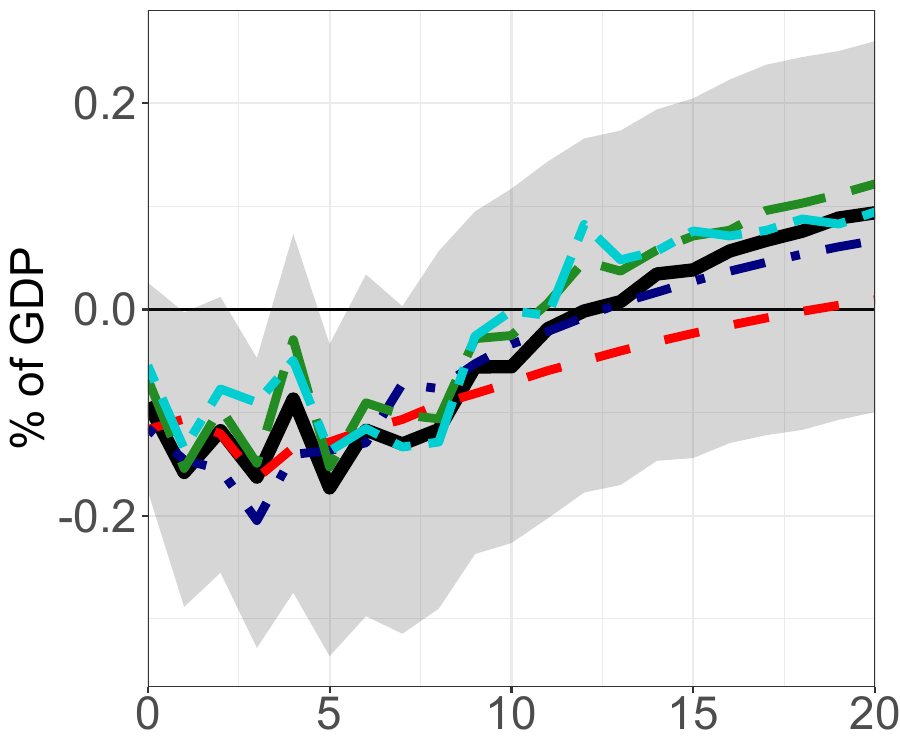}}

\vspace{-8mm}

\subfloat[]{\includegraphics[width=0.6\textwidth]{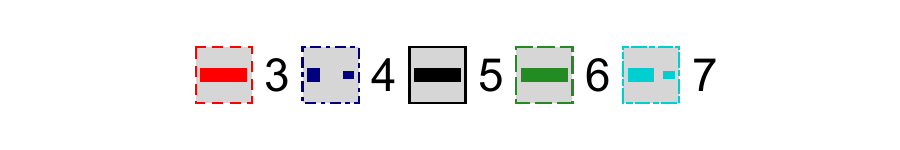}}\hfill

\vspace{-2mm}

\end{center}
{\footnotesize{\textit{Notes:} This figure plots SVAR impulse responses to fiscal shocks that either increase $g$ by 1\% of GDP or decrease $r$ by 1\% of GDP. Baseline VAR specification has 5 lags of $g$, $r$, $gdp$, $defl$, $rer$, $cab$, $srate$, $f_{\Delta g}$ and $f_{\Delta gdp}$ as defined in the text. $rer$ and $srate$ are considered as exogenous. Sample is 1999Q1-2019Q4. Black line and shaded area represent impulse responses and confidence intervals for baseline $CK$ identification while colored lines are IRFs using different lag lengths in the reduced form VAR. Moving block bootstrap 0.68 confidence intervals are for baseline specification (5 lags).  Horizontal axis has quarters from 0 to 20. \par}}
\end{figure}

Figures \ref{fig:lagrobust_can} and \ref{fig:lagrobust_eur} depict the impulse responses from SVARs with different lag lengths and different identification schemes for Canada and the euro area small open economies. Overall, it seems that the results are robust to choices related to reduced form VAR specification, for example, different lag lengths and whether the endogenous variables are in log levels or detrended.

\begin{figure}[!htpb]
\vspace{-2cm}
\captionsetup[subfigure]{labelformat=empty}
\begin{center}
\caption{SVAR impulse responses to 1\% of GDP fiscal shocks, Euro area countries.}\label{fig:leaveout_eur}
\subfloat[$g \rightarrow g$]{\includegraphics[width=0.25\textwidth]{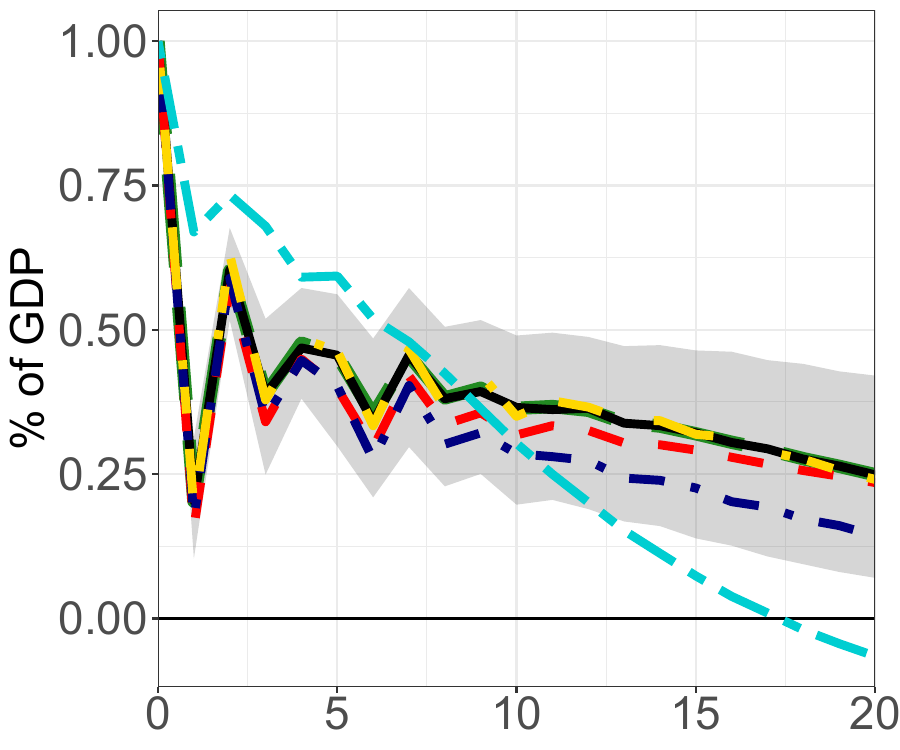}}\hfill
\subfloat[$g \rightarrow r$]{\includegraphics[width=0.25\textwidth]{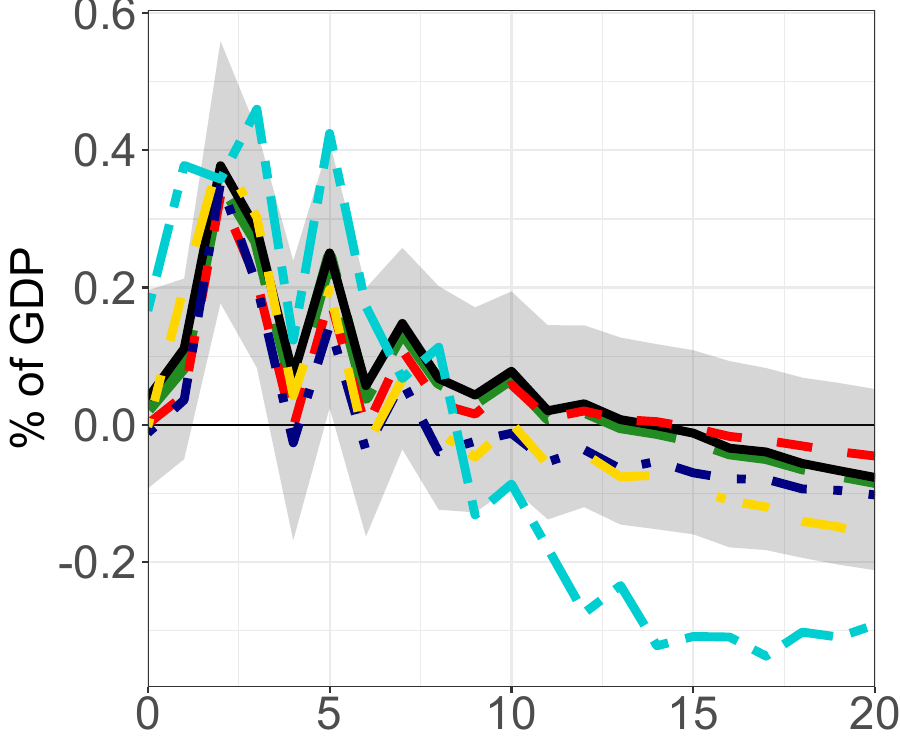}}\hfill
\subfloat[$g \rightarrow gdp$]{\includegraphics[width=0.25\textwidth]{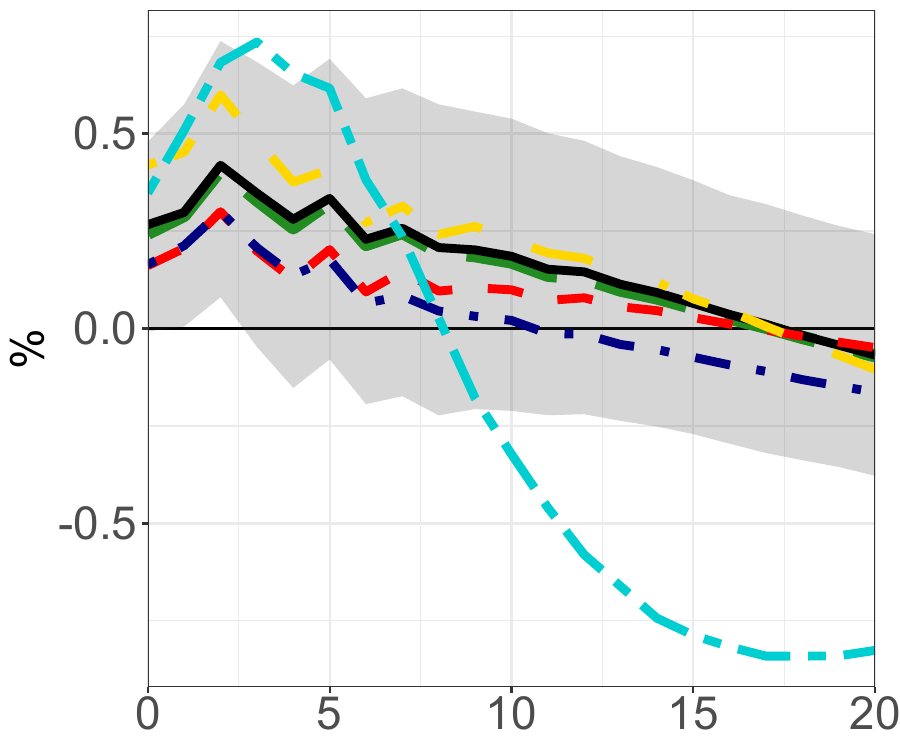}}
\vspace{1mm}
\subfloat[$g \rightarrow defl$]{\includegraphics[width=0.25\textwidth]{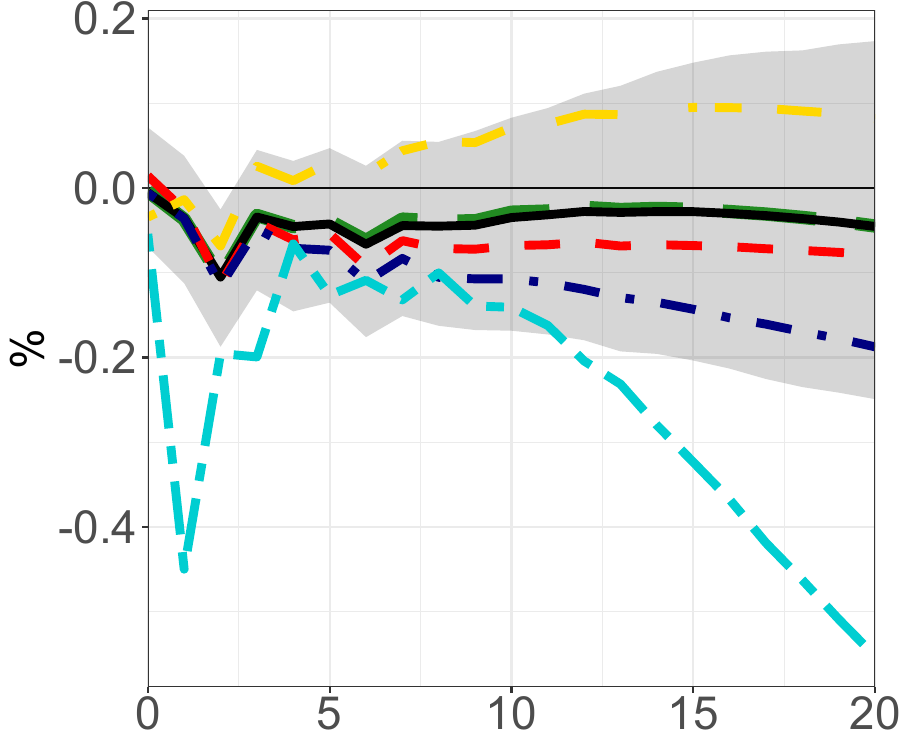}}\hfill
\subfloat[$g \rightarrow (r-g)$]{\includegraphics[width=0.25\textwidth]{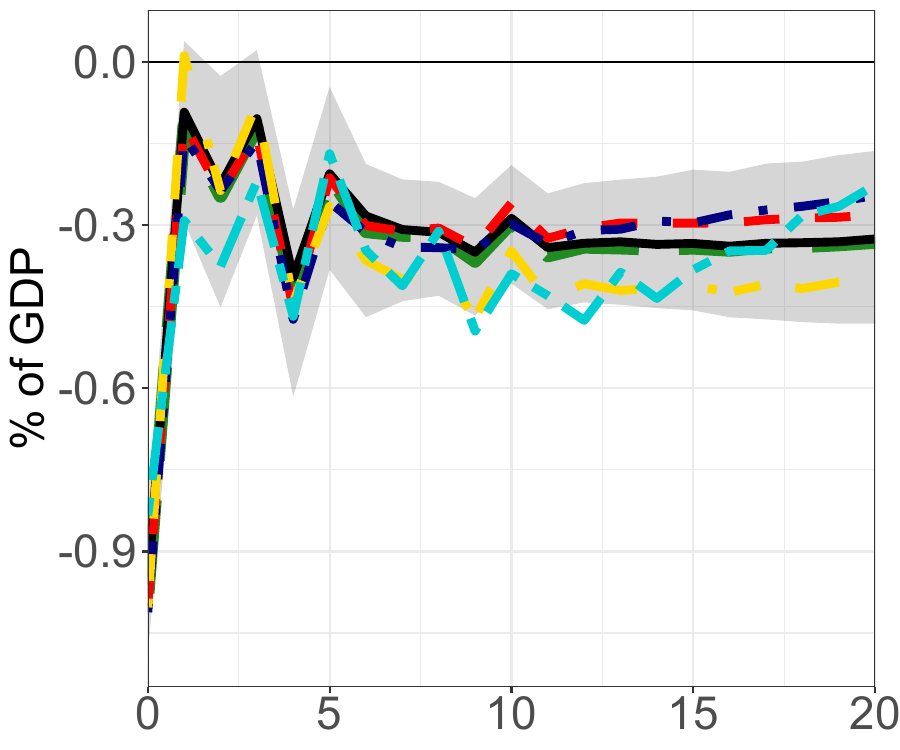}}\hfill
\subfloat[$g \rightarrow cab$]{\includegraphics[width=0.25\textwidth]{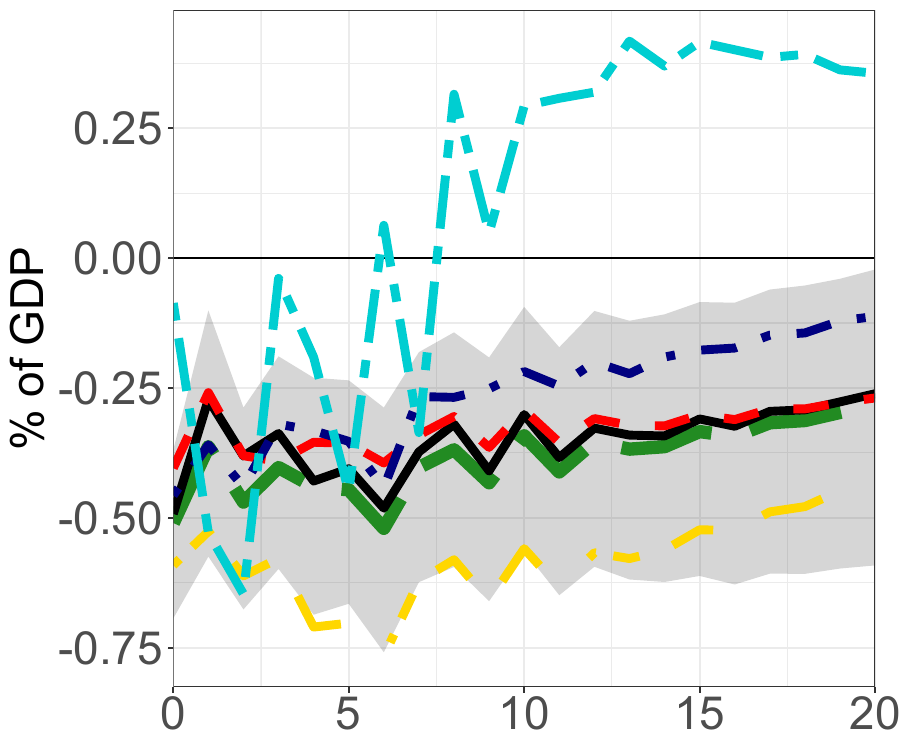}}

\vspace{8mm}

\subfloat[$r \rightarrow g$]{\includegraphics[width=0.25\textwidth]{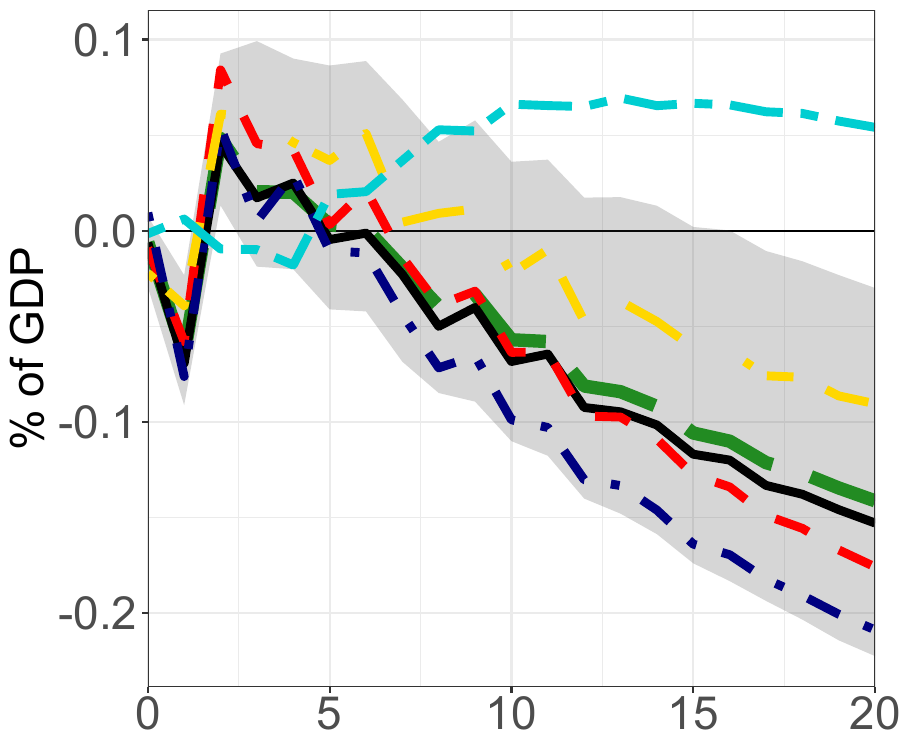}}\hfill
\subfloat[$r \rightarrow r$]{\includegraphics[width=0.25\textwidth]{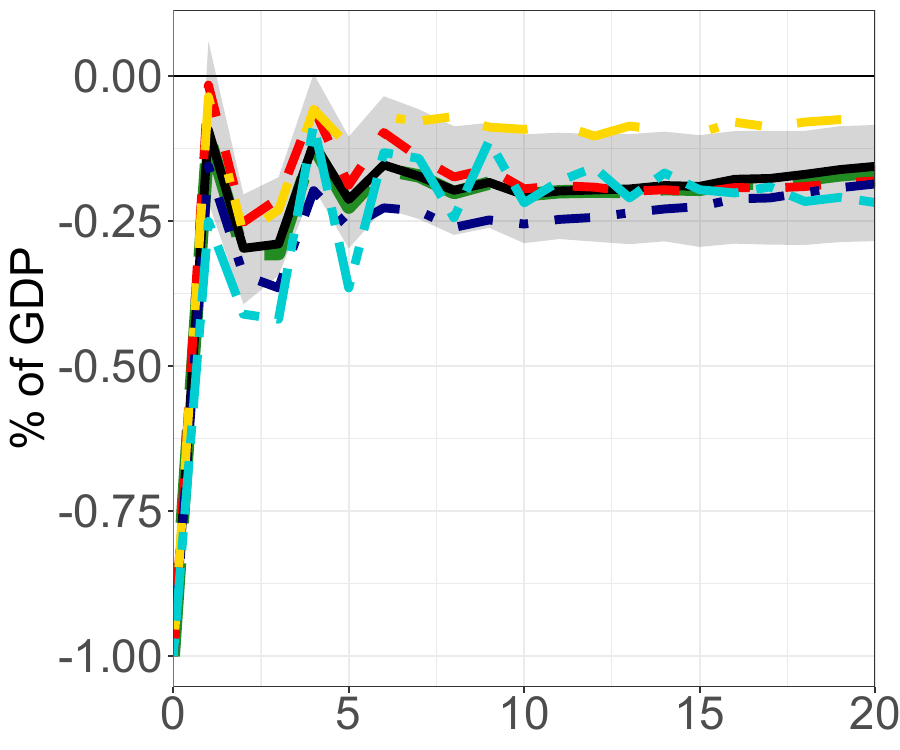}}\hfill
\subfloat[$r \rightarrow gdp$]{\includegraphics[width=0.25\textwidth]{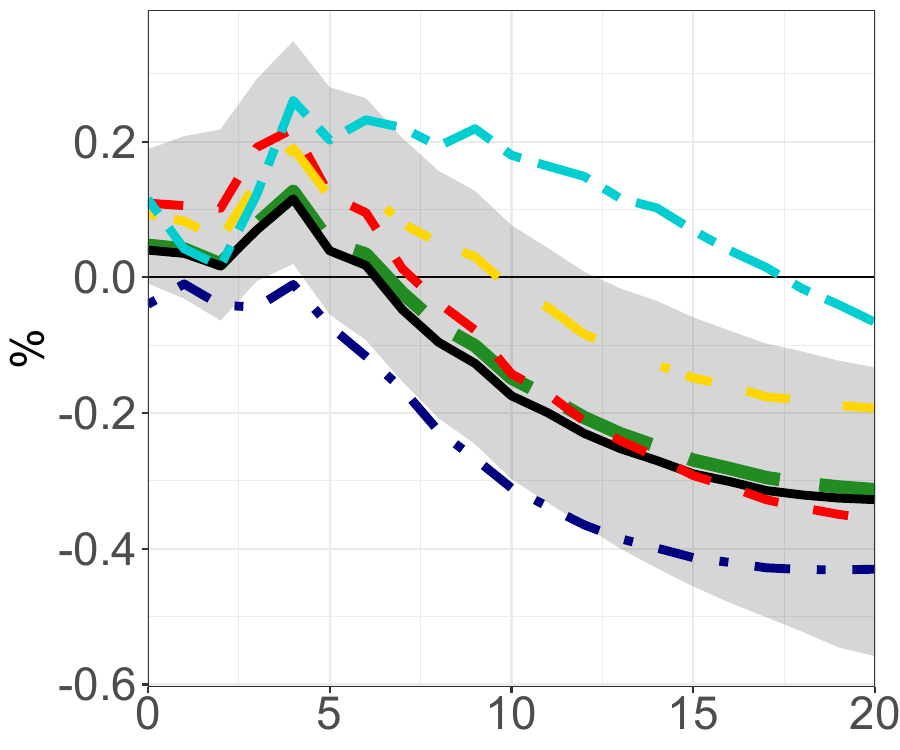}}
\vspace{1mm}
\subfloat[$r \rightarrow defl$]{\includegraphics[width=0.25\textwidth]{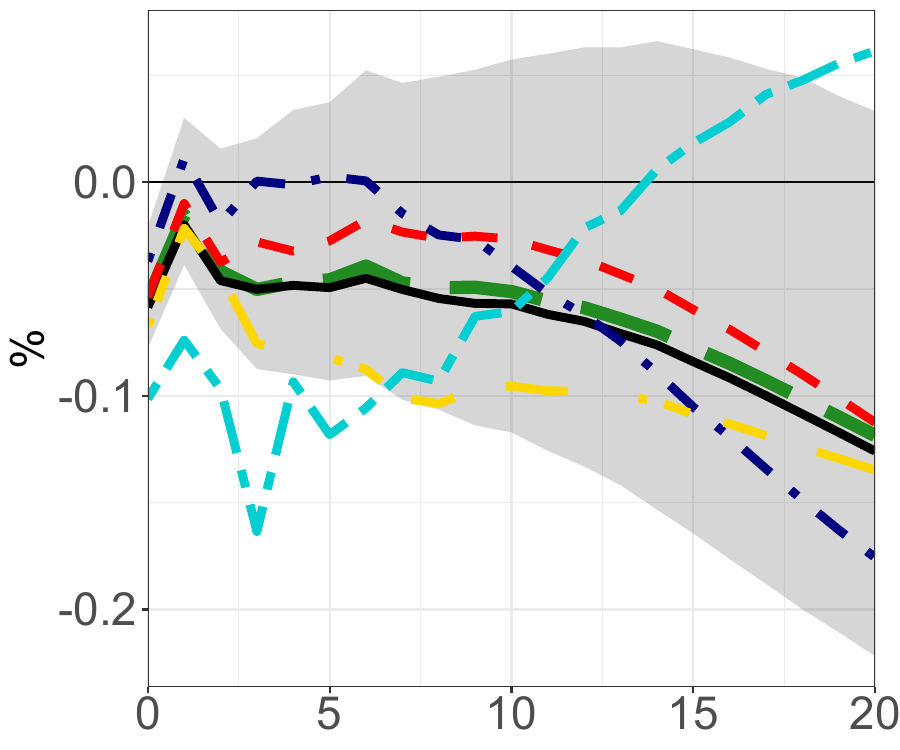}}\hfill
\subfloat[$r \rightarrow (r-g)$]{\includegraphics[width=0.25\textwidth]{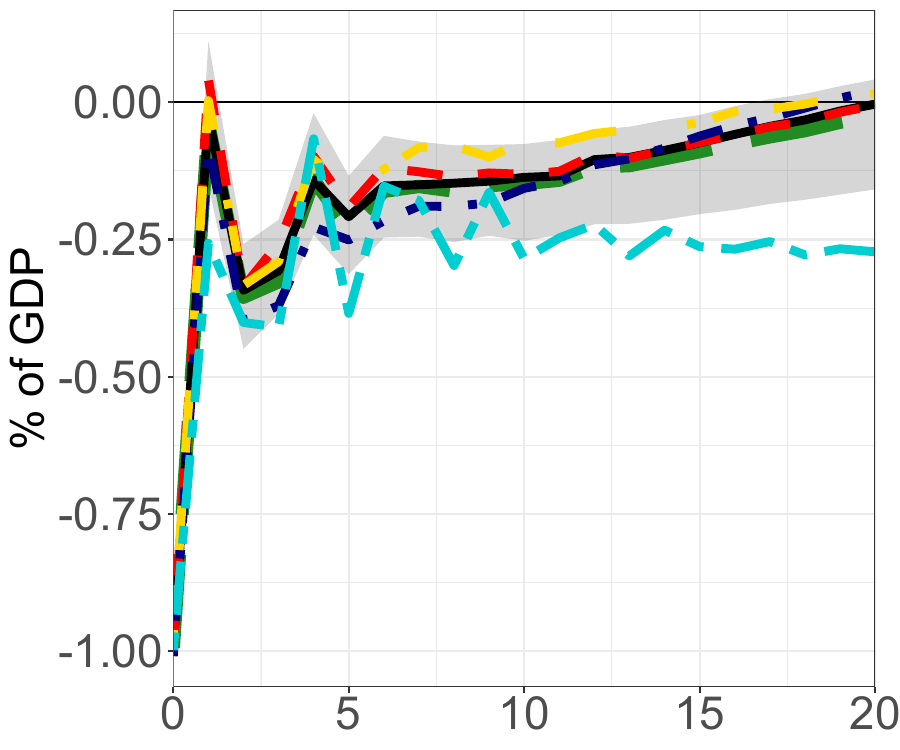}}\hfill
\subfloat[$r \rightarrow cab$]{\includegraphics[width=0.25\textwidth]{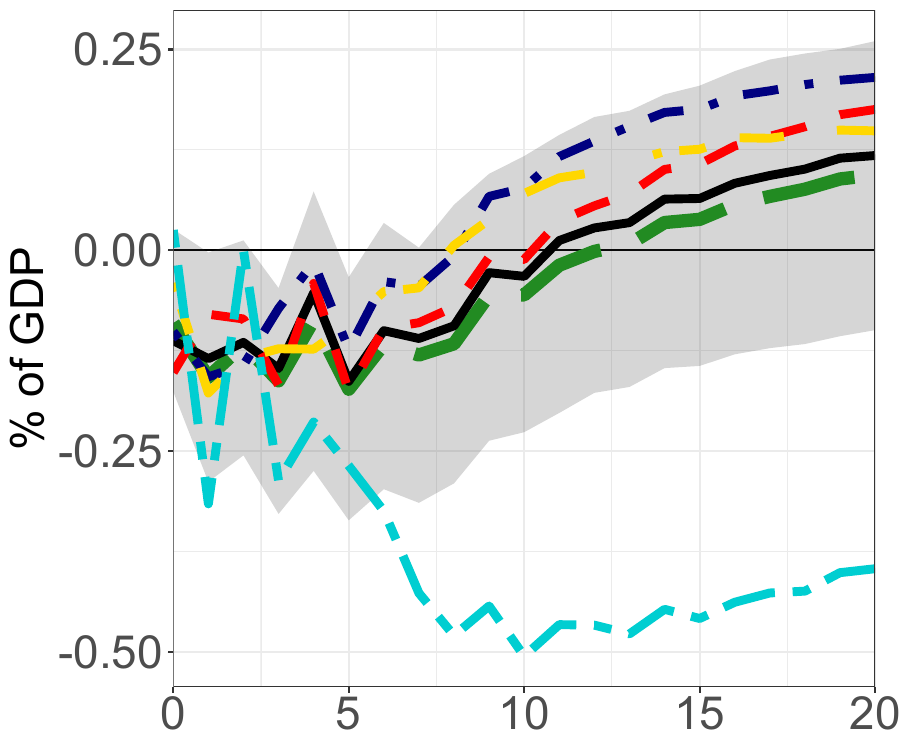}}

\vspace{-8mm}

\subfloat[]{\includegraphics[width=0.6\textwidth]{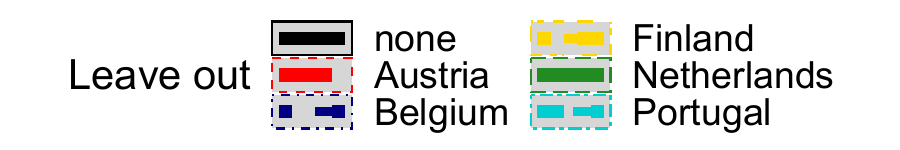}}\hfill

\vspace{-2mm}

\end{center}
{\footnotesize{\textit{Notes:} This figure plots SVAR impulse responses to fiscal shocks that either increase $g$ by 1\% of GDP or decrease $r$ by 1\% of GDP. Baseline VAR specification has 5 lags of $g$, $r$, $gdp$, $defl$, $rer$, $cab$, $srate$, $f_{\Delta g}$ and $f_{\Delta gdp}$ as defined in the text. $rer$ and $srate$ are considered as exogenous. Sample is 1999Q1-2019Q4. Black line and shaded area represent impulse responses and confidence intervals for baseline $CK$ identification while colored lines are IRFs when one of the Euro area countries are dropped from the sample. Moving block bootstrap 0.68 confidence intervals are for baseline specification (5 lags).  Horizontal axis has quarters from 0 to 20. \par}}
\end{figure}

\autoref{fig:leaveout_eur} depicts impulse responses to spending and revenue shocks in the cases where each country is excluded from the sample at a time. It turns out that, indeed, the observations for Portugal are influential. In all other cases, the impulse responses stay pretty much the same. This is troubling because it might be that Portugal's observations add variation to the sample that describes the representative small open economy of the monetary union, which the other country's observations lack due to short samples. On the other hand, Portugal, as one of the GIIPS countries, might react differently to fiscal shocks compared to the other countries in the sample.\footnote{The GIIPS countries are Greece, Ireland, Italy, Portugal, and Spain. All of these countries ran into trouble after the financial crisis due to weak economic and financial performances largely because these countries had developed high levels of debt before the financial crisis. Furthermore, the GIIPS countries were running a deficit already before joining the euro and the strong currency likely magnified their deficits due to impaired competitiveness.}

In studies focusing on the twin-deficit hypothesis in the Euro area, the GIIPS countries are often studied separately because of their distinct features, namely high budget deficits and high debt levels, see, for example, \citetsupp{algieri2013empirical} and \citetsupp{litsios2017empirical}. Considering the GIIPS countries, there is no consensus about the relationship between fiscal and current account deficits. However, some papers find, for example, \citetsupp{algieri2013empirical}, that there is no clear relation between budget and current account balances. This in turn is in line with the Ricardian theory. Considering VARs, for example, \cite{kim2008twin} also find evidence that a Ricardian move of private savings increasing in response to an increase in government deficit seems to partly explain the twin-divergence they find in the US data. \citetsupp{hurtgen2014} develop a small open economy model where at high debt-to-GDP levels private households facing uncertainty of future high taxes increase savings instead of accumulating debt to smooth consumption during economic downturns. Thereby they partially compensate current account deficits that result from budget deficits. 

The Ricardian theory could at least partly explain why Portugal's observations affect the response of GDP to a net revenue shock so strongly that it flips the sign of the response. Another explanation might be related to the austerity program Portugal was committed to by the EU Commission, ECB, and IMF after Portugal's bailout in 2011. \citetsupp{litsios2017empirical} finds supportive evidence that fiscal austerity can reduce current account balances in Greece, Portugal, and Spain. The fact that Portugal's observations are so influential in our panel suggests that austerity conducted by raising taxes might benefit the GIIPS countries because of rather small effects on the current account but positive effects on GDP in the long run possible because of the Ricardian effect. However, this aspect should be examined more carefully in future studies. 

As figure \autoref{fig:leaveout_eur} in the \hyperref[sec:online]{Online Appendix} depicts, when Portugal's observations are excluded, government spending shock first boosts GDP but, after roughly two years, starts to slow GDP down. The current account's impulse response is no longer negative for the whole period. And also, inflation slows down together with GDP roughly after two years. A revenue shock tends to raise government spending in the long run. Inflation first slows down but, in the long run, accelerates. The effect on the current account is negative for the whole period meaning that some of the household's excess disposable income is used on imports. While also the budget balance stays negative the whole period, a negative revenue shock slightly stimulates GDP. This is more in line with the presumption and previous findings. Considering a revenue shock, the results now align with the twin-deficits hypothesis. Whereas a shock to government spending results in a budget deficit, but while the current account starts negative, after roughly two years, it turns positive.

\newpage

\newpage

\newpage

\newpage

\newpage

\newpage

\newpage

\clearpage

\singlespacing
\setlength\bibsep{0pt}
\bibliographystylesupp{econ-aea}
\bibliographysupp{references}

\end{document}